\documentclass[secnumarabic,amssymb, nobibnotes, aps, prd]{revtex4-2}
\usepackage{graphicx}
\usepackage{amsmath}
\usepackage{subcaption}

\usepackage{multirow}

\begin{document}
	
	\title{Thermodynamic Topology of $D=4,5$  Horava Lifshitz Black Hole in Two Ensembles }%

	\author{Bidyut Hazarika$^1$}
	
	\email{$rs_bidyuthazarika@dibru.ac.in$}
	
	\author{Prabwal Phukon$^{1,2}$}
	\email{prabwal@dibru.ac.in}
	
	\affiliation{$1.$Department of Physics,Dibrugarh University, Dibrugarh,Assam,786004.\\$2.$Theoretical Physics Division, Centre for Atmospheric Studies, Dibrugarh University, Dibrugarh,Assam,786004.}

	\maketitle
	\tableofcontents

\section{Abstract}
	We study the thermodynamic topology of four and five dimensional Horava Lifshitz (HL) black holes in Horava gravity. These exotic black hole solutions belong to a special class of of black holes whose thermodynamics exhibit a line of (continuous) second order phase transitions known as $\lambda$ phase transitions akin to those observed in the superfluidity of  liquid $^{4}He$. To analyze their thermodynamic topology, we treat the Horava Lifshitz (HL) black holes as topological defects in their thermodynamic spaces and compute the winding numbers at those defects. We work in two different ensembles:  fixed $\epsilon$ ensemble and fixed $\zeta$ ensemble, where $\epsilon$ is a parameter of the HL black holes and $\zeta$ is its conjugate parameter. In the fixed $\epsilon$ ensemble, three different horizon types are considered : the spherical horizon for $k=+1$, the flat horizon for $k=0$, and  the hyperbolic horizon for $k=-1$. In the fixed $\zeta$ ensemble, two different horizon types are considered : the spherical horizon for $k=+1$, and  the hyperbolic horizon for $k=-1$. Fixed $\zeta$ ensemble could not be defined in case of flat horizon with $k=0$. In $D=5$, in the fixed $\epsilon$ ensemble, the total topological charge for a black hole with  spherical horizon is $+1$ for $\epsilon < 1$ and $0$ for $\epsilon \geq 1$. At the critical value $\epsilon_c = 0.9428$, a $\lambda$-line phase transition occurs. Similar observations are made in the $D=4$, fixed $\epsilon$ ensemble, spherical horizon case where the critical value is $\epsilon=0.9785$. The total topological charge for $5D$ and $4D$ HL black holes with flat horizon remains $1$, irrespective of the values of $\epsilon$ and pressure $P$. In case of 5D HL black hole with hyperbolic horizon in the fixed $\epsilon$ ensemble, the total topological charge is $0$ for $0\leq \epsilon < 1$ and is $1$ for $\epsilon \geq 1$. One facet of the study in fixed $\epsilon$ ensemble is that the change in pressure has no significant impact on determining the overall thermodynamic topology. It primarily depends on the parameter $\epsilon$ only. In the fixed $\zeta$ ensemble, pressure $P$ becomes a crucial parameter in determining the overall thermodynamic topology of HL black holes. In fixed $\zeta$ ensemble ,$5D$ black hole with spherical horizon posseses a total topological charge of $+1$ or $0$, depending on the values of $P$ and $\zeta$. Contrastingly, for $D=4$ black holes with spherical horizon, the total topological charge is always $1$ for all values of $\zeta$ and pressure. For $5D$ and $4D$ black hole with  hyperbolic horizon, the total topological charge is either $+1$ or $0$, depending on the values of $\zeta$ and pressure. Hence, we infer that the thermodynamic topology of Horava Lifshitz black holes in $D=4,5$ are highly influenced by the type of horizon, choice of ensemble, values of pressure and parameters $\epsilon$ or $\zeta$.

\section{Introduction:}
Since its inception in 1970s, black hole thermodynamics \cite{1,2,3,4,5,6,7,8,9} has continued to evolove and has remained an active area of interest in physics. One of the directions in which a lot of focus has been attributed to over the last decade or so is the framework of extended black hole thermodynamics \cite{10,11,12,13} . In this approach, the cosmological constant is treated as a counterpart to pressure in ordinary thermodynamics. For anti-de-Sitter (AdS) black hole with cosmological constant $\Lambda$, the thermodynamic pressure  is given by,
	$$P=-\frac{\Lambda}{8\pi}$$  
	
The extended black hole thermodynamics of a number of black holes has been extensively studied throwing new insights about the rich phase structure of a wide range of  black holes\cite{14,15,16,17,18,19,20,21,22,23,24,25,new}. A recent addition to the studies on black hole thermodynamics is the idea of thermodynamic topology. Introduced in  \cite{28} and further developed in \cite{29,64}, thermodynamic topology refers to the topology of the thermodynamic parameter space of a black hole. It is inspired mainly by the works of Duan \cite{26,27} in the context of a relativistic particle system.\\

In this approach, black hole solutions are treated as topological defects in their thermodynamic space and the associated local  and global topologies are studied by computing the winding numbers at these defects.The black holes are then classified according to their total winding number or topological charge. It has been widely reported in literature that based on  the topological number,  all black hole solutions can be classified into three topological classes. Moreover,  the thermal stability of a black hole has also been linked with the sign of the winding number.The studies on the thermodynamic topology have been extended to a number of interesting black hole systems \cite{30,31,32,33,34,35,36,37,38,39,40,41,42,43,44,45,46,47,48,49,50,51,52,53,54,55,56,57,58,59,60,61,62,63,65,66,67,68}.  The key concepts associated with thermodynamic topology that will be used in this work are stated below.\\

One of the fundamental ideas in topology is the concept of topological defects. If a vector field $\chi(\vec{a})$ is such that at some point, $\vec{a}=\vec{b}$ ,
\begin{equation}
\left(\chi(\vec{a})\right)_{\vec{a}=\vec{b}}=0
\end{equation} 
then, such a zero point of $\chi(\vec{a})$ can be connected to a topological defect. In \cite{29}, it was proposed that all physical black hole solutions can be interpreteted  as zero points of the tensor field $\gamma_{\mu \nu}$, where, $$\gamma_{\mu \nu}=G_{\mu \nu}-\frac{8 \pi G}{c^4} T_{\mu \nu}$$
 Accordingly, a topological charge can be defined for each black hole solution. In black hole thermodynamics, this idea is further extended by constructing a vector field using the generalized off-shell free energy.The off shell free energy for an black hole with arbitrary mass is given by,
	\begin{eqnarray}
	\mathcal{F}=E-\frac{S}{\tau} \label{8}  
\end{eqnarray}
Here, $E$ and $S$ denotes the energy and entropy of the black hole, respectively.The time scale parameter  $\tau$ can be thought of as the inverse of the temperature of the cavity that encloses the black hole. Here rather than the mass, the time parameter $\tau$ can change freely.The free energy takes its on-shell form when $\tau$ satisfies
\begin{equation}
	\tau=\frac{1}{T}
\end{equation}
where, $T$ is the equilibrium temperature at the surface of the cavity. This is the exact condition where a black hole solution identifies with a zero point of the tensor field $\gamma_{\mu \nu}$. Using this generalized free energy, a vector field is defined in the following way :
\begin{eqnarray}
	\phi=\left(\phi^r,\phi^\Theta \right)=\left(\frac{\partial\mathcal{F}}{\partial r_{+}},-\cot\Theta ~\csc\Theta  \right)
\end{eqnarray}
Through a straightforward computation, it can be demonstrated that the zero point of $\phi$ is $\theta=\pi/2$ and $\tau=1/T$. This validates the fact that a black hole solution corresponds to a zero point of the vector $\phi$.  Consequently, we can assign a topological charge to each black hole solution.The topological quantity that is linked to the zero point of a field is its winding number or topological charge.Topological charge can be calculated by constructing a topological current $j^\mu$ which is conserved i.e $\partial_\mu j^\mu=0$ \cite{28}.The expression for topological current is given by
\begin{equation}
j^\mu =\frac{1}{2 \pi} \epsilon^{\mu \nu \rho} \epsilon_{a b}\partial_{\nu} n^a \partial_{\rho} n^b \hspace{1.5cm} \mu,\nu,\rho=0,1,2 
\end{equation} 
where, $\partial_\nu=\frac{\partial}{\partial^\nu}$ and $x^{\nu}=(\tau,r_+,\Theta)$.The unit vectors $\left(n^1, n^2\right)$ are worked out as follows: 
$$n^1=\frac{\phi^r}{\sqrt{(\phi^r)^2+(\phi^\Theta)^2}}  \hspace{0.5cm}\text{and} \hspace{0.5cm} n^2=\frac{\phi^\theta}{\sqrt{(\phi^r)^2+(\phi^\Theta)^2}}$$

The expression for current  $j^\mu$ can be rewritten using the Jacobi tensor, $\epsilon^{ab}J^{\mu}\left(\frac{\phi}{x}\right)=\epsilon^{\mu\nu\rho}\partial_\nu \phi^a \partial_\rho \phi^b $ and the two dimensional Laplacian Green function, $\Delta_{\phi^a} \ln{||\phi||}=2\pi\delta^2\left(\phi\right)$ as, 
\begin{equation}
	j^\mu=\delta^2(\phi) J^\mu\left(\frac{\phi}{x}\right)
\end{equation}
it is evident from equation (6) that $j^\mu$ is nonzero only at $\phi^a(x^i)=0$ where $x^i$ is the $i^{th}$ solution of $\phi^a$. We denote these solutions as $z_{i}$. The $0^{th}$ component of topological current is the current density given by,
$$j^0=\sum_{i=1}^{N} \beta_{i}n_{i}\delta^2(\vec{x}-\vec{z_{i}})$$
where $\beta_{i}$ is the Hopf index. From the current density,the
corresponding total topological number or charge for a parameter region $\Sigma$ is obtained by,
	\begin{eqnarray}
	W=\int_{\Sigma}^{} j^{0}d^{2}x=\sum_{i=1}^{N} \beta_{i}n_{i}=\sum_{i=1}^{N} w_{i}
\end{eqnarray}
If the parameter region does not enclose any zero point, the total topological number or charge is $0$. Corresponding to a given value of $\tau$, first the zero points are located and  the corresponding winding numbers are computed. By summing over the winding numbers of all the zero points for a given $\tau$, the total topological number can be obtained.This method of calculating topological number or charge is called Duan's $\phi$ mapping technique. Another method for calculating topological number is suggested in \cite{64},where using residue theorem, the winding number$(w_{i})$ for each solution can be calculated as,
	\begin{eqnarray}
		w_{i}=\frac{Res\mathcal{R}(z_{i}}{|Res\mathcal{R}(z_{i}|}=Sign [Res\mathcal{R}(z_{i}]\label{residue1}
	\end{eqnarray}
	where the rational complex function can be written as 
	\begin{equation}
		\mathcal{R}(z)=\frac{1}{\tau - \mathcal{G}(z)}  \label{residue2}
	\end{equation}

where,$ \mathcal{G}(z)$ is the solution for$\tau$ satisfying the equation $(\frac{\partial\mathcal{F}}{\partial r_{+}}=0$ with $r_+$ replaced by $z$ . The total winding number or the topological charge, $W$ is given by $W=\sum_{i} w_{i}$. As we have mentioned earlier, based on the topological number $W$,  all black hole solutions can be classified into three topological classes.\\	
In this work, we extend the study of thermodynamic topology to $D=4, 5$ Horava-Lifshitz blac holes in Horava gravity. These special black holes are also known as superfluid black holes whose thermodynamics exhibit a line of (continuous) second order phase transitions known as $\lambda$ phase transitions similar to those observed in the superfluidity of  liquid $^{4}He$. Horava gravity is considered to be a probable candidate for quantum gravity at extremely high energies \cite{69}. There has been a lot of interest in the Horava-Lifshitz (HL) black hole solutions, thermodynamics, and phase transitions \cite{70,71,72,73,74,75,76,77,78,79,80} due to their rich phase structures. In this paper, we address the issues related to the dependence of thermodynamic topology on dimension, type of horizon thermodynamic parameters and the choice of ensemble. \\

This paper is organized into the following sections: in Section \textbf{III}, we briefly introduce the Horava Lifshitz  black hole and its thermodynamic quantities. In Section \textbf{IV}, we analyze the thermodynamic topology of $D=4,5$ HL black holes in fixed $\epsilon$ ensemble, where $\epsilon$ is a parameter of the black hole solution. We do so for three types of horizon : spherical, flat and hyperbolic. In Section \textbf{V}, we extend our analysis to  examine fixed $\zeta$ ensemble, where $\zeta$ is a parameter conjugate to $\epsilon$. Finally, the conclusions are presented in Section\textbf{VI}.
	\section{ Horava Lifshitz black hole}
	The space-time metric for $D=d+1$ dimensional Horava Lifshitz AdS black hole is given by\cite{71,79},

	$$ds^2=-f(r)dt^2+\frac{dr^2}{f(r)}+r^2 d\Omega^2_{d-1,k}$$
	
	$$f(r)=k+\frac{32 \pi  P r^2}{(d-1) d \left(1-\epsilon ^2\right)}-4 r^{2-\frac{d}{2}} \sqrt{\frac{64 \pi ^2 P^2 \epsilon ^2 r^d}{(d-1)^2 d^2 \left(1-\epsilon ^2\right)^2}+\frac{\pi  (d-2) M P}{d \left(1-\epsilon ^2\right)}}$$
	where, $ k$ can take three values $+1,0,-1$ corresponding to spherical, flat and hyperbolic horizons respectively. $d$ is the spatial dimension. $M$ and $P$  are the mass and thermodynamic pressure of the black hole respectively. Here, $\epsilon$ is a constant parameter that appears in the action for HL gravity \cite{79}. While $\epsilon=0$ corresponds to the  detailed balance condition, $\epsilon=1$ takes the action back to the general relativity regime. Mass, $M$ is given by the condition $f\left(r_+\right)=0$, where  $r=r_+$ is the location of the horizon.
	\begin{equation}
		M=\frac{r_{+}^{d-4} \left(d^2 k (\epsilon +1)-d k (\epsilon +1)+32 \pi  P r_{+}^2\right) \left(d^2 (k-k \epsilon )+d k (\epsilon -1)+32 \pi  P r_{+}^2\right)}{16 \pi  (d-2) (d-1)^2 d P}
	\end{equation}
	
	where $r_{+}$ is the event horizon radius
	The entropy is computed to be
	\begin{equation}
		S=
		\begin{cases}
			4 \pi  r_{+}^2 \left(1+\frac{d k \left(1-\epsilon ^2\right) \ln (r_{+})}{8 \pi  P r_{+}^2}\right)+S_{0}& \text{d=3}\\
			\frac{16 \pi  r_{+}^{d-1} \left(\frac{d (d-1)^2 k \left(1-\epsilon ^2\right)}{32 \pi  (d-3) P r_{+}^2}+1\right)}{(d-2) (d-1)^2}+S_{0}& \text{$d\geq 3$}
			
		\end{cases}
	\end{equation}
Where, $S_{0}$ is an integration constant. In this work, we fix $S_{0}=0$ for our analysis.

	
	\section{Thermodynamic Topology of $D=4, 5$ HL black hole in Fixed $\epsilon$ ensemble}
	
In this section, we study the thermodynamic topology of $D=4,5$ Horava Lifshitz black holes in an ensemble where the parameter $\epsilon$ is kept fixed. The section is organized into three subsections corresponding to spherical, hyperbolic and flat horizon cases. 
	
	\subsection{For Spherical Horizon $\left(k=1\right)$}
	
	\subsubsection{\textbf{Case I : For D=5}}
	
\noindent For five dimensional HL black hole with spherical horizon,  the free energy, $\mathcal{F}=E-S/\tau$, is given by
	\begin{equation}
		\mathcal{F}=\frac{64 \pi ^2 P r_{+}^3 (P r_{+} \tau -1)+24 \pi  r_{+} \left(2 P r_{+} \tau +3 \epsilon ^2-3\right)-9 \tau  \left(\epsilon ^2-1\right)}{72 \pi  P \tau }
		\label{fe5}
	\end{equation}
	
      \noindent  The components of the vector $\phi$ are found to be 
	\begin{equation}
		\phi^{r}=\frac{\partial\mathcal{F}}{\partial r_{+}}=\frac{12 k P r_{+} \tau +9 k \epsilon ^2-9 k+32 \pi  P^2 r_{+}^3 \tau -24 \pi  P r_{+}^2}{9 P \tau }
	\end{equation}
	\begin{equation}
		\phi^\Theta=-\cot\Theta ~\csc\Theta 
	\end{equation}
\noindent The unit vectors $\left(n^1, n^2\right)$ are computed using the following prescription : 
	
$$n^1=\frac{\phi^r}{\sqrt{(\phi^r)^2+(\phi^\Theta)^2}}  \hspace{0.5cm}\text{and} \hspace{0.5cm} n^2=\frac{\phi^\theta}{\sqrt{(\phi^r)^2+(\phi^\Theta)^2}}$$

\noindent The expression for $\tau$ corresponding to zero points is obtained by setting $\phi^r=0$.
\begin{equation}
\tau=\frac{3 \left(8 \pi  P r_{+}^2-3 \epsilon ^2+3\right)}{4 P r_{+} \left(8 \pi  P r_{+}^2+3\right)}
\label{eos}
\end{equation}

\begin{figure}[t]
		\centering
	\includegraphics[width=11cm,height=6cm]{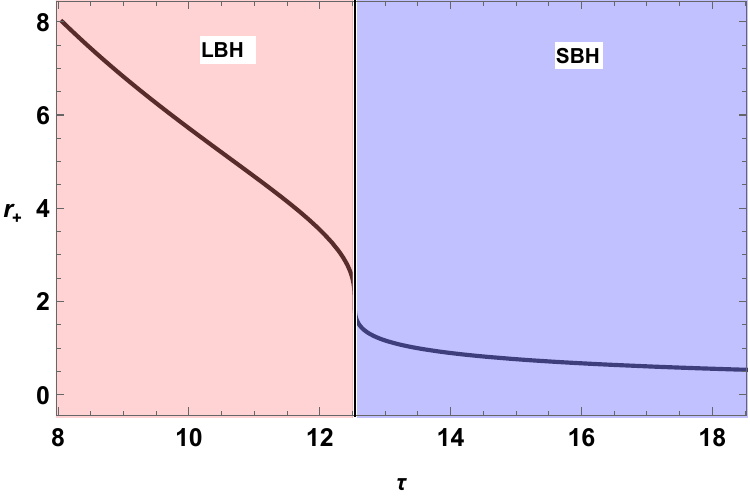}
		\caption{$\tau$ vs $r_+$ plot for $5 D$ HL black hole with spherical horizon at $\epsilon=\epsilon_c=0.9428$ with $ P=0.01$. The $LBH$ and $SBH$ regions represent the large and small black hole branches respectively.}
		\label{1}
	\end{figure}

\noindent Critical points are computed by applying the following conditions : 
\begin{equation}
\frac{d\tau}{dr_{+}}=0 \hspace{0.4cm},\hspace{0.4cm} \frac{d^2 \tau}{dr_{+}^2}=0
\end{equation}  
These conditions result in the two equations stated below: 
	\begin{equation}
		64 \pi  P^2 r_{+}^2 \text{$\tau $}+4 P \left(8 \pi  P r_{+}^2+3\right) \text{$\tau $}-48 \pi  P r_{+}=0
	\end{equation}
	\begin{equation}
		192 \pi  P^2 r_{+} \text{$\tau $}-48 \pi  P=0
	\end{equation}
By solving the above equations, we end up with the expressions for critical pressure, $P_{c}$  and  critical time scale parameter,$\tau_{c}$ given by,
	\begin{equation}
		P_{c} =\frac{1}{8\pi r_{+}^2}\hspace{0.4cm},\hspace{0.4cm} \tau_{c}=2\pi r_{+}
	\end{equation}
Putting these expressions in eq.$\left(\ref{eos}\right)$, the critical value of the parameter $\epsilon$ is obtained
	$$\epsilon_{c}=\frac{2 \sqrt{2}}{3} \approx 0.9428$$
The horizon radius $r_+$ is plotted against $\tau$ at this critical  value, $\epsilon_c=0.9428$ in Figure.\ref {1}. The pressure, $P$ is set equal to $0.01$. We can clearly see a $\lambda$ line transition between a small black hole and a large black hole.\\
\begin{figure}[h]	
\centering
		\begin{subfigure}{0.32\textwidth}
			\includegraphics[width=\linewidth]{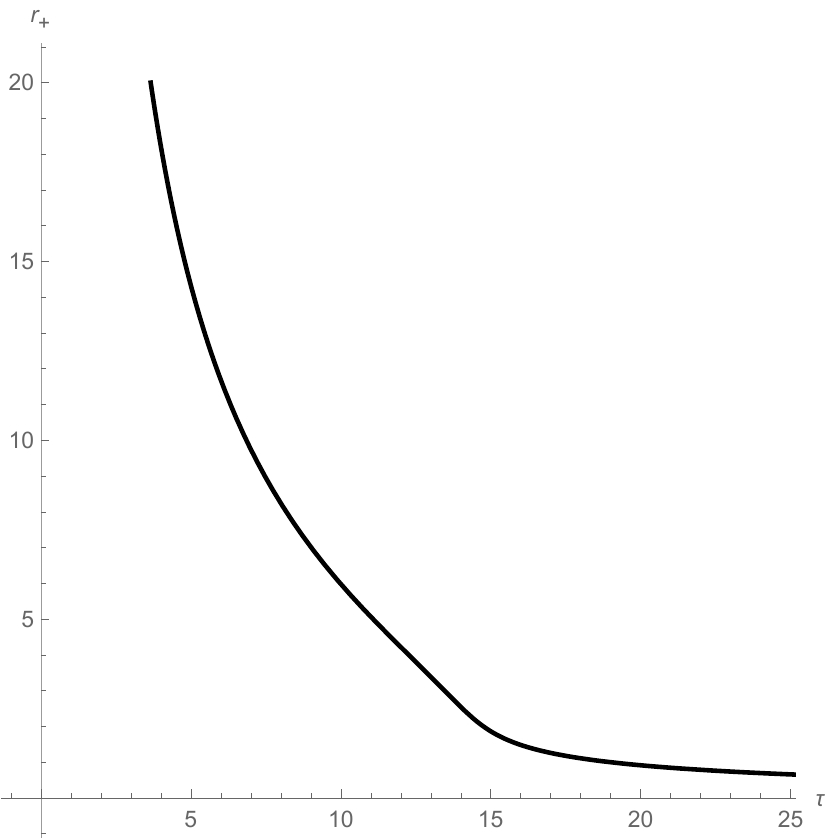}
			\caption{}
			\label{2a}
		\end{subfigure}
		\begin{subfigure}{0.32\textwidth}
			\includegraphics[width=\linewidth]{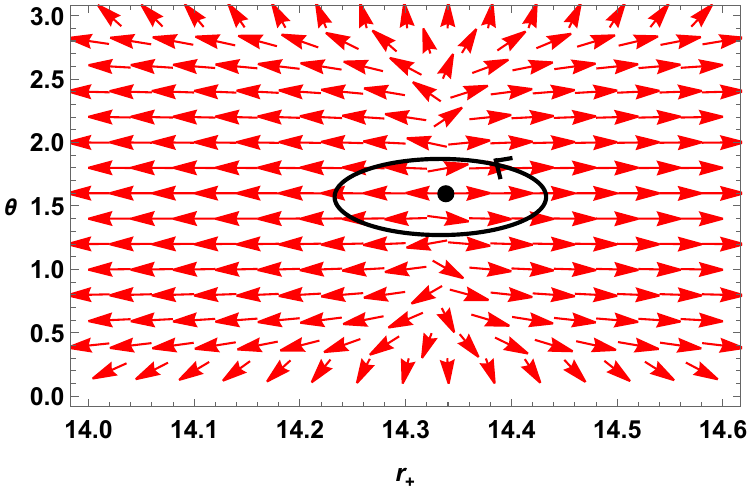}
			\caption{}
			\label{2b}
		\end{subfigure}
		\begin{subfigure}{0.32\textwidth}
			\includegraphics[width=\linewidth]{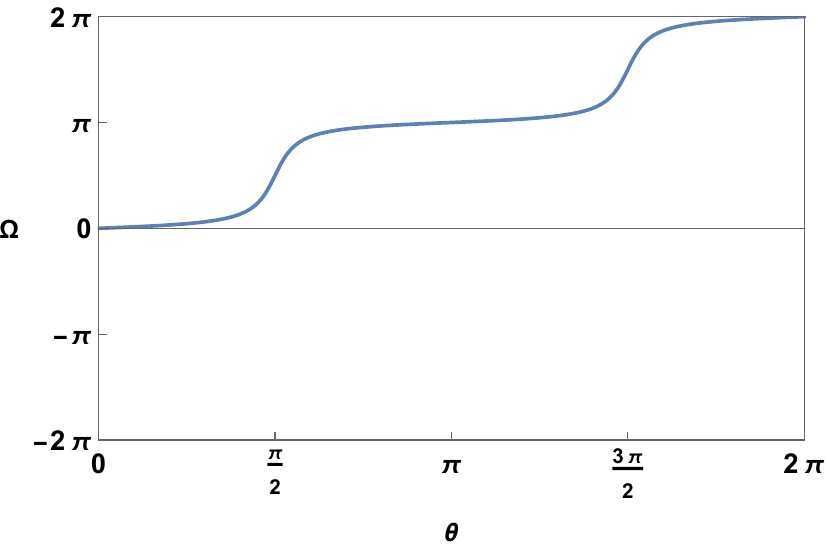}
			\caption{}
			\label{2c}
		\end{subfigure}
		\caption{ Plots for $5D$ HL black hole with spherical horizon at $\epsilon=0.9 $ with $ P=0.01$. Figure $\left(a\right)$ shows $\tau$ vs $r_+$ plot,  figure $\left(b\right)$  is the plot of vector field $n$ on a portion of $r_+-\theta$ plane for $\tau=5$ . The red arrows represent the vecor field $n$. The zero point is located at $r_+=14.3328. $ In figure $\left(c\right)$, computation of the winding number for the contour around the zero point $\tau=5$ and $r_+=14.3328$ is  shown.}
		\label{2}
	\end{figure}
	
At the same value of pressure, $P=0.01$, but at $\epsilon=0.90$ which is below the value of  $\epsilon_c$,  the horizon radius $r_+$ is plotted  against $\tau$  in Figure.\ref {2a}. Here, we observe a single black hole branch whose points are also the zero points of $\phi$. For $\tau=5$, the zero point is located at $r_+=14.3328$. This is also confirmed from the vector plot of $n$ in the $r_+-\theta$ plane as shown in  Figure.\ref {2b}. For finding out the winding number/topological number associated with this zero point, we perform a contour integration around $r_+=14.3328$ which is shown in Figure.\ref {2c}. It reveals that the topological charge in this case is equal to $+1$. We have explicitly verified that the topological charge of any zero point on the black hole branch remains the same and is equal to $+1$.\\

\begin{figure}[h]
\centering
		\begin{subfigure}{0.32\textwidth}
			\includegraphics[width=\linewidth]{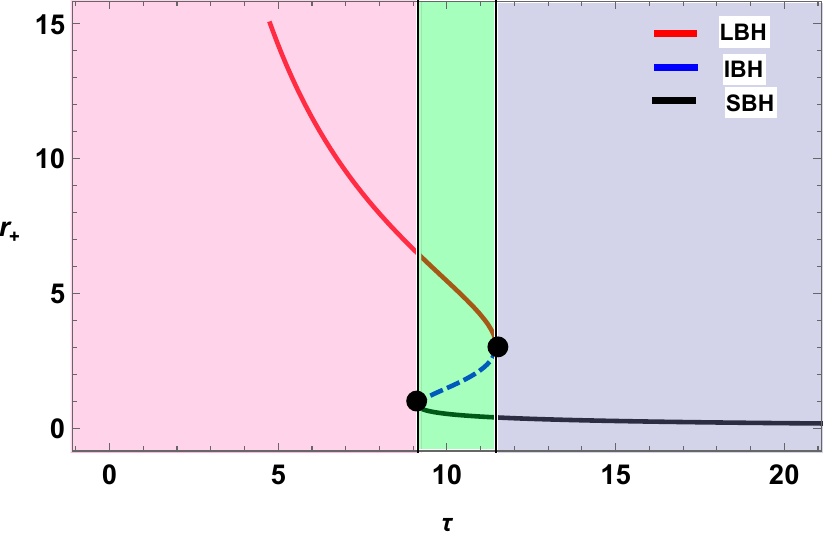}
			\caption{}
			\label{3a}
		\end{subfigure}
		\begin{subfigure}{0.32\textwidth}
			\includegraphics[width=\linewidth]{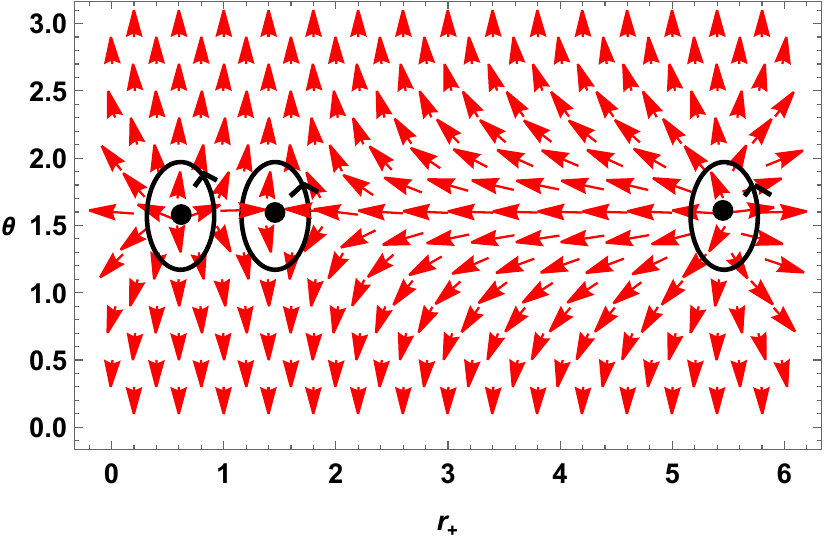}
			\caption{}
			\label{3b}
		\end{subfigure}
		\begin{subfigure}{0.32\textwidth}
			\includegraphics[width=\linewidth]{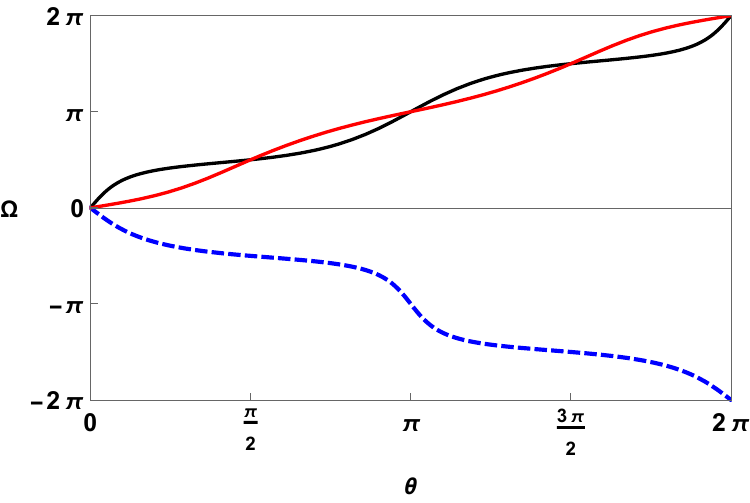}
			\caption{}
			\label{3c}
		\end{subfigure}
		\caption{ Plots for $5D$ HL black hole with spherical horizon at $\epsilon=0.975 $ with $ P=0.01$. Figure $\left(a\right)$ shows $\tau$ vs $r_+$ plot,  figure $\left(b\right)$  is the plot of vector field $n$ on a portion of $r_+-\theta$ plane for $\tau=10$ . The zero points are located at $r_+=0.5408, 1.495, 5.463$. In figure $\left(c\right)$, computation of the the winding numbers for the contours around the zero points $r_+=0.5408, 1.495, 5.463$ are shown in black colored solid line, blue colored  dashed line and red colored solid line respectively.}
		\label{3}
	\end{figure}
	
We repeat the analysis keeping  pressure constant at  $P=0.01$ with $\epsilon=0.975$ which is above the value of  $\epsilon_c$. The horizon radius $r_+$ is plotted  against $\tau$  in Figure.\ref {3a}. Here, we observe three black hole branches: a small, an intermediate and a large black hole branch. For $\tau=10$, three zero points are located a t$r_+=0.5408, 1.495, 5.463$. This is again confirmed from the vector plot of $n$ in the $r_+-\theta$ plane as shown in  Figure.\ref {3b}. As shown in Figure.\ref {3c}, the winding numbers corresponding to $r_+=0.5408, 1.495, 5.463$ (represented by the  black colored solid line, the blue colored  dashed line and the red colored solid line respectively) are found to be $+1$, $-1$ and $+1$ respectively. In fact, all the points on the small black hole and large black hole branches are found to have winding number equal to $+1$. The winding number of the intermediate branch is found to be $-1$. The sum of the winding numbers of the three branches gives us the total topological charge of the black hole, which in this case equals $1-1+1=1$. From Figure.\ref {3a}, we can also see a generation point at $\tau=11.453, r_+= 3.0695$ and  an annihilation point at $\tau=9.143, r_+= 0.864$ which are shown as black dots. \\

\begin{figure}[h]
\centering
		\begin{subfigure}{0.32\textwidth}
			\includegraphics[width=\linewidth]{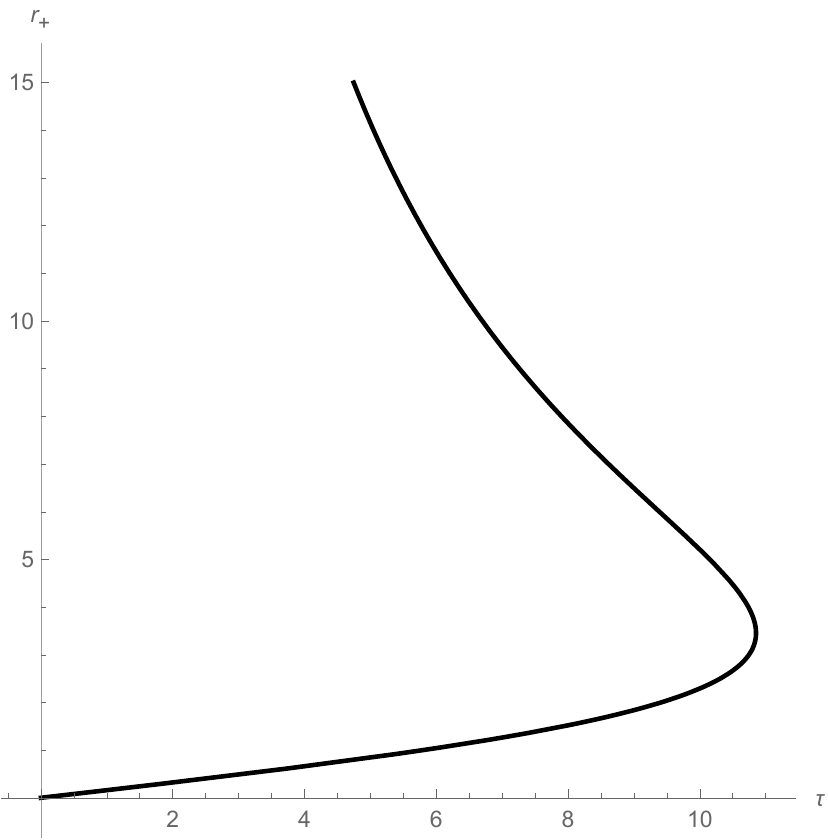}
			\caption{}
			\label{4a}
		\end{subfigure}
		\begin{subfigure}{0.32\textwidth}
			\includegraphics[width=\linewidth]{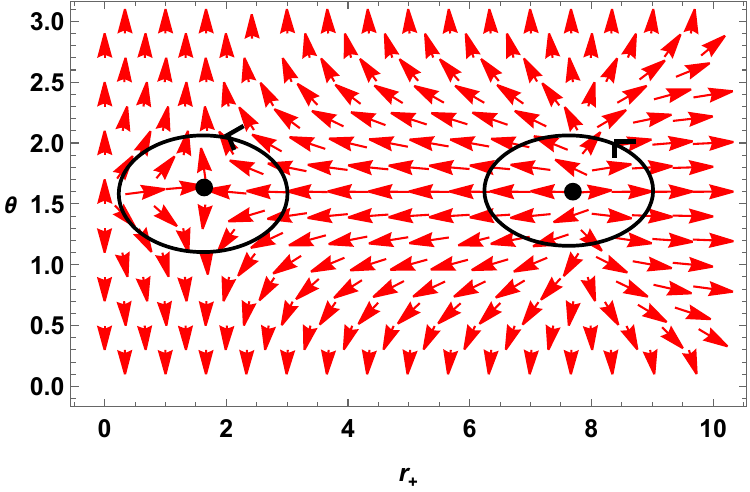}
			\caption{}
			\label{4b}
		\end{subfigure}
		\begin{subfigure}{0.32\textwidth}
			\includegraphics[width=\linewidth]{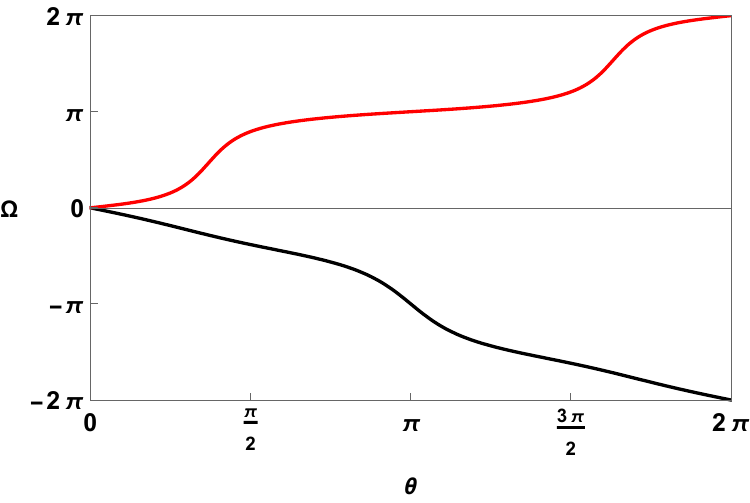}
			\caption{}
			\label{4c}
		\end{subfigure}
		\caption{ Plots for $5D$ HL black hole with spherical horizon at $\epsilon=1 $ with $ P=0.01$. Figure $\left(a\right)$ shows $\tau$ vs $r_+$ plot,  figure $\left(b\right)$  is the plot of vector field $n$ on a portion of $r_+-\theta$ plane for $\tau=8$ . The zero points are located at $r_+= 1.519, 7.855$. In figure $\left(c\right)$, computation of the the winding numbers for the contours around the zero points $r_+= 1.519, 7.855$  are shown in black colored and red colored lines respectively.}
		\label{4}
	\end{figure}

We now consider the GR limit by setting $\epsilon=1$ . The pressure is again kept at  $P=0.01$. The horizon radius $r_+$ is plotted  against $\tau$  in Figure.\ref {4a}. Here, we observe two black hole branches: a small and a large black hole branch. For $\tau=8$, the zero points are located at $r_+= 1.519, 7.855$.  The vector plot of $n$ in the $r_+-\theta$ plane  in  Figure.\ref {4b} demonstrates the same. As shown in Figure.\ref {4c}, the winding numbers corresponding to $r_+= 1.519, 7.855$ (black colored and red colored lines respectively) are found to be $-1$ and $+1$ respectively. Hence, the total topological charge  equals $1-1=0$.  We locate one generation point at $\tau=10.854$ and $r_+=3.454$ as seen in Figure.\ref {4a}.\\

For, all the cases discussed above, we also studied the dependence of topological charge on pressure. We found that, for a given value of $\epsilon$, the topological charge of $5D$ HL black hole with spherical horizon remains constant at different values of pressure. \\ 

Therefore, for $5D$ HL black hole with spherical horizon in the fixed $\epsilon$ ensemble, below a critical value of $\epsilon=\epsilon_c=0.9428$, we find one black hole branch with topological charge $+1$. Above $\epsilon_c$, but below the GR limit, $\epsilon=1$, we observe three black hole branches but the topological charge still remains $+1$. In the GR limit with $\epsilon\ge1$, two black hole branches are found with total topological charge equaling $0$. Variation of pressure has no impact on the topological charge. We summarize our results in the table below:

\begin{table}[ht]
\caption{Summary of results for $5D$ HL black hole with spherical horizon in fixed $\epsilon$ ensemble.} 
\centering 
\begin{tabular}{|c| c |c| c| c| c|} 
\hline 
$\epsilon$ & $P$ & No of black hole branches & Topological charge &  No of generation points & No of annihilation points\\ [0.5ex] 
\hline 
0.90 & 0.005 & 1 & 1 & 0 & 0  \\ 
0.90 & 0.01 & 1 & 1& 0 & 0   \\
0.90 & 0.1 & 1 & 1 & 0 & 0  \\
0.975 & 0.005 & 3 & 1-1+1=1 & 1  & 1 \\
0.975 & 0.01 & 3  & 1-1+1=1 & 1 & 1   \\
0.975 &  0.1 & 3  & 1-1+1=1 &1 &1  \\
1 & 0.005 & 2 & -1+1=0 & 1 & 0   \\
1 & 0.01 & 2 & -1+1=0 & 1 & 0   \ \\
1 & 0.1 & 2 &- 1+1=0 & 1 & 0   \   \\ [1ex] 
\hline 
\end{tabular}
\label{table:nonlin} 
\end{table}

		\vspace{0.7cm}
	\subsubsection{\textbf{Case II : For D=4}}
	\vspace{0.7cm}
	\noindent For four dimensional HL black hole with spherical horizon,  the free energy is given by
	\begin{equation}
		\mathcal{F}=\frac{4 \pi   r_{+}^2 (4 P r_{+} \tau -3)}{3 \tau }+\frac{3 \left(\epsilon ^2-1\right) \log ( r_{+})}{2 P \tau }-\frac{3 \left(\epsilon ^2-1\right)}{16 \pi  P  r_{+}}+2  r_{+}
		\label{fe4}
	\end{equation}
	
	\noindent  The components of the vector $\phi$ are found to be 
	\begin{equation}
		\phi^{r}=\frac{3 k^2 \tau  \left(\epsilon ^2-1\right)+8 \pi  k r_{+} \left(4 P r_{+} \tau +3 \epsilon ^2-3\right)+128 \pi ^2 P r_{+}^3 (2 P r_{+} \tau -1)}{16 \pi  P r_{+}^2 \tau}
	\end{equation}
	\begin{equation}
		\phi^\Theta=-\cot\Theta ~\csc\Theta 
	\end{equation}
	
	\noindent The expression for $\tau$ corresponding to zero points is obtained as:
	\begin{equation}
		\tau=\frac{8 \pi  \left(-3 k r_{+} \epsilon ^2+3 k r_{+}+16 \pi  P r_{+}^3\right)}{3 k^2 \epsilon ^2-3 k^2+32 \pi  k P r_{+}^2+256 \pi ^2 P^2 r_{+}^4}
		\label{eos4}
	\end{equation}
	
\begin{figure}[t]
		\centering
		\includegraphics[width=11cm,height=6cm]{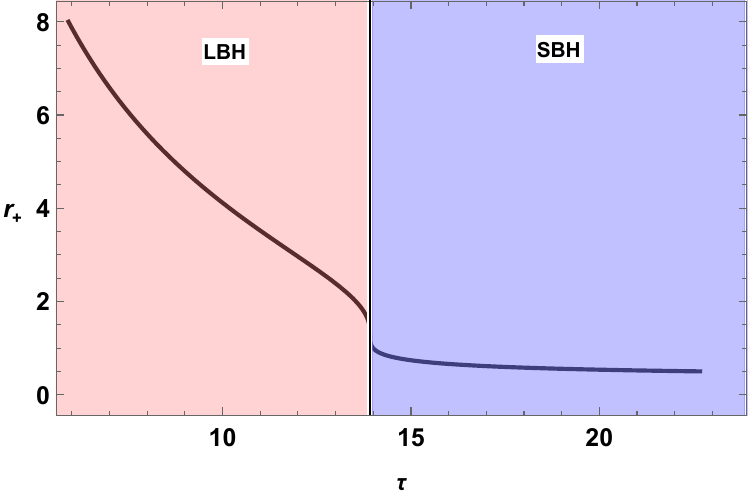}
		\caption{$\tau$ vs $r_+$ plot for $4 D$ HL black hole with spherical horizon at $\epsilon=\epsilon_c=0.9785$ with $ P=0.01$. The $LBH$ and $SBH$ regions represent the large and small black hole branches respectively.}
		\label{5}
	\end{figure}

	\noindent Critical points can be computed by following the same procedure as explained in the case of 5D HL black holes,resulting in the equations:
	\begin{equation}
		P_{c} =P_{c} =-\frac{k \left(-3 \epsilon ^2 \pm \sqrt{\epsilon ^2 \left(9 \epsilon ^2-8\right)}+2\right)}{32 \pi  r_{+}^2},\hspace{0.4cm} \tau_{c} =\frac{\pm 3 \pi  r_{+} \left(3 \epsilon \sqrt{  \left(9 \epsilon ^2-8\right)}+ \left(9 \epsilon ^2-4\right)\right)}{2 k}
	\end{equation}
	Putting these expressions in eq.$\left(\ref{eos4}\right)$, the critical value of the parameter $\epsilon$ is obtained
	$$\epsilon_{c}=\sqrt{\frac{4}{9}+\frac{8}{9 \sqrt{3}}} \approx 0.978593$$
	The horizon radius $r_+$ is plotted against $\tau$ at this critical  value, $\epsilon_c=0.9785$ in Figure.\ref {5}. The pressure, $P$ is set equal to $0.01$. We see a $\lambda$ line transition between a small black hole and a large black hole for the case of $D=4$ case also.\\
	\begin{figure}[h]	
		\centering
		\begin{subfigure}{0.32\textwidth}
			\includegraphics[width=\linewidth]{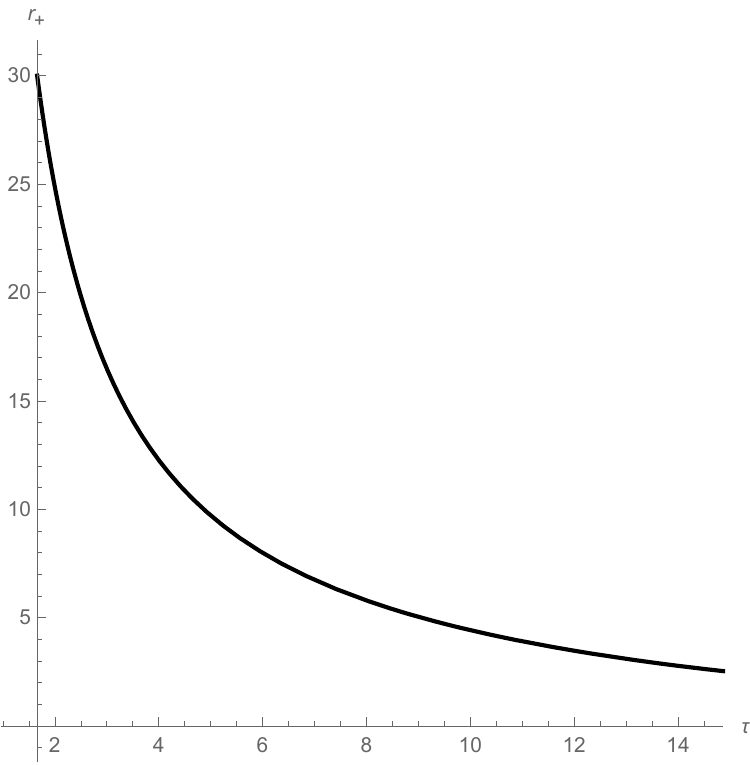}
			\caption{}
			\label{6a}
		\end{subfigure}
		\begin{subfigure}{0.32\textwidth}
			\includegraphics[width=\linewidth]{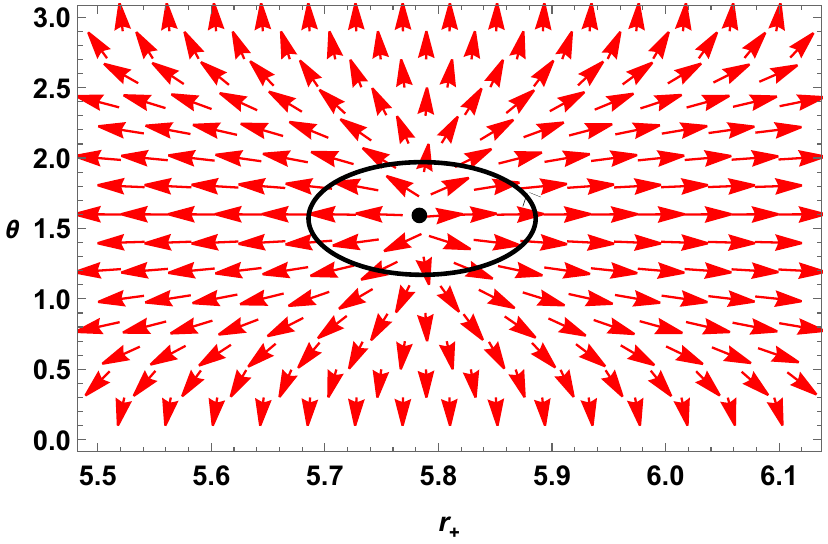}
			\caption{}
			\label{6b}
		\end{subfigure}
		\begin{subfigure}{0.32\textwidth}
			\includegraphics[width=\linewidth]{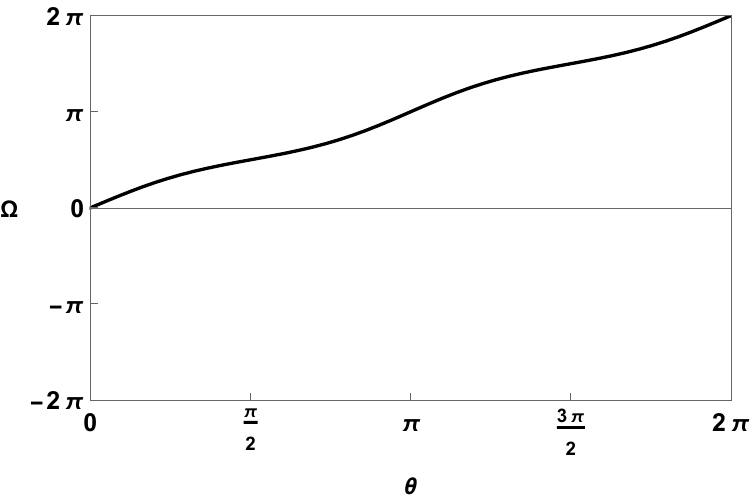}
			\caption{}
			\label{6c}
		\end{subfigure}
		\caption{ Plots for $4D$ HL black hole with spherical horizon at $\epsilon=0.9 $ with $ P=0.01$. Figure $\left(a\right)$ shows $\tau$ vs $r_+$ plot,  figure $\left(b\right)$  is the plot of vector field $n$ on a portion of $r_+-\theta$ plane for $\tau=8$ . The red arrows represent the vecor field $n$. The zero point is located at $r_+=5.785. $ In figure $\left(c\right)$, computation of the the winding number for the contour around the zero point $\tau=8$ and $r_+=5.785$ is  shown.}
		\label{2}
	\end{figure}
	
	At the same value of pressure, $P=0.01$, but at $\epsilon=0.90$ which is below the value of  $\epsilon_c$,  the horizon radius $r_+$ is plotted  against $\tau$  in figure.\ref {6a}. Here, we observe a single black hole branch whose points are also the zero points of $\phi$. For $\tau=8$, the zero point is located at $r_+=5.785$. This is also confirmed from the vector plot of $n$ in the $r_+-\theta$ plane as shown in  figure \ref {6b}. By performing contour integration around $r_+=5.785$ which is shown in figure \ref {6c} we get the topological charge equal to $+1$.\\
	
	\begin{figure}[h]
		\centering
		\begin{subfigure}{0.32\textwidth}
			\includegraphics[width=\linewidth]{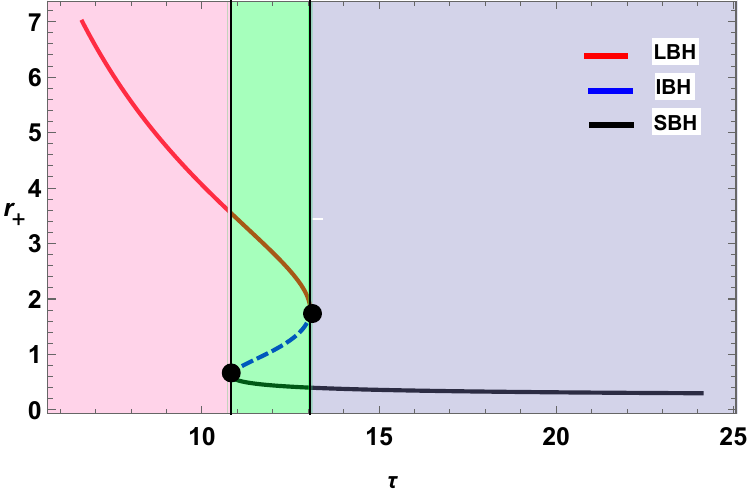}
			\caption{}
			\label{7a}
		\end{subfigure}
		\begin{subfigure}{0.32\textwidth}
			\includegraphics[width=\linewidth]{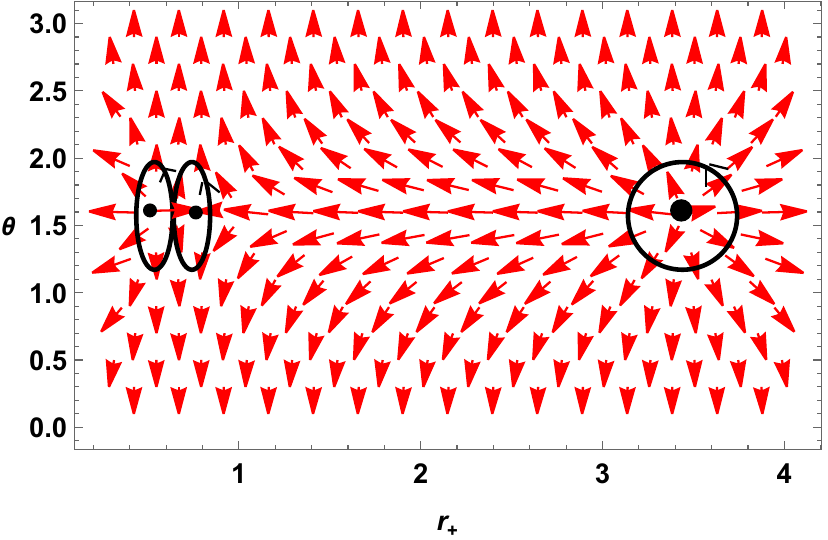}
			\caption{}
			\label{7b}
		\end{subfigure}
		\begin{subfigure}{0.32\textwidth}
			\includegraphics[width=\linewidth]{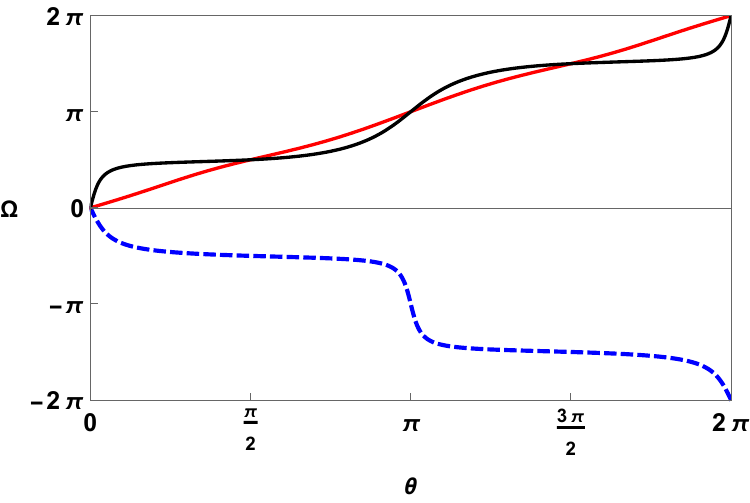}
			\caption{}
			\label{7c}
		\end{subfigure}
		\caption{ Plots for $4D$ HL black hole with spherical horizon at $\epsilon=0.99 $ with $ P=0.01$. Figure $\left(a\right)$ shows $\tau$ vs $r_+$ plot,  figure $\left(b\right)$  is the plot of vector field $n$ on a portion of $r_+-\theta$ plane for $\tau=11$ . The zero points are located at $r_+=0.5358, 0.7419, 3.44$. In figure $\left(c\right)$, computation of the the winding numbers for the contours around the zero points $r_+=0.5358, 0.7419, 3.44$ are shown in black colored solid line, blue colored  dashed line and red colored solid line respectively.}
		\label{7}
	\end{figure}
	
	We repeat the analysis keeping  pressure constant at  $P=0.01$ with $\epsilon=0.99$ which is above the value of  $\epsilon_c$. The horizon radius $r_+$ is plotted  against $\tau$  in figure \ref {7a}. Here, we observe three black hole branches: a small, an intermediate and a large black hole branch. For $\tau=11$, three zero points are located a t$r_+=0.5358, 0.7419, 3.44$. This is again confirmed from the vector plot of $n$ in the $r_+-\theta$ plane as shown in  Figure.\ref {7b}. As shown in figure \ref {7c}, the winding numbers corresponding to $r_+=0.5358, 0.7419, 3.44$ (represented by the  black colored solid line, the blue colored  dashed line and the red colored solid line respectively) are found to be $+1$, $-1$ and $+1$ respectively. Here also all the points on the small black hole and large black hole branches are found to have winding number equal to $+1$. and the winding number of the intermediate branch is found to be $-1$. The total topological charge of the 4D black hole is equals to $1-1+1=1$. From figure \ref {7a}, we can also see a generation point at $\tau=13.0563, r_+= 1.8047$ and  an annihilation point at $\tau=10.8284, r_+= 0.6225$ which are shown as black dots. \\

	\begin{figure}[h]
		\centering
		\begin{subfigure}{0.32\textwidth}
			\includegraphics[width=\linewidth]{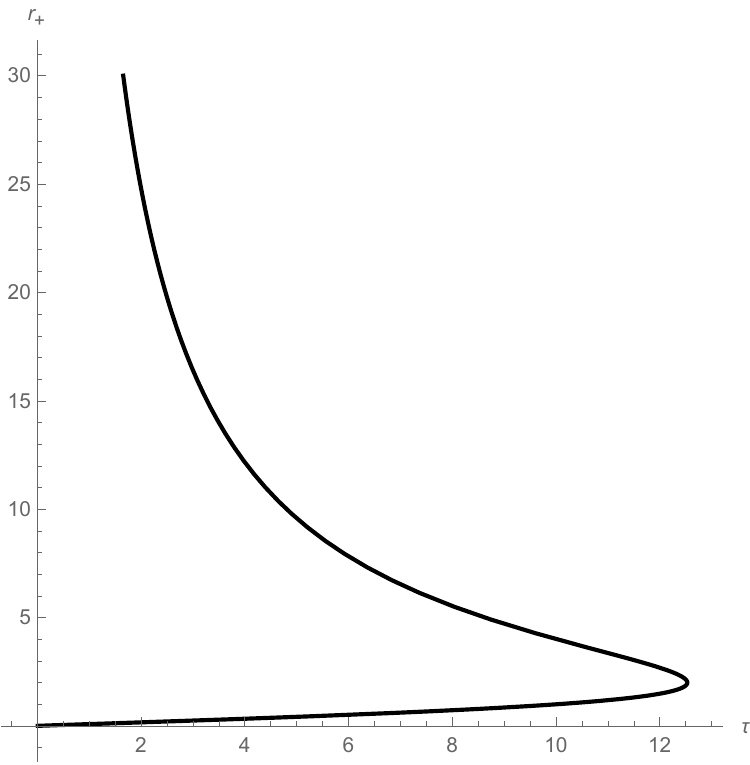}
			\caption{}
			\label{8a}
		\end{subfigure}
		\begin{subfigure}{0.32\textwidth}
			\includegraphics[width=\linewidth]{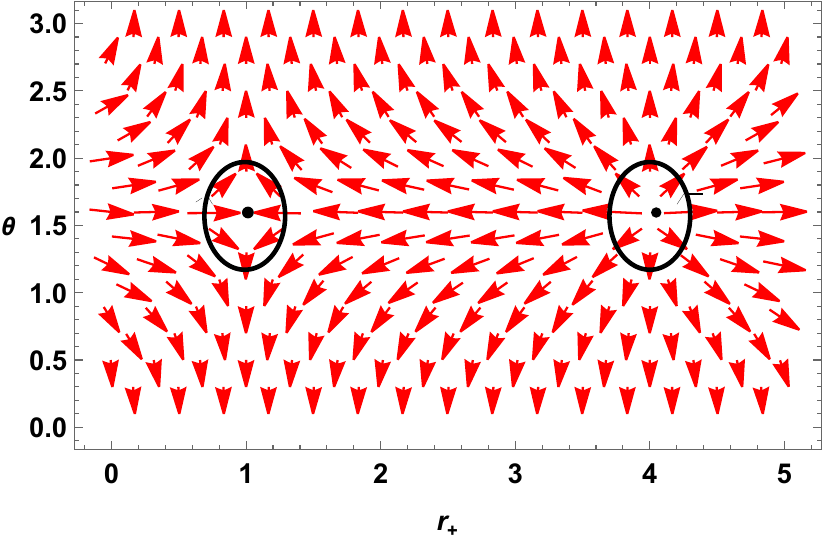}
			\caption{}
			\label{8b}
		\end{subfigure}
		\begin{subfigure}{0.32\textwidth}
			\includegraphics[width=\linewidth]{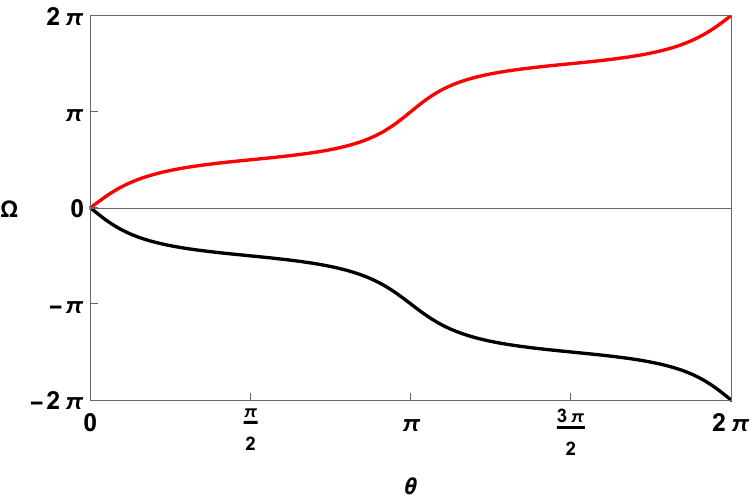}
			\caption{}
			\label{8c}
		\end{subfigure}
		\caption{ Plots for $4D$ HL black hole with spherical horizon at $\epsilon=1 $ with $ P=0.01$. Figure $\left(a\right)$ shows $\tau$ vs $r_+$ plot,  figure $\left(b\right)$  is the plot of vector field $n$ on a portion of $r_+-\theta$ plane for $\tau=10$ . The zero points are located at $r_+= 0.9929, 4.007$. In figure $\left(c\right)$, computation of the winding numbers for the contours around the zero points $r_+= 0.9929, 4.007$  are shown in black colored and red colored lines respectively.}
		\label{8}
	\end{figure}
	
	We now consider the GR limit by setting $\epsilon=1$ . The pressure is again kept fixed at $P=0.01$. The horizon radius $r_+$ is plotted  against $\tau$ in figure \ref {8a}. Here, we observe two black hole branches: a small and a large black hole branch. For $\tau=10$, the zero points are located at $r_+= 0.9929, 4.007$. The vector plot of $n$ in the $r_+-\theta$ plane in  figure \ref {8b} demonstrates the same. As shown in figure \ref {8c}, the winding numbers corresponding to $r_+= 0.9929, 4.007$ (black colored and red colored lines respectively) are found to be $-1$ and $+1$ respectively. Hence, the total topological charge  equals $1-1=0$.  We locate one generation point at $\tau=12.5331$ and $r_+=1.9947$ as seen in Figure.\ref {8a}.\\
	Similar to the outcomes in the case of a 5D HL black hole, pressure does not significantly influence the determination of the topological number for various values of $\epsilon$. \\ 
	Therefore it can be inferred that for $4D$ HL black hole with spherical horizon in the fixed $\epsilon$ ensemble, below a critical value of $\epsilon=\epsilon_c=0.9785$, we find one black hole branch with topological charge $+1$. Above $\epsilon_c$, but below the GR limit, $\epsilon=1$, we observe three black hole branches but the topological charge still remains $+1$. In the GR limit with $\epsilon\ge1$, two black hole branches are found with total topological charge equaling $0$. We summarize our results in the table below:

	\begin{table}[ht]
		\caption{Summary of results for $4D$ HL black hole with spherical horizon in fixed $\epsilon$ ensemble.} 
		\centering 
		\begin{tabular}{|c| c |c| c| c| c|} 
			\hline 
			$\epsilon$ & $P$ & No of black hole branches & Topological charge &  No of generation points & No of annihilation points\\ [0.5ex] 
			\hline 
			0.90 & 0.008 & 1 & 1 & 0 & 0  \\ 
			0.90 & 0.05 & 1 & 1& 0 & 0   \\
			0.90 & 0.2 & 1 & 1 & 0 & 0  \\
			0.99 & 0.008 & 3 & 1-1+1=1 & 1  & 1 \\
			0.99 & 0.05 & 3  & 1-1+1=1 & 1 & 1   \\
			0.99 &  0.2 & 3  & 1-1+1=1 &1 &1  \\
			1 & 0.008 & 2 & -1+1=0 & 1 &0   \\
			1 & 0.05 & 2 & -1+1=0 & 1 & 0   \ \\
			1 & 0.2 & 2 &- 1+1=0 & 1 &  0   \   \\ [1ex] 
			\hline 
		\end{tabular}
		\label{table:nonlin} 
	\end{table}
	\subsection{For flat horizon (k=0)}

\subsubsection{\textbf{Case I : For D=5}}

For flat horizon , we put $k=0$ in the expression for mass and entropy .Finally obtained the free energy for $5D$ HL black hole with flat horizon as : 
$$\mathcal{F}=\frac{8 \pi  r_{+}^3 (P r \tau -1)}{9 \tau }$$
followed by-
$$\phi_{r}=\frac{8 \pi  r_{+}^2 (4 P r_{+} \tau -3)}{9 \tau } \hspace{1cm } \text{and} \hspace{1cm}\tau=\frac{3}{4 P r_{+}}$$
For $5D$ HL black hole with flat horizon ,from the expression for $\tau$ it is clear that $\tau$ is independent of $\epsilon$ ,hence no criticality will be observed for this case.
\begin{figure}[h]
	\centering
	\begin{subfigure}{0.32\textwidth}
		\includegraphics[width=\linewidth]{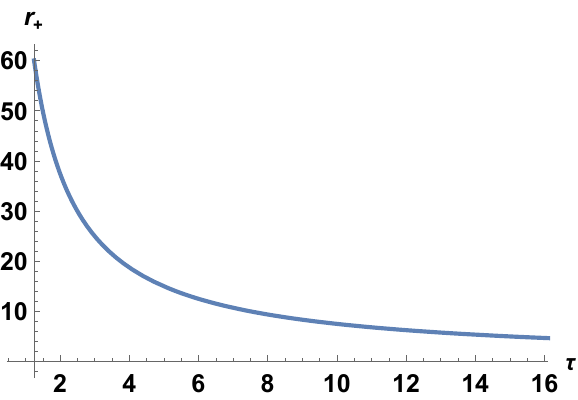}
		\caption{}
		\label{9a}
	\end{subfigure}
	\begin{subfigure}{0.32\textwidth}
		\includegraphics[height=4cm,width=5cm]{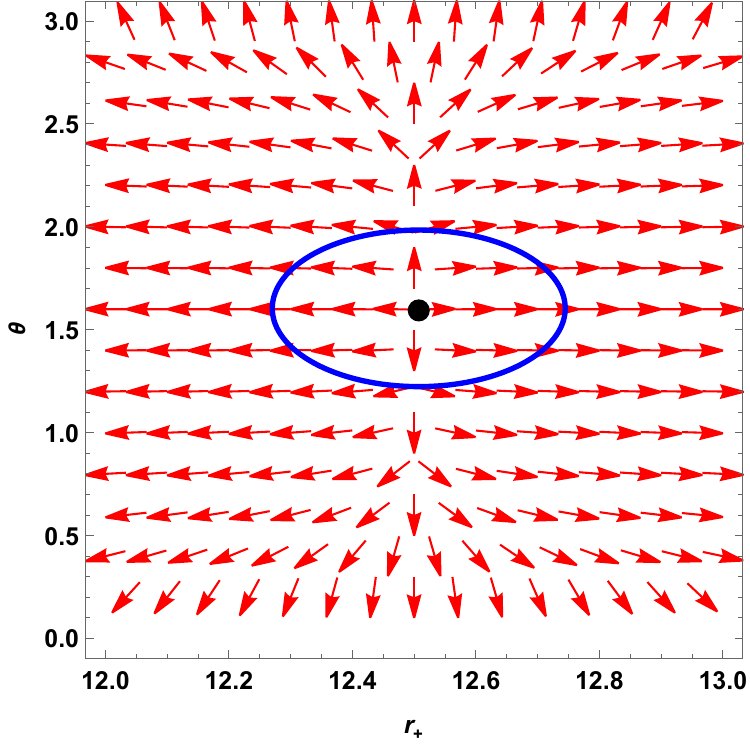}
		\caption{}
		\label{9b}
	\end{subfigure}
	\begin{subfigure}{0.32\textwidth}
		\includegraphics[width=\linewidth]{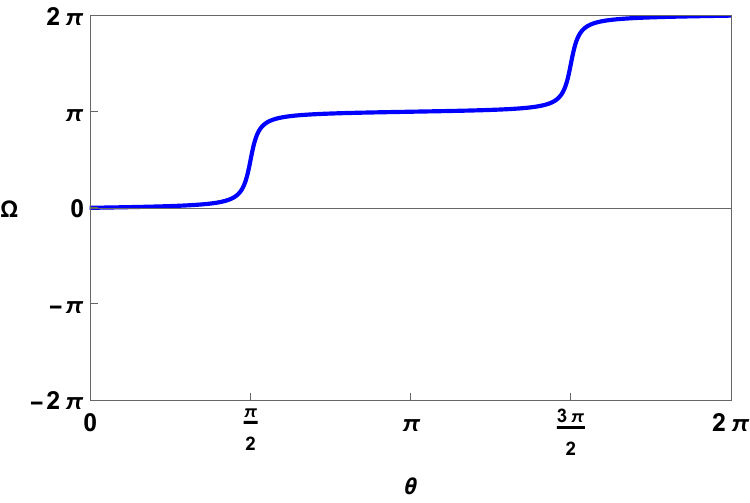}
		\caption{}
		\label{9c}
	\end{subfigure}
	\caption{ Plots for $5D$ HL black hole with flat horizon at $\epsilon=0.96 $ with $ P=0.01$. Figure $\left(a\right)$ shows $\tau$ vs $r_+$ plot,  figure $\left(b\right)$  is the plot of vector field $n$ on a portion of $r_+-\theta$ plane for $\tau=6$ . The red arrows represent the vecor field $n$. The zero point is located at $r_+=12.5. $ in figure $\left(c\right)$, computation of the winding number for the contour around the zero point $\tau=6$ and $r_+=12.5$ is shown.}
	\label{9}
\end{figure}
Here, we observe a single black hole branch whose points are also the zero points of $\phi$. For $\tau=6$, the zero point is located at $r_+=12.5$. This is also confirmed from the vector plot of $n$ in the $r_+-\theta$ plane as shown in  figure \ref {9b}. By performing contour integration around $r_+=12.5$ which is shown in figure \ref {9c} we get the topological charge equal to $+1$.Infact for any value of pressure and $\epsilon$ the topological charge remains constant.\\

\subsubsection{\textbf{Case II: For D=4}}

For $D=4$, we get the expression for free energy of 4D HL black hole with flat horizon as:
$$\mathcal{F}=\frac{4 \pi  r_{+}^2 (4 P r \tau -3)}{3 \tau }$$
followed by-
$$\phi^r=\frac{8 \pi  r_{+}^2 (4 P r_{+} \tau -3)}{9 \tau } \hspace{1cm } \text{and} \hspace{1cm}\tau=\frac{3}{4 P r_{+}}$$
For $4D$ HL black hole with flat horizon ,the $\tau$ is independent of $\epsilon$ similar to the case of 5D HL black hole ,hence no criticality will be observed for this case also.
\begin{figure}[h]
	\centering
	\begin{subfigure}{0.32\textwidth}
		\includegraphics[width=\linewidth]{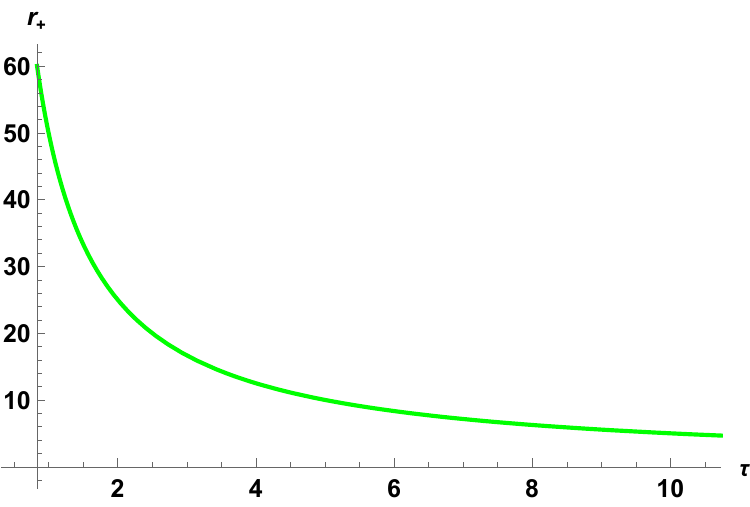}
		\caption{}
		\label{10a}
	\end{subfigure}
	\begin{subfigure}{0.32\textwidth}
		\includegraphics[height=4cm,width=5cm]{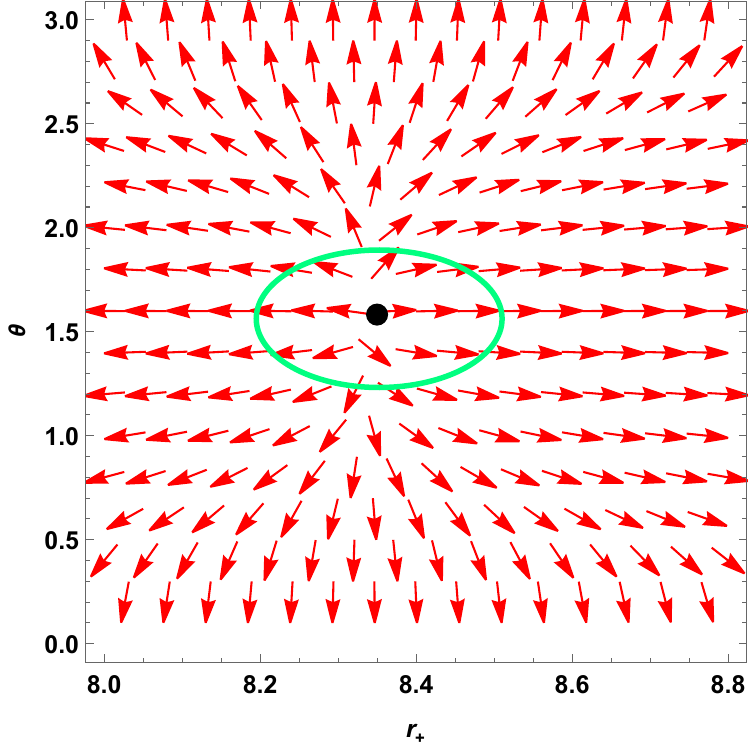}
		\caption{}
		\label{10b}
	\end{subfigure}
	\begin{subfigure}{0.32\textwidth}
		\includegraphics[width=\linewidth]{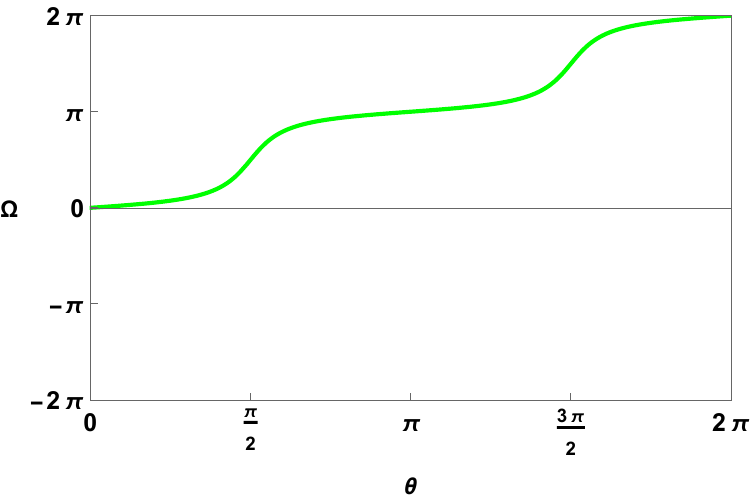}
		\caption{}
		\label{10c}
	\end{subfigure}
	\caption{ Plots for $4D$ HL black hole with flat horizon at $\epsilon=0.96 $ with $ P=0.01$. Figure $\left(a\right)$ shows $\tau$ vs $r_+$ plot,  figure $\left(b\right)$  is the plot of vector field $n$ on a portion of $r_+-\theta$ plane for $\tau=6$ . The red arrows represent the vecor field $n$. The zero point is located at $r_+=8.333. $ In figure $\left(c\right)$, computation of the winding number for the contour around the zero point $\tau=6$ and $r_+=8.333$ is  shown.}
	\label{10}
\end{figure}
Here also, we observe a single black hole branch whose points are also the zero points of $\phi$. For $\tau=6$, the zero point is located at $r_+=8.333$. This is also confirmed from the vector plot of $n$ in the $r_+-\theta$ plane as shown in  Figure.\ref {10b}.By performing contour integration around $r_+=12.5$ which is shown in Figure.\ref {10c}we get the topological charge equal to $+1$.Here also for any value of pressure and $\epsilon$ the topological charge remains constant.

	\subsection{For hyperbolic horizon ( k=-1)}
	\subsubsection{\textbf{Case I : For D=5}}
\begin{figure}[h]
		\centering
		\includegraphics[width=7cm,height=5cm]{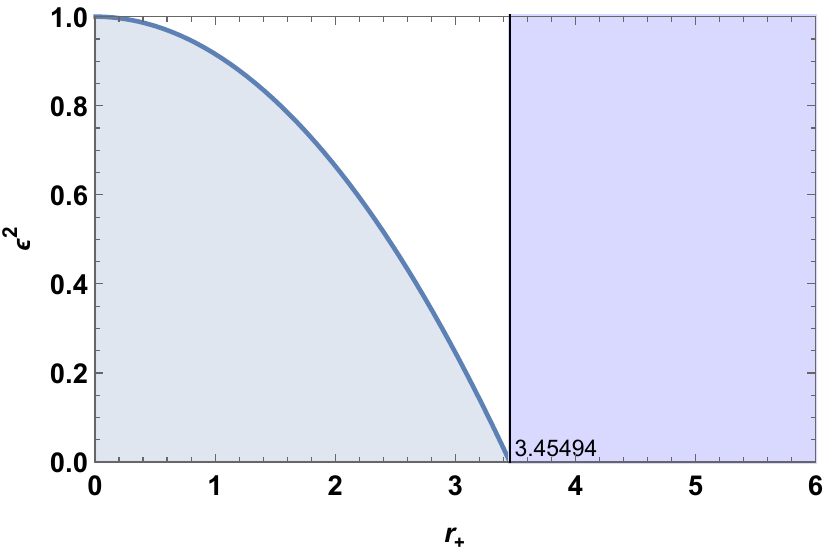}
		
		\caption{The relations between $\epsilon^2$
			and $r$ for the positive temperature for $k=-1$ in $D=5.$Temperature is positive only on the shaded portions.We have taken $P= 0.01$.}
		\label{11}
	\end{figure}

	The free energie expression for 5D HL black hole with hyperbolic horizon is found to be:
	$$\mathcal{F}=\frac{64 \pi ^2 P r_{+}^3 (P r_{+} \tau -1)-24 \pi  r_{+} \left(2 P r_{+} \tau +3 \epsilon ^2-3\right)-9 \tau  \left(\epsilon ^2-1\right)}{72 \pi  P \tau }$$
	and the expression for $\phi_{r}$ can be found out as-
	$$\phi_{r}=\frac{32 \pi  P^2 r_{+}^3 \tau -12 P r_{+} (2 \pi  r_{+}+\tau )-9 \epsilon ^2+9}{9 P \tau }$$
	The zero points of the vector is obtained by setting $\phi_{r_{+}}=0$ and followed by getting a relation between $\tau$ and $r_{+}$ as-
	$$\tau=\frac{3 \left(8 \pi  P r_{+}^2+3 \epsilon ^2-3\right)}{4 P r_{+} \left(8 \pi  P r_{+}^2-3\right)}$$ 
	For 5D HL black hole with hyperbolic horizon,we do not get any critical conditions hence there is no superfluid $\lambda$ phase transition observed.It is important to note that while calculating the winding number some restrictions regarding the values of $r_+$ and $\epsilon$ ,are needed to be taken care of \cite{77}. From Figure \ref{11} it can be observed that we can have values of r and $\epsilon$ only from the shaded portions due to positive temperature condition.
	Thus while calculating winding number from the $\tau$ vs $r_+$ curve for a particular value of $\epsilon^2$ and pressure $P$  we will have to take values of $r_+$ from the allowed range(shaded portions).\\
	\begin{figure}[h]
		\centering
		\begin{subfigure}{0.32\textwidth}
			\includegraphics[width=\linewidth]{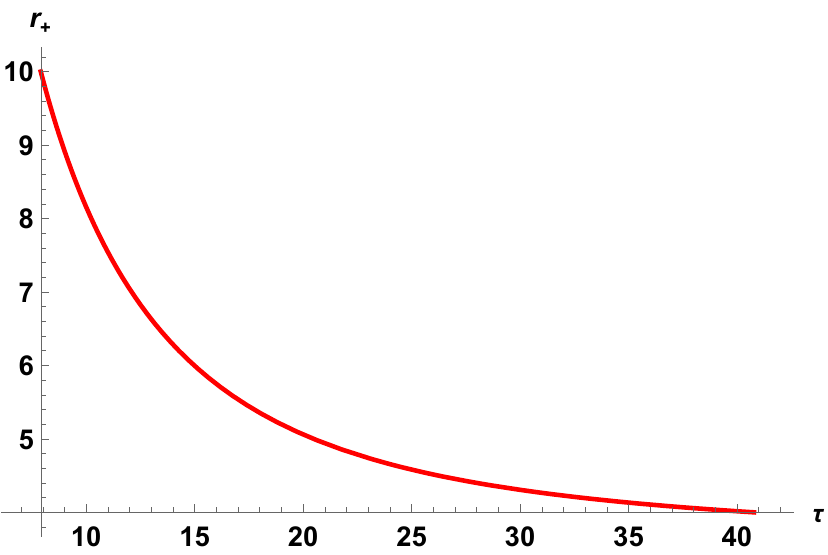}
			\caption{}
			\label{12a}
		\end{subfigure}
		\begin{subfigure}{0.32\textwidth}
			\includegraphics[width=\linewidth]{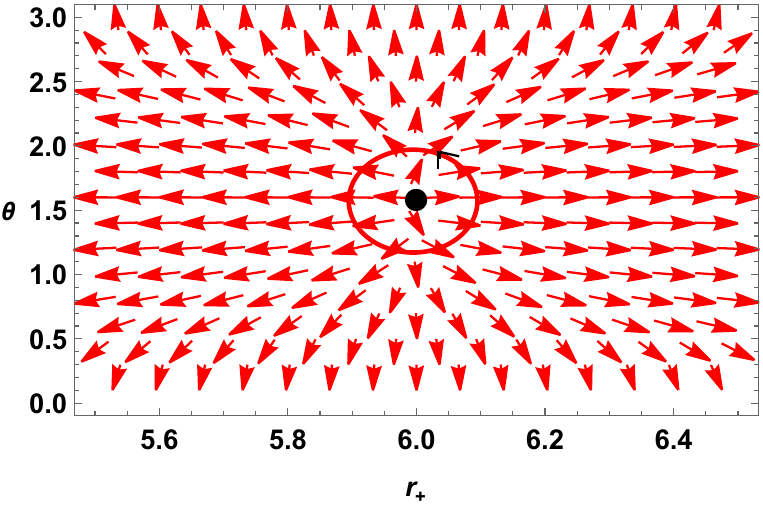}
			\caption{}
			\label{12b}
		\end{subfigure}
		\begin{subfigure}{0.32\textwidth}
			\includegraphics[width=\linewidth]{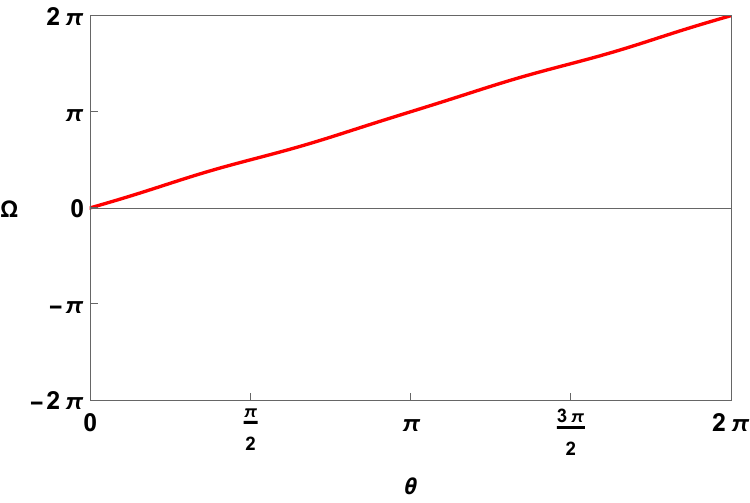}
			\caption{}
			\label{12c}
		\end{subfigure}
		
		\medskip
		
		\begin{subfigure}{0.32\textwidth}
			\includegraphics[width=\linewidth]{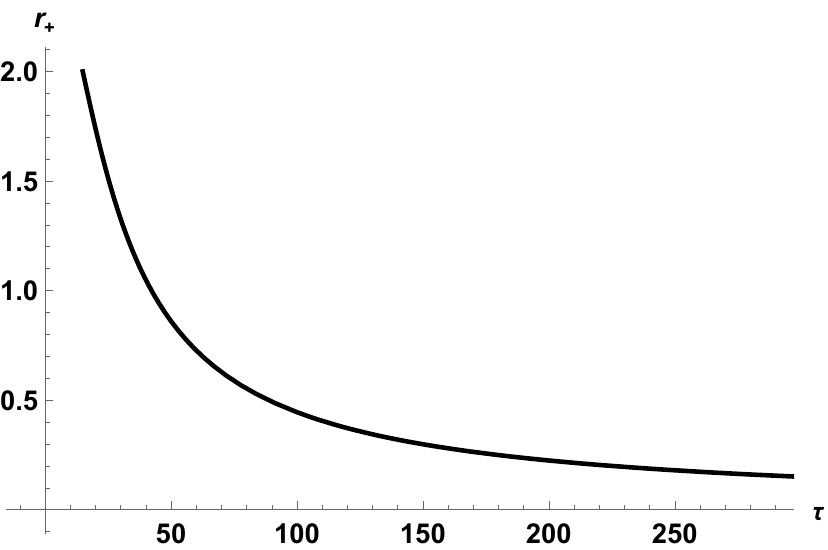}
			\caption{}
			\label{12d}
		\end{subfigure}
		\begin{subfigure}{0.32\textwidth}
			\includegraphics[width=\linewidth]{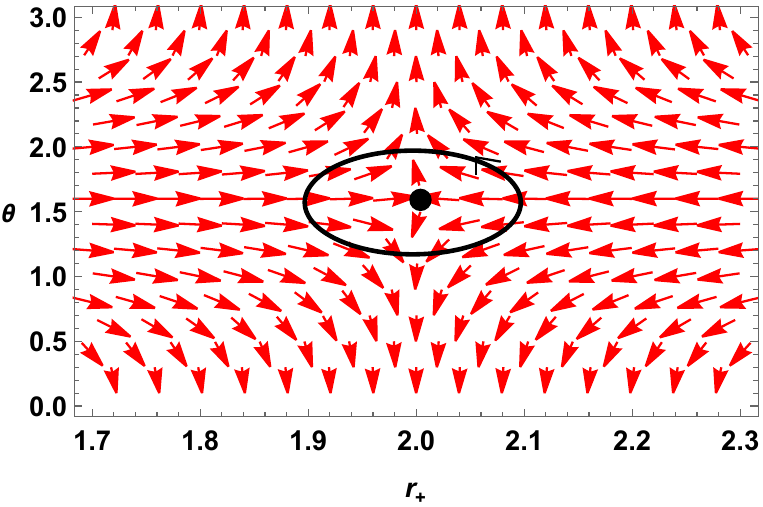}
			\caption{}
			\label{12e}
		\end{subfigure}
		\begin{subfigure}{0.32\textwidth}
			\includegraphics[width=\linewidth]{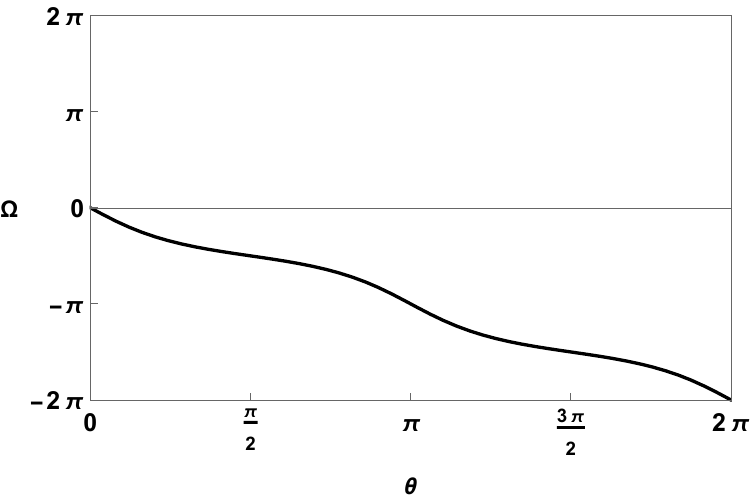}
			\caption{}
			\label{12f}
		\end{subfigure}
		
		\caption{Plots for $5D$ HL black hole with hyperbolic horizon at $\epsilon^2=0.4 $ with $ P=0.01$. Figure $\left(a\right)$ shows $\tau$ vs $r_+$ plot for $0 \leq r_+ \leq 2.67619$ range and $\left(d\right)$ shows the same for the range $3.4549 \leq r \leq 10$.figure $\left(b\right)$ and figure $\left(b\right)$ is the plot of vector field $n$ on a portion of $r_+-\theta$ plane for $\tau=15$ in the range $0 \leq r_+ \leq 2.67619$ and $3.4549 \leq r_+ \leq 10$ respectively.The zero points are located at $r_+=5.994 $ and $r_+=1.996$ . In figure $\left(c\right)$ and $\left(f\right)$, computation of the contours around the zero points for $r_+=5.994$ and $r_+=1.996$ are shown in red and black colored solid lines respectively.}
		\label{12}
	\end{figure}

\begin{figure}[h]
		\centering
		\begin{subfigure}{0.32\textwidth}
			\includegraphics[width=\linewidth]{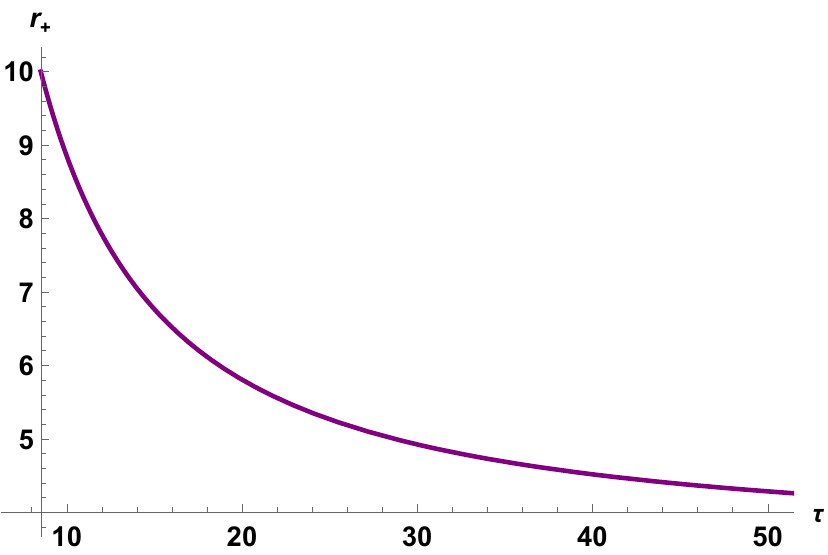}
			\caption{}
			\label{E1a}
		\end{subfigure}
		\begin{subfigure}{0.32\textwidth}
			\includegraphics[width=\linewidth]{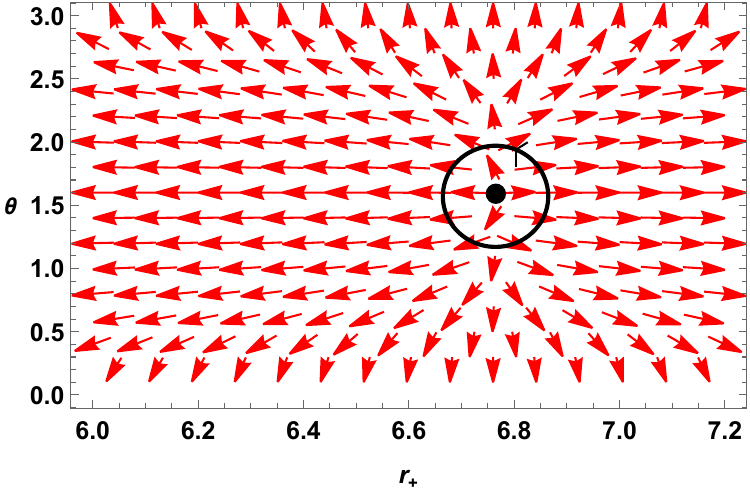}
			\caption{}
			\label{E1b}
		\end{subfigure}
		\begin{subfigure}{0.32\textwidth}
			\includegraphics[width=\linewidth]{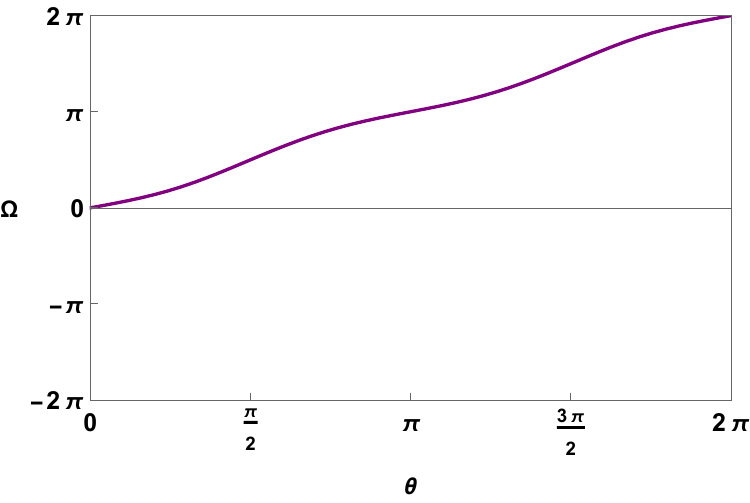}
			\caption{}
			\label{E1c}
		\end{subfigure}
		\caption{ Plots for $5D$ HL black hole with hyperbolic horizon at $\epsilon=1 $ with $ P=0.01$. Figure $\left(a\right)$ shows $\tau$ vs $r_+$ plot,  figure $\left(b\right)$  is the plot of vector field $n$ on a portion of $r_+-\theta$ plane for $\tau=15$ . The zero point is located at $r_+=6.76458 $. In figure $\left(c\right)$, computation of the contours around the zero point $r_+=6.76458$ is shown in purple colored solid lines.}
		\label{13}
		
	\end{figure}

	The horizon radius $r_+$ is plotted  against $\tau$  in Figure.\ref {12a} and Figure.\ref {12b}. Here, we observe two separate or discontinous black hole branches. One for the range $0 \leq r \leq 2.67619$(Figure \ref{12a}) and other for the range $3.4549 \leq r \leq 10$(Figure.\ref {12b}).We will do the analysis for calculating the winding numbers by keeping the pressure $P$ ,$\epsilon^2$ and $\tau$ fixed at  $P=0.01$,$\epsilon^2=0.4$ and $\tau=15$.For this combination of values we will observe two zero point of vector field $n$ at $r_+=1.996$ and $r_+=5.994$ which lied in two distinct shaded portion as shown in Figure \ref{11}. Figure.\ref {12c} and Figure \ref{12f} suggests that the winding numbers corresponding to $r_+=1.9968,5.9947$ (represented by the  black colored solid line and the red colored solid line respectively) are found to be $+1$, and $-1$ respectively. The sum of the winding numbers of the two discontinous branches gives us the total topological charge of the black hole, which in this case equals $1-1=0$.in conclusion of the above study we can say there will be always two allowed range of $r_+$ values for any $\epsilon^2$ value,for which we will get two branches.For the lower branch we will have winding number $-1$ and for the upper branch the winding number will be $1$.The total topological charge is always $0$.Pressure do not play any significant role here also similar to the case of 5D black hole with spherical horizon\\
	We now consider the GR limit by setting $\epsilon=1$ . The pressure is again kept at  $P=0.01$. The horizon radius $r_+$ is plotted  against $\tau$ considering the allowed range for $r_+$.  in Figure.\ref {E1a}. Here, we observe only one black hole branches.For $\tau=15$, the zero point is located at $r_+= 6.76458$.  The vector plot of $n$ in the $r_+-\theta$ plane  in  figure\ref {E1b} demonstrates the same. As shown in Figure.\ref {15c}, the winding number corresponding to $r_+= 6.76458$ (purple colored line) are found to be $+1$. Hence, the total topological charge  equals $1$.It is seen that similar to the spherical horizon case ,in GR limit the topological charge changes.For 5D HL black hole with hyperbolic horizon the topological charge changes from $0$ to $1.$ which is opposite to the transition $1 \to 0$ in case of  5D HL black hole with spherical horizon.The results are summarized in the table\ref{t4}. \\

\begin{table}[ht]
	\caption{Summary of results for $5D$ HL black hole with hyperbolic horizon in fixed $\epsilon$ ensemble.} 
	\centering 
	\begin{tabular}{|c| c |c| c|} 
		\hline 
		$\epsilon^2$ & $P$ & No of black hole branches & Topological charge  \\ [0.5ex] 
		\hline 
		0.40 & 0.008 & 2 & -1+1=0 \\ 
		0.40 & 0.05 & 2 & -1+1=0   \\
		0.40 & 0.1 & 2 & -1+1=0 \\
		0.6 & 0.008 & 2 & -1+1=0 \\
		0.6 & 0.05 & 2 & -1+1=0   \\
		0.6 &  0.1 & 2  & -1+1=0   \\
		1 & 0.008 & 1 & 1   \\
		1 & 0.05 & 1& 1   \ \\
		1 & 0.1 & 1&  1      \   \\ [1ex] 
		\hline 
	\end{tabular}
	\label{t4} 
\end{table}

	  \subsubsection{\textbf{Case II : For D=4}}
\begin{figure}[h]
	  	\centering
	  	\includegraphics[width=7cm,height=5cm]{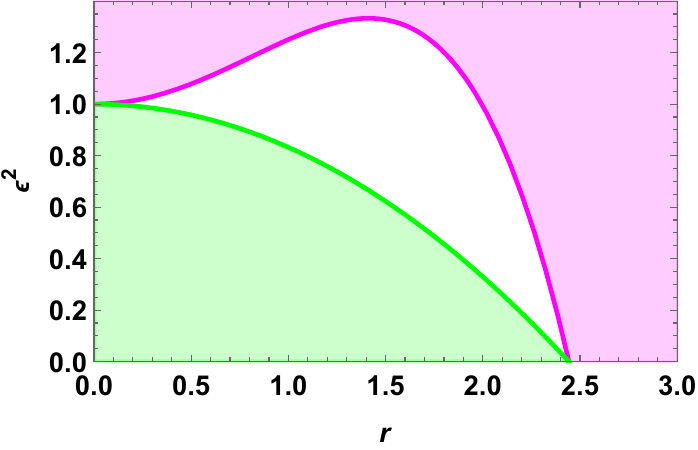}
	  	
	  	\caption{The relations between $\epsilon^2$
	  		and $r$ for the positive temperature for $k=-1$ in $D=4.$Temperature is positive only on the shaded portions.We have taken $P= 0.01$.}
	  	\label{14}
	  \end{figure}

	  The free energie expression for 4D HL black hole with hyperbolic horizon is found to be:
	 $$\mathcal{F}=\frac{16}{3} \pi  P  r_+^3-\frac{3 \left(\epsilon ^2-1\right) \log ( r_+)}{2 P \tau }-\frac{3 \left(\epsilon ^2-1\right)}{16 \pi  P  r_+}-\frac{2  r_+ (2 \pi   r_+ +\tau )}{\tau }$$
	 The expression for $\phi_{r}$ can be found out as-
	 $$\phi_{r}=\frac{128 \pi ^2 P r_+^3 (2 P r \tau -1)-8 \pi  r \left(4 P  r_+ \tau +3 \epsilon ^2-3\right)+3 \tau  \left(\epsilon ^2-1\right)}{16 \pi  P  r_+^2 \tau }$$
	 The zero points of the vector is obtained by setting $\phi_{r}=0$ and followed by getting a relation between $\tau$ and $r$ as-
	 $$\tau=\frac{8 \pi  \left(16 \pi  P  r_+^3+3 r \epsilon ^2-3  r_+\right)}{256 \pi ^2 P^2  r_+^4-32 \pi  P  r_+^2+3 \epsilon ^2-3}$$
	  For 4D HL black hole  with hyperbolic horizon also,we do not get any critical conditions hence there is no phase transition observed.While calculating the winding number some restrictions regarding the values of $r_+$ and $\epsilon$ are observed here as well  From figure \ref{14} it can be observed that we can have values of r and $\epsilon$ only from the shaded portions due to positive temperature condition.
	  Thus while calculating winding number from the $\tau$ vs $r_+$ curve for a particular value of $\epsilon$ and pressure $P$  we will have to take values of $r$ from the allowed range(shaded portions).\\
	  \begin{figure}[h]
	  	\centering
	  	\begin{subfigure}{0.32\textwidth}
	  		\includegraphics[width=\linewidth]{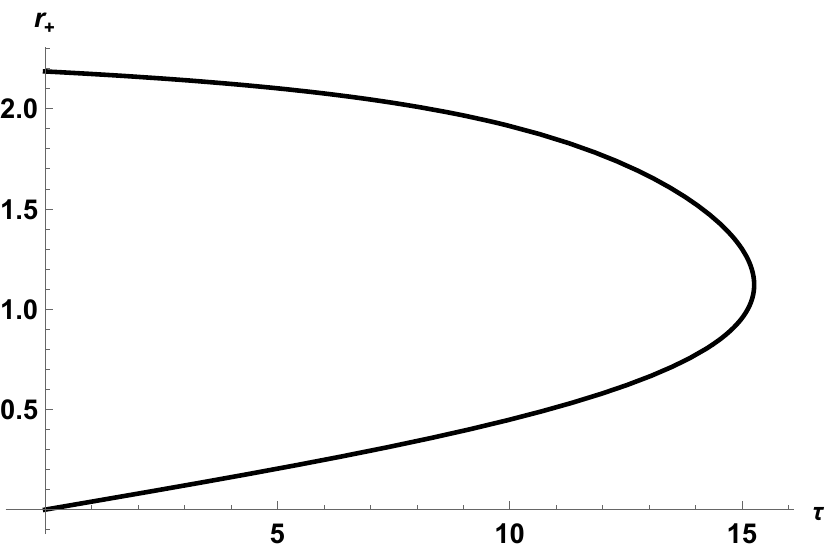}
	  		\caption{}
	  		\label{15a}
	  	\end{subfigure}
	  	\begin{subfigure}{0.32\textwidth}
	  		\includegraphics[width=\linewidth]{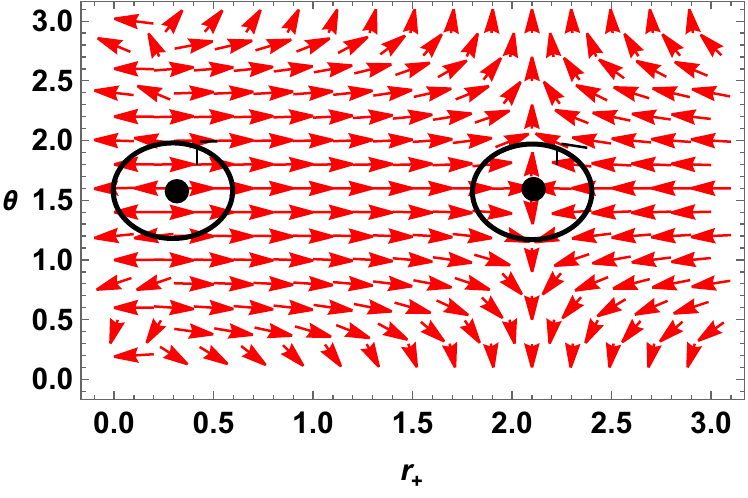}
	  		\caption{}
	  		\label{15b}
	  	\end{subfigure}
	  	\begin{subfigure}{0.32\textwidth}
	  		\includegraphics[width=\linewidth]{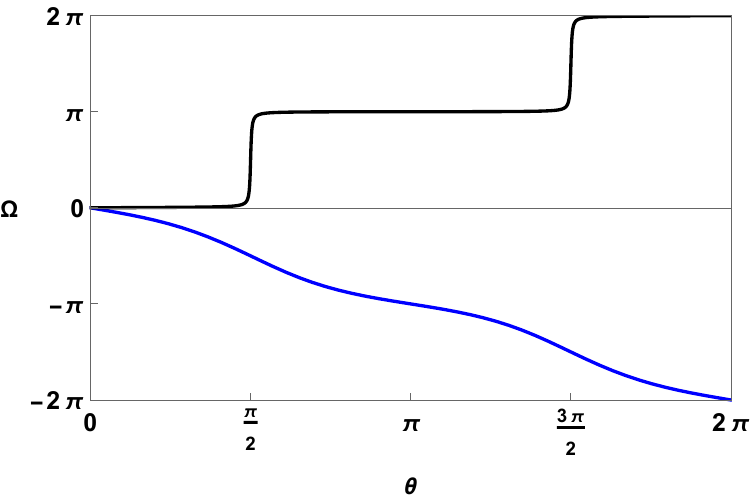}
	  		\caption{}
	  		\label{15c}
	  	\end{subfigure}
	  	
	  	\medskip
	  	
	  	\begin{subfigure}{0.32\textwidth}
	  		\includegraphics[width=\linewidth]{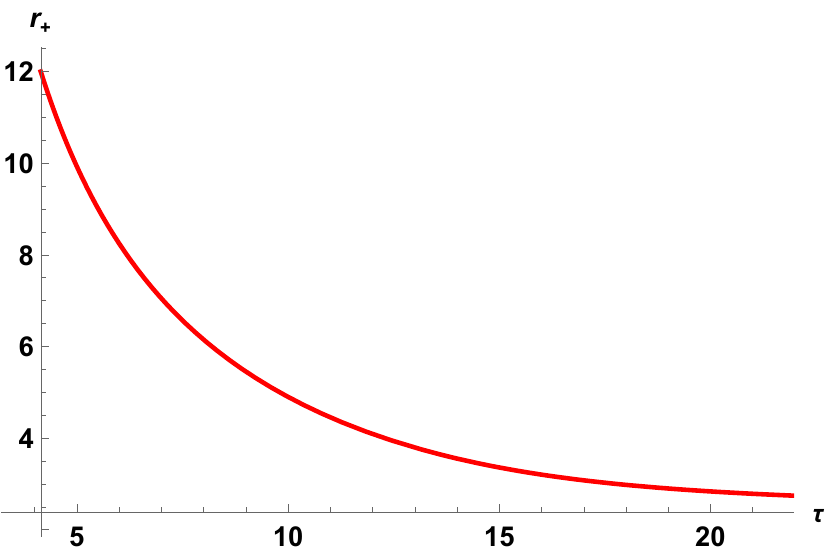}
	  		\caption{}
	  		\label{15d}
	  	\end{subfigure}
	  	\begin{subfigure}{0.32\textwidth}
	  		\includegraphics[width=\linewidth]{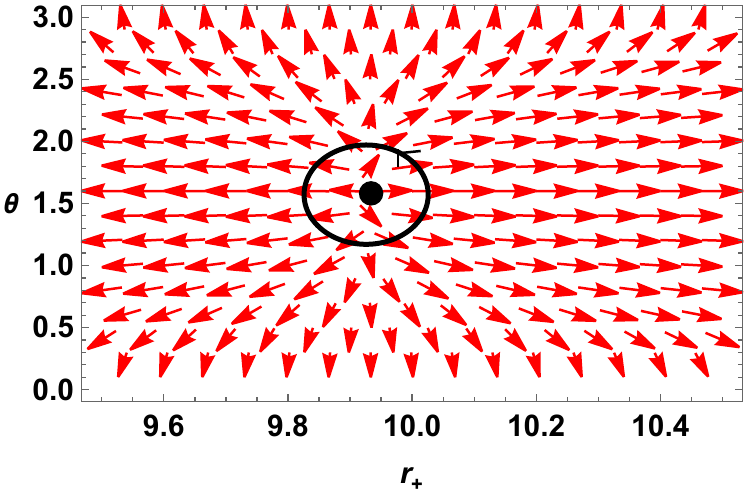}
	  		\caption{}
	  		\label{15e}
	  	\end{subfigure}
	  	\begin{subfigure}{0.32\textwidth}
	  		\includegraphics[width=\linewidth]{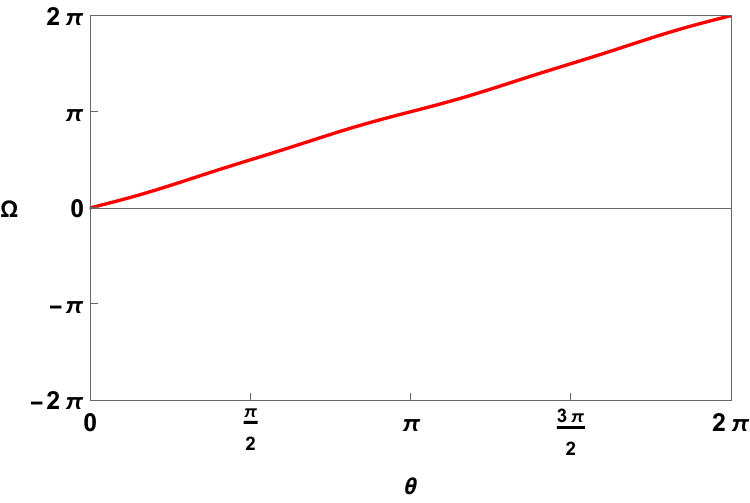}
	  		\caption{}
	  		\label{15f}
	  	\end{subfigure}
	  	
	  	\caption{Plots for $4D$ HL black hole with hyperbolic horizon at $\epsilon^2=0.2 $ with $ P=0.01$. Figure $\left(a\right)$ shows $\tau$ vs $r_+$ plot for $0 \leq r_+ \leq 2.18509$ range and $\left(d\right)$ shows the same for the range $2.37861 \leq r \leq 12$.figure $\left(b\right)$ and figure $\left(b\right)$ is the plot of vector field $n$ on a portion of $r_+-\theta$ plane for $\tau=5$ in the range $0 \leq r_+ \leq 2.18509$ and $2.37861 \leq r_+ \leq 12$ respectively.The zero points are located at $r_+=0.2041 $ and $r_+=2.10093$ for the range $0 \leq r_+ \leq 2.18509$ and $r_+=9.9259$ for the range  $2.37861 \leq r_+ \leq 12$. In figure $\left(c\right)$ computation of the contours around the zero points for $r_+=0.2041$ and $r_+=2.10093$ are shown in blue and black colored solid lines respectively.In $\left(f\right)$ computation of the contour around the zero points for $r_+=9.9259$ is shown in red colored solid line.}
	  	\label{15}
	  \end{figure}
	  For $P=0.01$,$\epsilon^2=0.4$ the horizon radius $r_+$ is plotted  against $\tau$  in Figure.\ref{15a} and Figure.\ref{15b}. Here, we observe two separate or discontinous black hole branches. One for the range $0 \leq r_+ \leq 2.18509$(Figure \ref{15a}) and other for the range $2.37861 \leq r_+ \leq 12$(Figure.\ref {15b}).For the case of 4D HL black hole depending upon the value of $\tau$ there will be one or three solution of the vector field $n$.Firstly we will do the analysis for calculating the winding numbers by keeping the pressure $P$ ,$\epsilon^2$ and $\tau$ fixed at $P=0.01$,$\epsilon^2=0.2$ and $\tau=5$ where three solution of the vector field is observed.For this combination of values we will observe three zero point of vector field $n$ at $r_+=0.2041$, $r_+=2.10093$ and $r_+=9.9259$ which lied in two distinct shaded portion as shown in Figure \ref{14}. Figure.\ref {15c} and Figure \ref{15f} suggests that the winding numbers corresponding to $r_+=0.2041$, $r_+=2.10093$ and $r_+=9.9259$  (represented by the  black,blueband the red colored solid line respectively) are found to be $+1$, and $-1$ and $+1$ respectively. The sum of the winding numbers of the two discontinous branches gives us the total topological charge of the black hole, which in this case equals $1-1+1=1$.Again for the same set values of $\epsilon^2$ and $P$ but for different $\tau$ value where there exist only one solution of vector field $n$ ,the winding number/topological charge is still 1 as all the single zero points will be on the red curve shown in Figure\ref{15d} and for that curve ,the topological charge is $1$.In conclusion of the above study we can say there will be always two allowed range of $r_+$ values for any $\epsilon^2$ value,for which we will get two branches for some values of $\tau$ and one branch for some values of $\tau$.For the lower branch we will have winding number $0$ and for the upper branch the winding number will be $1$.The total topological charge is always $1$.\\
	  	\begin{figure}[h]
	  	\centering
	  	\begin{subfigure}{0.32\textwidth}
	  		\includegraphics[width=\linewidth]{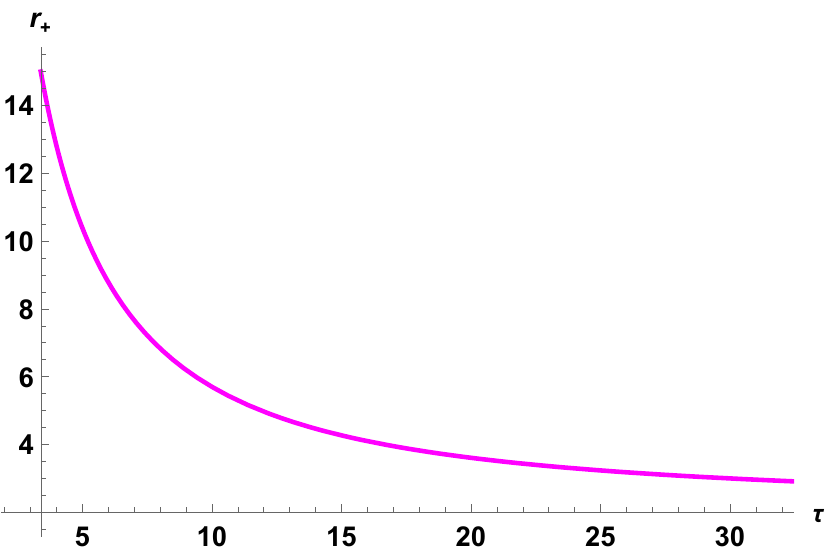}
	  		\caption{}
	  		\label{16a}
	  	\end{subfigure}
	  	\begin{subfigure}{0.32\textwidth}
	  		\includegraphics[width=\linewidth]{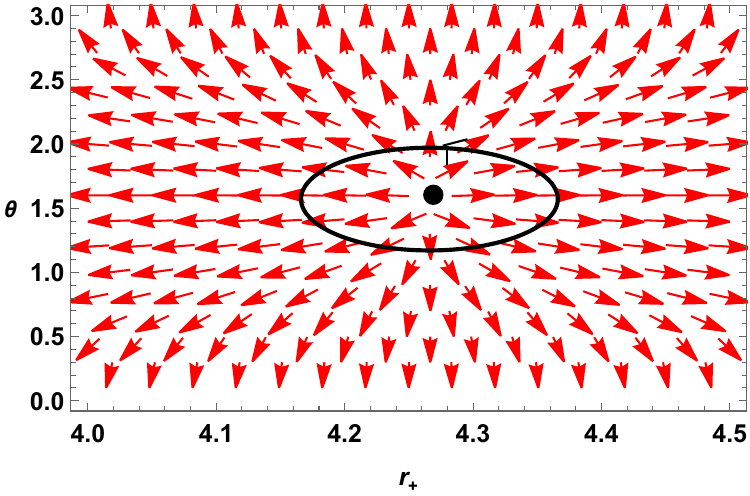}
	  		\caption{}
	  		\label{16b}
	  	\end{subfigure}
	  	\begin{subfigure}{0.32\textwidth}
	  		\includegraphics[width=\linewidth]{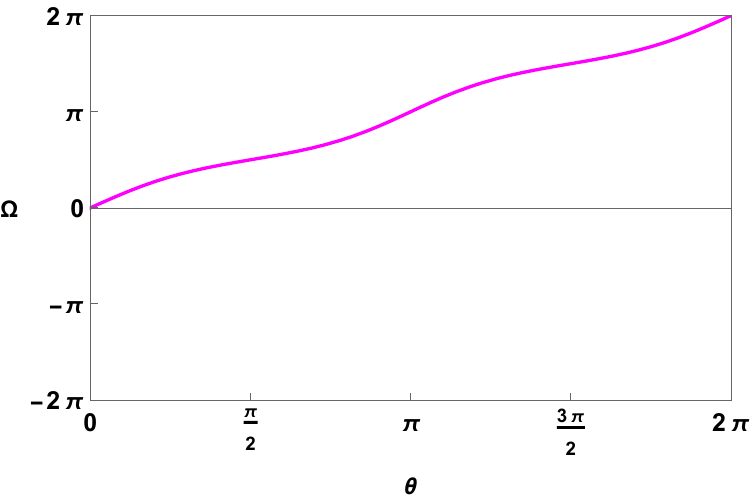}
	  		\caption{}
	  		\label{16c}
	  	\end{subfigure}
	  	\caption{ Plots for $4D$ HL black hole with hyperbolic horizon at $\epsilon=1 $ with $ P=0.01$. Figure $\left(a\right)$ shows $\tau$ vs $r_+$ plot,  figure $\left(b\right)$  is the plot of vector field $n$ on a portion of $r_+-\theta$ plane for $\tau=15$ . The zero point is located at $r_+=4.266 $. In figure $\left(c\right)$, computation of the contours around the zero point $r_+= 4.266$ is shown in magenta colored solid lines.}
	  	\label{16}
	  \end{figure}
	  We now consider the GR limit by setting $\epsilon=1$ . The pressure is again kept at  $P=0.01$. The horizon radius $r_+$ is plotted  against $\tau$ considering the allowed range for $r_+$.  in Figure.\ref {16a}. Here, we observe only one black hole branches.For $\tau=15$, the zero point is located at $r_+= 4.266$.  The vector plot of $n$ in the $r_+-\theta$ plane  in  Figure.\ref {16b} demonstrates the same. As shown in Figure.\ref {16c}, the winding number corresponding to $r_+= 4.266$ (magenta colored line) are found to be $+1$. Hence, the total topological charge  equals $1$.Strikingly it is seen that unlike 5D black holes with any horizon types,in GR limit the topological charge does not change for 4D HL black hole with hyperbolic horizon.Pressure do not play any significant role here also similar to the case of 4D black hole with spherical horizon.The results are summarized in the table \ref{t5} \\
	  \begin{table}[ht]
	  	\caption{Summary of results for $4D$ HL black hole with hyperbolic horizon in fixed $\epsilon$ ensemble.} 
	  	\centering 
	  	\begin{tabular}{|c| c |c| c|} 
	  		\hline 
	  		$\epsilon^2$ & $P$ & No of black hole branches & Topological charge  \\ [0.5ex] 
	  		\hline 
	  		0.2 & 0.008 & 3 & -1+1+1=1 \\ 
	  		0.2 & 0.05 & 3 & -1+1+1=1  \\
	  		0.2 & 0.1 & 3 & -1+1+1=1 \\
	  		0.4 & 0.008 & 3 & -1+1+1=1 \\
	  		0.4 & 0.05 & 3 & -1+1+1=1   \\
	  		0.4 &  0.1 & 3  & -1+1+1=1   \\
	  		1 & 0.008 & 1 & 1   \\
	  		1 & 0.05 & 1& 1   \ \\
	  		1 & 0.1 & 1&  1      \   \\ [1ex] 
	  		\hline 
	  	\end{tabular}
	  	\label{t5} 
	  \end{table}

	\section{Thermodynamic topology in fixed $\zeta$ ensemble}
	\subsection{For spherical horizon (k=1)}
	\subsubsection{\textbf{Case I : For D=5}}
	In fixed $\zeta$ ensemble, we define a  parameter $\zeta$ conjugate to $\epsilon$ and keep it fixed. The relation between these two can be found out by solving $\zeta=\frac{dM}{d\epsilon}$. The new mass of the system is given by the following transformation-
	$$\Tilde{M}=M-\zeta \epsilon$$
	Using equation (5) and substituting  $\epsilon=-\frac{4 \pi  P \zeta }{k^2}$ we get
	\begin{equation}
		\tilde{M}=\frac{k^2}{8 \pi  P}+\frac{2 k r_{+}^2}{3}+\frac{8}{9} \pi  P r_{+}^4+\frac{2 \pi  P \zeta ^2}{k^2}
	\end{equation}
	The new entropy for D=5 is given by-
	\begin{equation}
		\tilde{S}=\frac{k r_{+}}{P}+\frac{8 \pi  r_{+}^3}{9}-\frac{16 \pi ^2 P r_{+} \zeta ^2}{k^3}
	\end{equation}
	The modified free energy can be calculated using-
	\begin{equation}
		\tilde{\mathcal{F}}=\tilde{M}-\frac{\tilde{S}}{\tau}
	\end{equation}
	or
	\begin{equation}
		\tilde{\mathcal{F}}=\frac{16 \pi ^2 P r_{+} \zeta ^2}{k^3 \tau }+\frac{2 \pi  P \zeta ^2}{k^2}+\frac{k^2}{8 \pi  P}+k \left(\frac{2 r_{+}^2}{3}-\frac{r_{+}}{P \tau }\right)+\frac{8 \pi  r_{+}^3 (P r_{+} \tau -1)}{9 \tau }
	\end{equation}
The components of the vector field are obtained as:
\begin{equation}
	\phi^r=\frac{\partial\tilde{\mathcal{F}}}{\partial r_+}=\frac{32 \pi  k^3 P^2 r_+^3 \tau -24 \pi  k^3 P r_+^2+12 k^4 P r_+ \tau -9 k^4+144 \pi ^2 P^2 \zeta ^2}{9 k^3 P \tau }
	\label{phi5}
\end{equation}
and
\begin{equation}
	\phi^\theta=-\cot\Theta\csc\Theta
\end{equation}
	Finally we obtain the expression for $\tau$ as-
	\begin{equation}
		\tau=\frac{3 \left(8 \pi  k^3 P r_{+}^2+3 k^4-48 \pi ^2 P^2 \zeta ^2\right)}{4 k^3 P r_{+} \left(3 k+8 \pi  P r_{+}^2\right)}
		\label{eqsz5}
	\end{equation} 
	for $k=1$ i.e for spherical horizon, equation \ref{eqsz5} takes the form 
	$$\tau=\frac{3 \left(8 \pi  k^3 P r_{+}^2+3 k^4-48 \pi ^2 P^2 \zeta ^2\right)}{4 k^3 P r_{+} \left(3 k+8 \pi  P r_{+}^2\right)}$$
	It is important to note that in this paper we have only considered the range $0\leq\epsilon^2\leq 1$,If values of $\epsilon^2$ is equal to 1 then HL gravity returns back to general relativity and $\epsilon$=0 corresponds to the  detailed-balance condition.Based on this restriction a range is defiend for $\zeta$ as $0 \leq \zeta^2 \leq 63.389$(pressure is kept fixed as P=0.01).\\
	In fixed $\zeta$ ensemble the $\lambda$ phase transition property disappears and other interesting phenomenon observed is : while transforming to grand canonical ensemble , we lost one topological horizon i.e the flat horizon as k can not take zero as a value here.\\
	\begin{figure}[h]	
		\centering
		\begin{subfigure}{0.32\textwidth}
			\includegraphics[width=\linewidth]{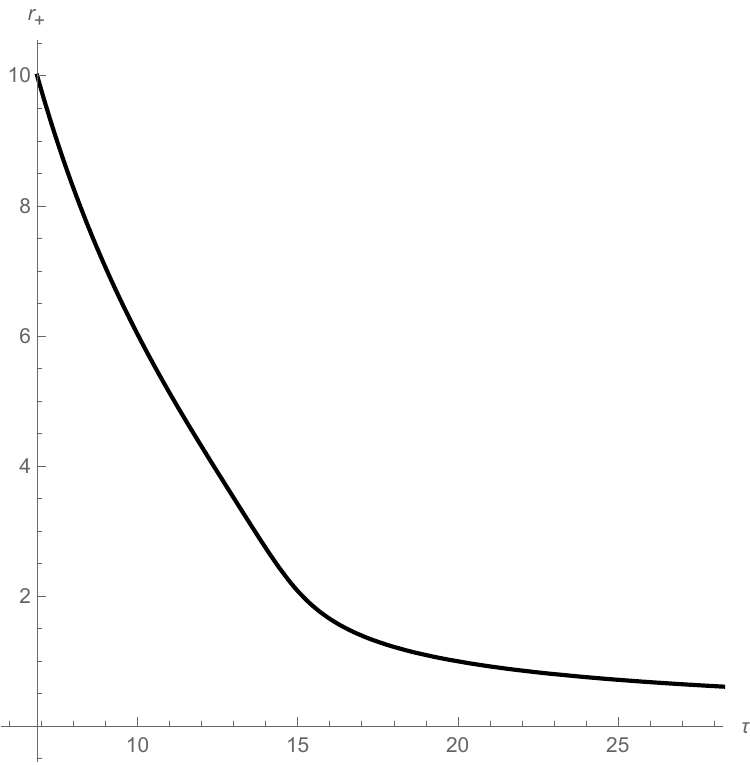}
			\caption{}
			\label{17a}
		\end{subfigure}
		\begin{subfigure}{0.32\textwidth}
			\includegraphics[width=\linewidth]{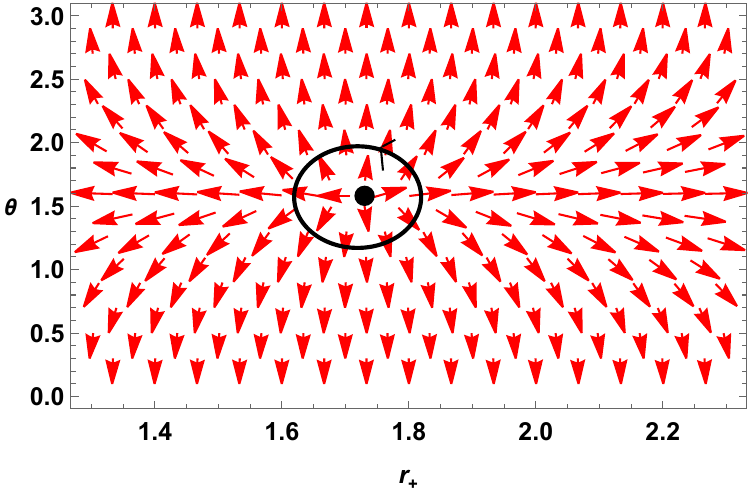}
			\caption{}
			\label{17b}
		\end{subfigure}
		\begin{subfigure}{0.32\textwidth}
			\includegraphics[width=\linewidth]{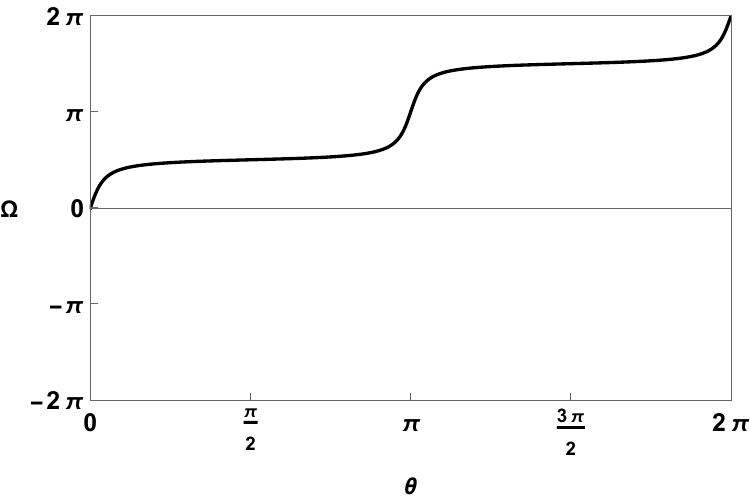}
			\caption{}
			\label{17c}
		\end{subfigure}
		\caption{ Plots for $5D$ HL black hole with spherical horizon in fixed $\zeta$ ensemble at $\zeta=7.2 $ with $ P=0.01$. Figure $\left(a\right)$ shows $\tau$ vs $r_+$ plot,  figure $\left(b\right)$  is the plot of vector field $n$ on a portion of $r_+-\theta$ plane for $\tau=15$ . The red arrows represent the vecor field $n$. The zero point is located at $r_+=1.719. $ In figure $\left(c\right)$, computation of the contour around the zero point $\tau=15$ and $r_+=1.719$ is  shown.}
		\label{17}
	\end{figure}
	
	Firstly the horizon radius $r_+$ is plotted  against $\tau$ for different values $\zeta$ keeping the pressure $P$ fixed.It is seen that depending on the values of $\zeta$ 5D black hole exhibit different number of branches.For $\zeta$ value smaller than a critical value it shows only one branch and no phase transition.In Figure.\ref {17a}one such scenario is being demonstrated. Here, we observe a single black hole branch whose points are also the zero points of $\phi$. For $\tau=15$, the zero point is located at $r_+=1.719$. This is also confirmed from the vector plot of $n$ in the $r_+-\theta$ plane as shown in  Figure.\ref {17b}. For finding out the winding number/topological number associated with this zero point, we perform a contour integration around $r_+=1.719$ which is shown in Figure.\ref {17c}. It reveals that the topological charge in this case is equal to $+1$. We have explicitly verified that the topological charge of any zero point on the black hole branch remains the same and is equal to $+1$.\\

	\begin{figure}[h]
		\centering
		\begin{subfigure}{0.32\textwidth}
			\includegraphics[width=\linewidth]{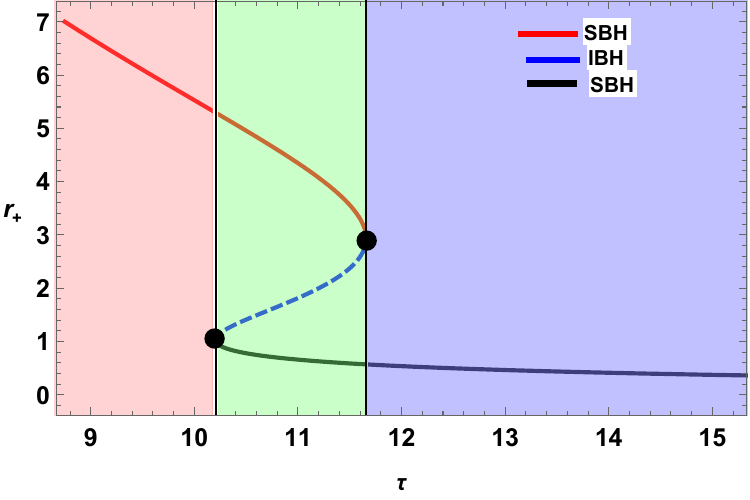}
			\caption{}
			\label{18a}
		\end{subfigure}
		\begin{subfigure}{0.32\textwidth}
			\includegraphics[width=\linewidth]{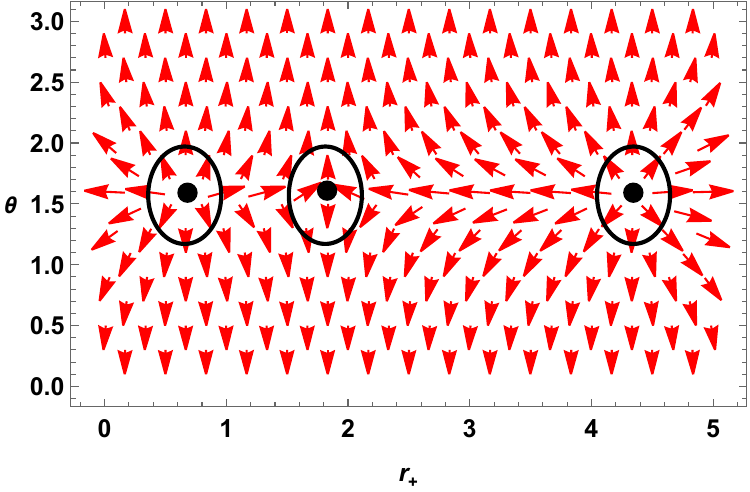}
			\caption{}
			\label{18b}
		\end{subfigure}
		\begin{subfigure}{0.32\textwidth}
			\includegraphics[width=\linewidth]{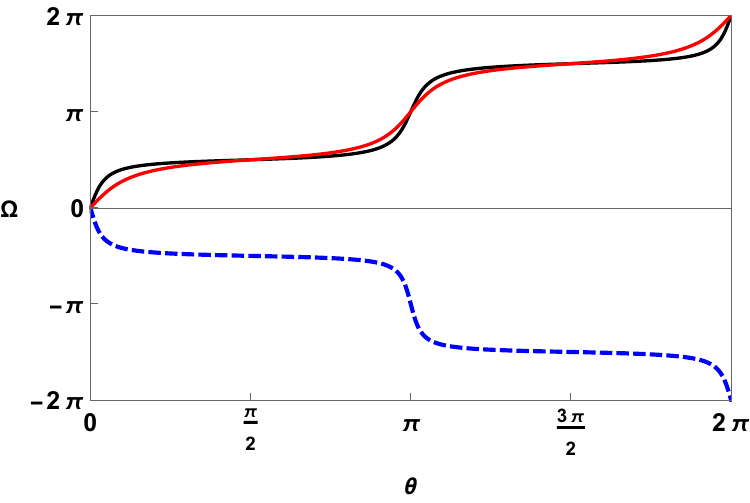}
			\caption{}
			\label{18c}
		\end{subfigure}
		\caption{ Plots for $5D$ HL black hole with spherical horizon in fixed $\zeta$ ensemble at $\zeta=7.7 $ with $ P=0.01$. Figure $\left(a\right)$ shows $\tau$ vs $r_+$ plot,  figure $\left(b\right)$  is the plot of vector field $n$ on a portion of $r_+-\theta$ plane for $\tau=11$ . The zero points are located at $r_+=$. In figure $\left(c\right)$, computation of the contours around the zero points $r_+=0.657, 1.813, 4.346$ are shown in black colored solid line, blue colored  dashed line and red colored solid line respectively.}
		\label{18}
	\end{figure}
	
	We repeat the analysis keeping  pressure constant at  $P=0.01$ with $\zeta=7.7$ which is above the critical value.The horizon radius $r_+$ is plotted  against $\tau$  in Figure.\ref {17a}. Here, we observe three black hole branches: a small, an intermediate and a large black hole branch. For $\tau=10$, three zero points are located a t$r_+=0.5408, 1.495, 5.463$. This is again confirmed from the vector plot of $n$ in the $r_+-\theta$ plane as shown in  Figure.\ref {17b}. As shown in Figure.\ref {17c}, the winding numbers corresponding to  $r_+=0.657, 1.813, 4.346$ (represented by the  black colored solid line, the blue colored  dashed line and the red colored solid line respectively) are found to be $+1$, $-1$ and $+1$ respectively.The sum of the winding numbers of the three branches gives us the total topological charge of the black hole, which in this case equals $1-1+1=1$. From Figure.\ref {3a}, we can also see a generation point at $\tau=11.6571, r_+= 2.9323$ and  an annihilation point at $\tau=10.2052, r_+= 1.02765$ which are shown as black dots. \\

	\begin{figure}[h]
		\centering
		\begin{subfigure}{0.32\textwidth}
			\includegraphics[width=\linewidth]{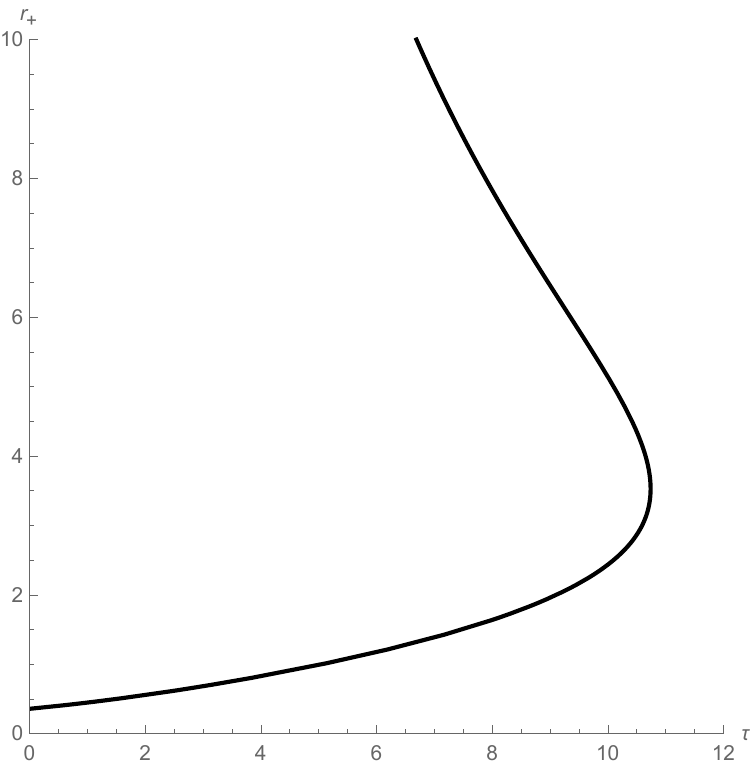}
			\caption{}
			\label{19a}
		\end{subfigure}
		\begin{subfigure}{0.32\textwidth}
			\includegraphics[width=\linewidth]{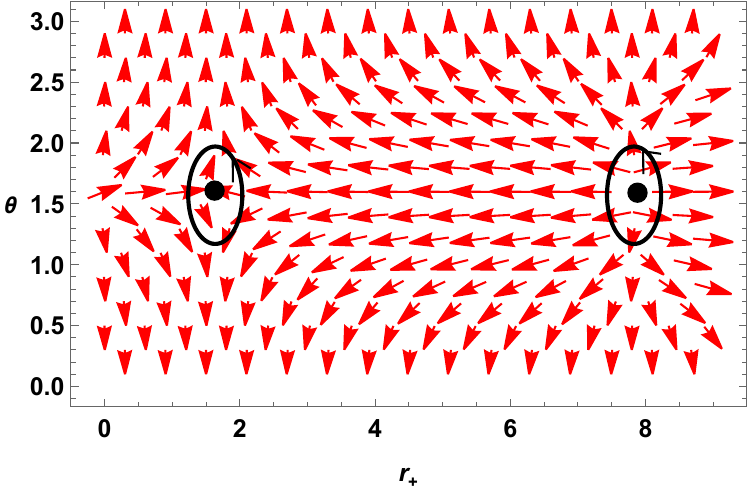}
			\caption{}
			\label{19b}
		\end{subfigure}
		\begin{subfigure}{0.32\textwidth}
			\includegraphics[width=\linewidth]{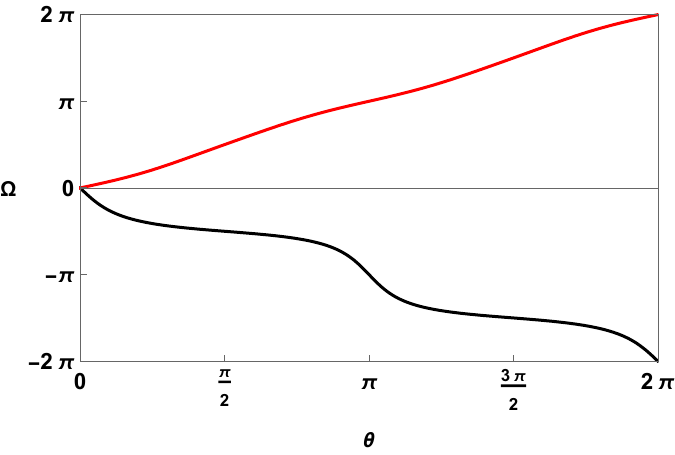}
			\caption{}
			\label{19c}
		\end{subfigure}
		\caption{ Plots for $5D$ HL black hole with spherical horizon in fixed $\zeta$ ensemble at $\zeta=8 $ with $ P=0.01$. Figure $\left(a\right)$ shows $\tau$ vs $r_+$ plot,  figure $\left(b\right)$  is the plot of vector field $n$ on a portion of $r_+-\theta$ plane for $\tau=8$ . The zero points are located at $r_+= 1.6366,7.8313$. In figure $\left(c\right)$, computation of the contours around the zero points $r_+= 1.6366,7.8313$  are shown in black colored and red colored lines respectively.}
		\label{19}
	\end{figure}
	
	We now consider the case $\zeta=8$ which is beyond the defined limit of $0 \leq \zeta^2 \leq 63.389$.The pressure is again kept at  $P=0.01$. The horizon radius $r_+$ is plotted  against $\tau$  in Figure.\ref {19a}. Here, we observe two black hole branches: a small and a large black hole branch. For $\tau=8$, the zero points are located at $r_+=1.6366,7.8313 $.  The vector plot of $n$ in the $r_+-\theta$ plane  in  Figure.\ref {19b} demonstrates the same. As shown in Figure.\ref {19c}, the winding numbers corresponding to $r_+= 1.6366,7.8313$ (black colored and red colored lines respectively) are found to be $-1$ and $+1$ respectively. Hence, the total topological charge  equals $1-1=0$.\\
	\newpage
	\begin{figure}[h]
		\centering
		\begin{subfigure}{0.32\textwidth}
			\includegraphics[width=\linewidth]{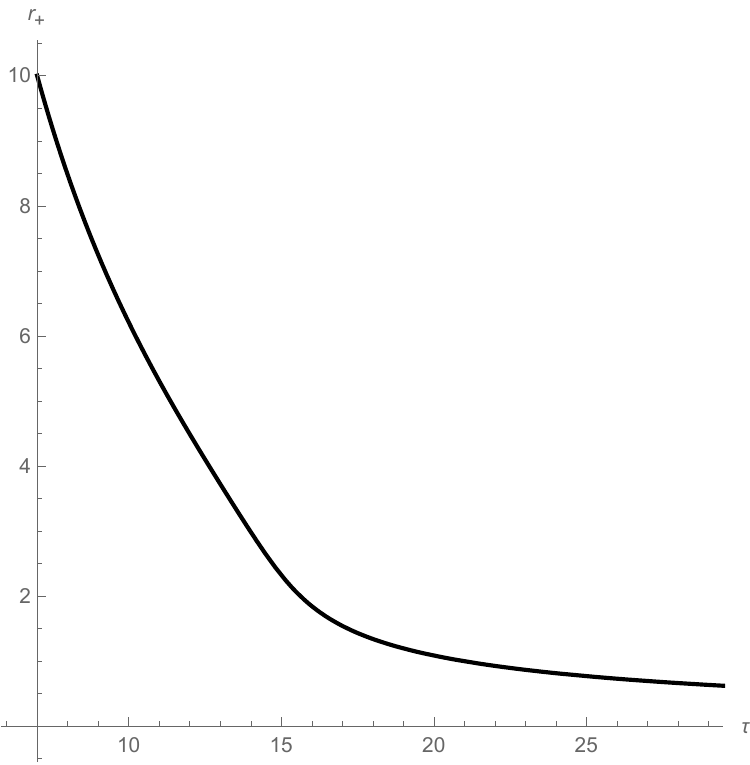}
			\caption{}
			\label{20a}
		\end{subfigure}
		\begin{subfigure}{0.32\textwidth}
			\includegraphics[width=\linewidth]{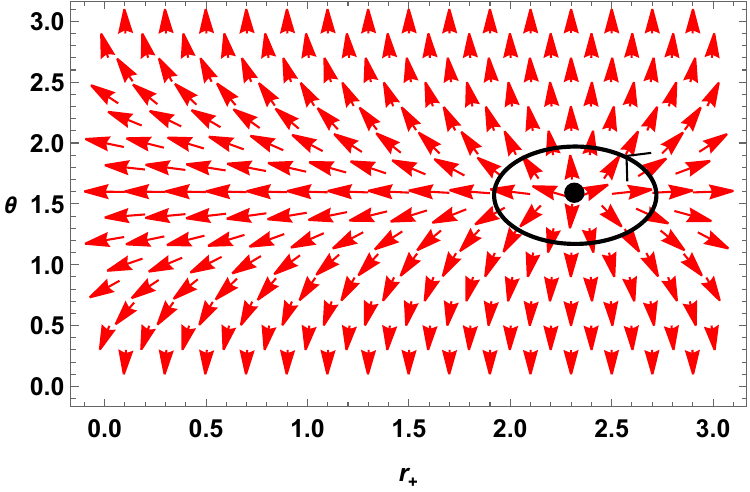}
			\caption{}
			\label{20b}
		\end{subfigure}
		\begin{subfigure}{0.32\textwidth}
			\includegraphics[width=\linewidth]{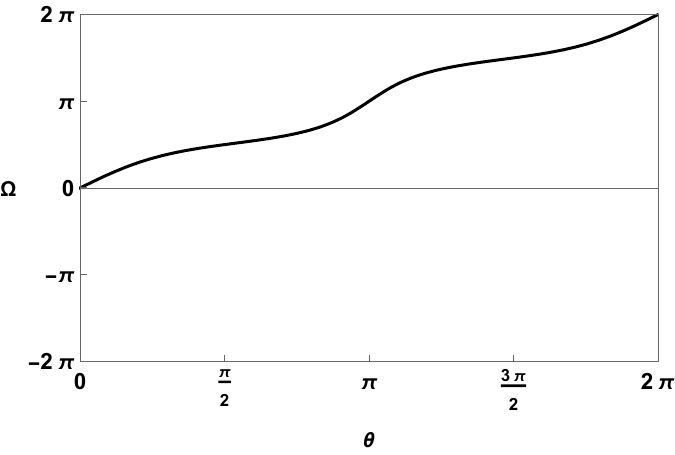}
			\caption{}
			\label{20c}
		\end{subfigure}
		\begin{subfigure}{0.32\textwidth}
		\includegraphics[width=\linewidth]{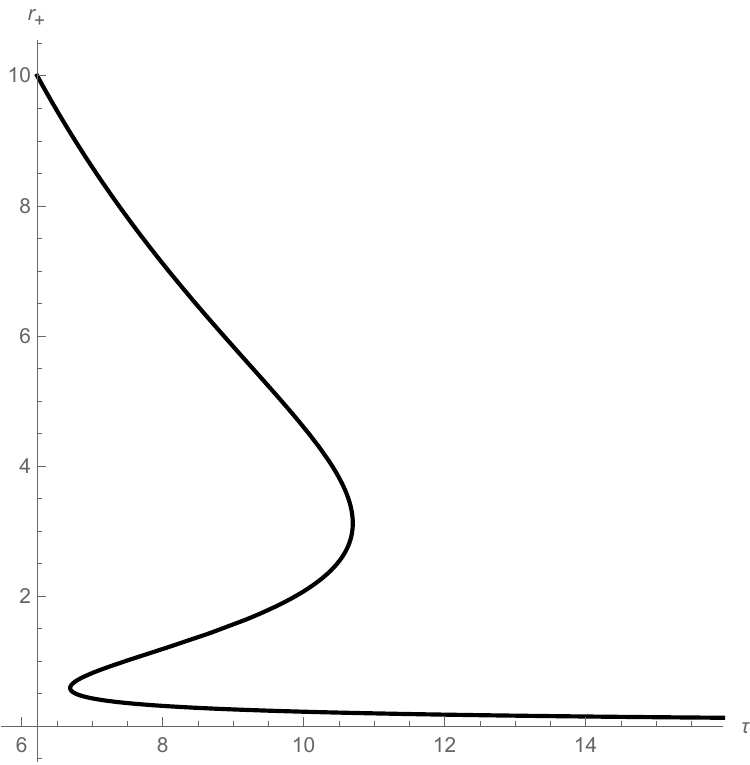}
		\caption{}
		\label{20d}
	\end{subfigure}
	\begin{subfigure}{0.32\textwidth}
	\includegraphics[width=\linewidth]{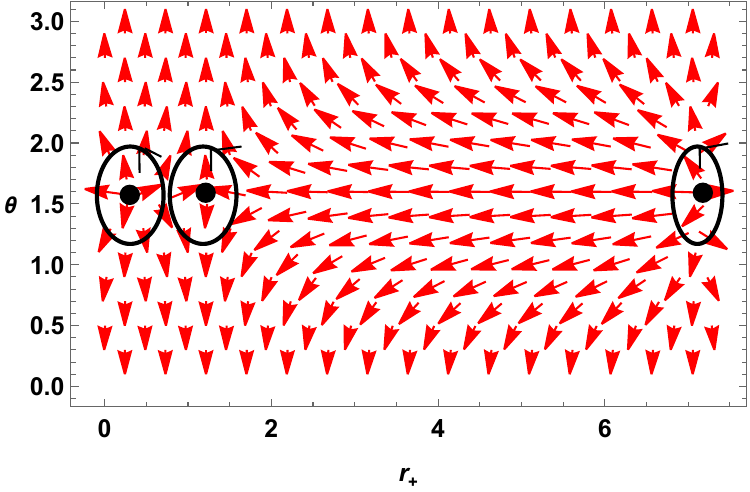}
	\caption{}
	\label{20e}
\end{subfigure}
	\begin{subfigure}{0.32\textwidth}
	\includegraphics[width=\linewidth]{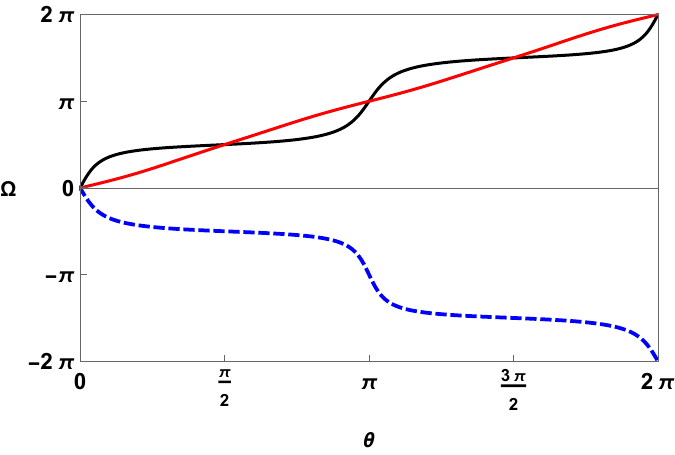}
	\caption{}
	\label{20f}
\end{subfigure}
	\begin{subfigure}{0.32\textwidth}
	\includegraphics[width=\linewidth]{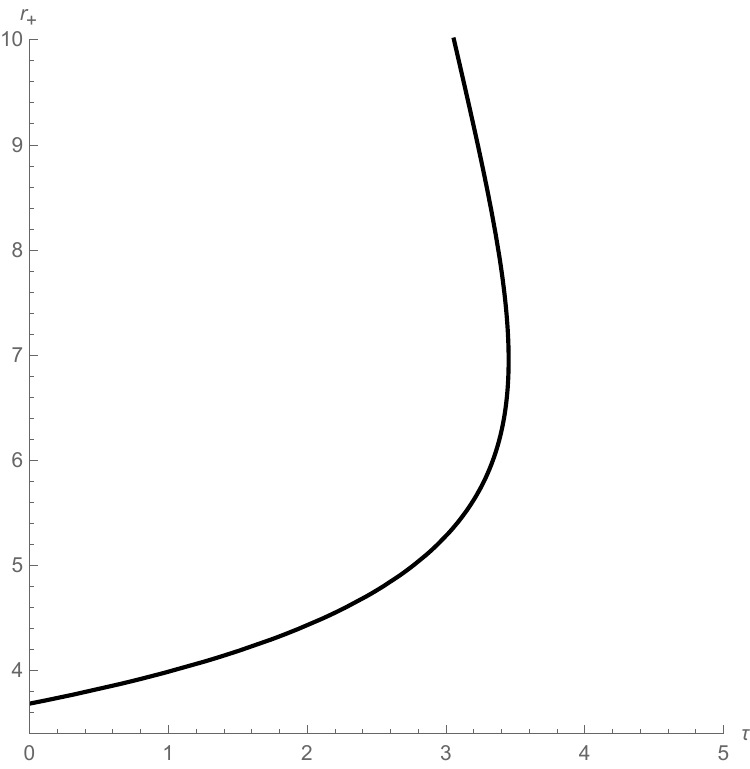}
	\caption{}
	\label{20g}
\end{subfigure}
	\begin{subfigure}{0.32\textwidth}
	\includegraphics[width=\linewidth]{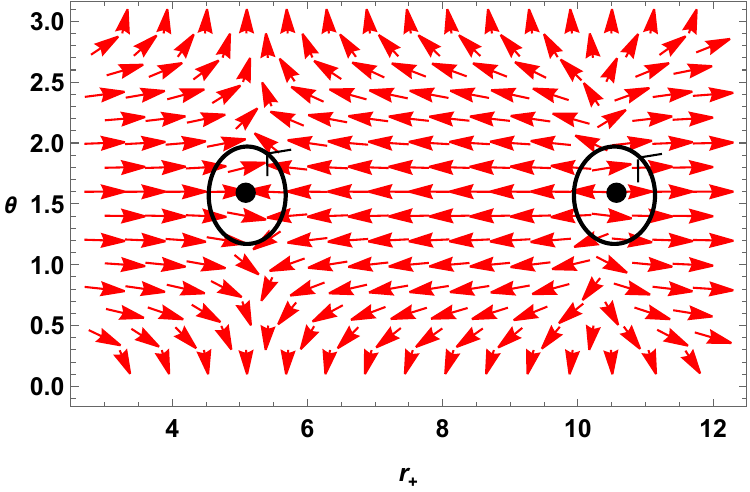}
	\caption{}
	\label{20h}
\end{subfigure}
	\begin{subfigure}{0.32\textwidth}
	\includegraphics[width=\linewidth]{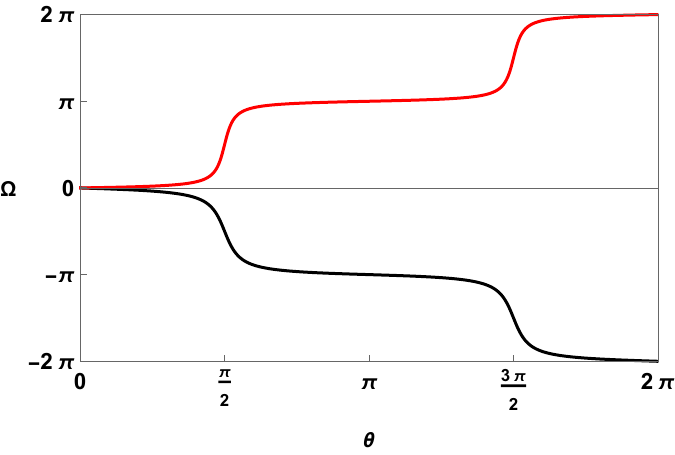}
	\caption{}
	\label{20i}
\end{subfigure}
\caption{Effect of variation of pressure on number black hole branches and total topological charge.The $\zeta$ value is kept fixed at $\zeta=7.2$ and for different values of pressure $r_+$ vs $\tau$ graphs are plotted from which we calculate the totaltopological charge.First ,second and third row corresponds to $P=0.0098,0.0109,0.02$ respectively.  }
\label{20}
\end{figure}
Above analysis is followed by plotting the horizon radius $r_+$  against $\tau$ for different values pressure $P$ keeping $\zeta$ value fixed. Unlike fixed $\epsilon$ ensemble ,here pressure becomes an important quantity.The number of black hole branch along with the total topological charge is changed when pressure is increased and $\zeta$ is kept fixed.Figure \ref{20} confirms the same where variation of pressure is shown and $\zeta$ is kept fixed at $\zeta=7.2$.It is seen that when the pressure is lowered from $P=0.01$ to $P=0.0098$ then the number of branch(Figure \ref{20a}) and total topological remains the same(Figure \ref{20c}) .When pressure is slightly increased to $P=0.0109$ then first order phase transition takes place(Figure \ref{20d}).However the total topological charge remains the same i.e $1$(Figure \ref{20f}).When pressure is increased further to $P=0.02$ then the number of black hole branch decreases to two(Figure \ref{20g}) and the total topological charge becomes zero(Figure \ref{20i}).Variations of the topological charge and number of black hole branches with $\zeta$ and $P$ are shown in Table \ref{t6}	
\begin{table}[ht]
	\caption{Summary of results for $5D$ HL black hole with spherical horizon in fixed $\zeta$ ensemble.} 
	\centering 
	\begin{tabular}{|c| c |c| c| c| c|} 
		\hline 
		$\zeta$ & $P$ & No of black hole branches & Topological charge &  No of generation points & No of annihilation points\\ [0.5ex] 
		\hline 
		7.1 & 0.008 & 1 & 1 & 0 & 0  \\ 
		7.1 & 0.0109 & 3 &1-1+1=1& 1 & 1   \\
		7.1 & 0.02 & 2 & -1+1=0 & 1 & 0  \\
		7.5 & 0.008 & 1 & 1 & 0  & 0 \\
		7.5 & 0.0105 & 3  & 1-1+1=1 & 1 & 1   \\
		7.5 &  0.0109 & 2  & 1-1=0 &1 &0  \\
		7.8 & 0.008 & 1 & 1 & 0 &0   \\
		7.8 & 0.0101 & 3 & -1+1+1=1 & 1 & 1   \ \\
		7.8 & 0.0105 & 2 &- 1+1=0 & 1 &  0   \   \\ [1ex] 
		\hline 
	\end{tabular}
	\label{t6} 
\end{table}	
	\subsubsection{\textbf{Case II : For D=4}}
	
	In this case ,substituting  $\epsilon=-\frac{8 \pi  P r_{+} \zeta }{3 k^2}$ we get the new mass as-
	\begin{equation}
		\tilde{M}=\frac{3 k^2}{16 \pi  P r_{+}}+2 k r_{+}+\frac{16}{3} \pi  P r_{+}^3+\frac{4 \pi  P r_{+} \zeta ^2}{3 k^2}
	\end{equation}
	Also,the entropy for fixed $zeta$ ensemble in 4D is given by-
	\begin{equation}
		\tilde{S}=\log (r_{+}) \left(\frac{3 k}{2 P}-\frac{32 \pi ^2 P r_{+}^2 \zeta ^2}{3 k^3}\right)+4 \pi  r_{+}^2
	\end{equation}
	The modified free energy is-
	
	\begin{equation}
		\tilde{\mathcal{F}}=\frac{1}{48} \left(\frac{8 \log (r_{+}) \left(64 \pi ^2 P^2 r_{+}^2 \zeta ^2-9 k^4\right)}{k^3 P \tau }+\frac{64 \pi  P r_{+} \zeta ^2}{k^2}+\frac{9 k^2}{\pi  P r_{+}}+96 k r_{+}+\frac{64 \pi  r_{+}^2 (4 P r_{+} \tau -3)}{\tau }\right)
	\end{equation}
The components of the vector field are obtained as:
\begin{equation}
	\phi^r=\frac{1}{48} \left(\frac{8 \left(64 \pi ^2 P^2 r_+^2 \zeta ^2-9 k^4\right)}{k^3 P r_+ \tau }-\frac{9 k^2}{\pi  P r_+^2}+\frac{1024 \pi ^2 P r_+ \zeta ^2 \log (r)}{k^3 \tau }+\frac{64 \pi  P \zeta ^2}{k^2}+96 k+256 \pi  P r_+^2+\frac{128 \pi  r_+ (4 P r_+ \tau -3)}{\tau }\right)
	\label{phi4}
\end{equation}
and
\begin{equation}
	\phi^\theta=-\cot\Theta\csc\Theta
\end{equation}
	And finally the expression for $\tau$ as-
	\begin{equation}
		\tau=\frac{8 \pi  \left(-48 \pi  k^3 P r_{+}^3-9 k^4 r_{+}+64 \pi ^2 P^2 r_{+}^3 \zeta ^2+128 \pi ^2 P^2 r_{+}^3 \zeta ^2 \log (r_{+})\right)}{k \left(-768 \pi ^2 k^2 P^2 r_{+}^4-96 \pi  k^3 P r_{+}^2+9 k^4-64 \pi ^2 P^2 r_{+}^2 \zeta ^2\right)}
		\label{eqsz4}
	\end{equation} 
	For spherical horizon 
	$$\tau=\frac{8 \pi  r_{+} \left(64 \pi ^2 P^2 r_{+}^2 \zeta ^2+128 \pi ^2 P^2 r_{+}^2 \zeta ^2 \log (r)-48 \pi  P r_{+}^2-9\right)}{64 \pi ^2 P^2 r_{+}^2 \left(12 r_{+}^2+\zeta ^2\right)+96 \pi  P r_{+}^2-9}$$
	
	Similar to 5D case, the study is confined within the range $0 \leq \zeta^2 \leq 63.389$(for pressure $ P=0.01$) and here also the $\lambda$ phase transition property disappears.In 4D also the flat horizon does not exit.\\
	\begin{figure}[h]	
		\centering
		\begin{subfigure}{0.32\textwidth}
			\includegraphics[width=\linewidth]{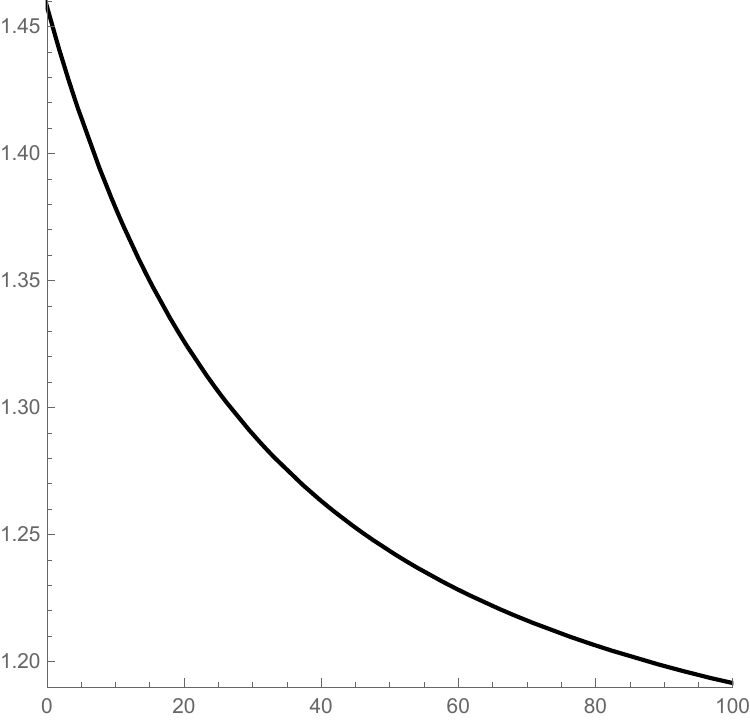}
			\caption{}
			\label{21a}
		\end{subfigure}
		\begin{subfigure}{0.32\textwidth}
			\includegraphics[width=\linewidth]{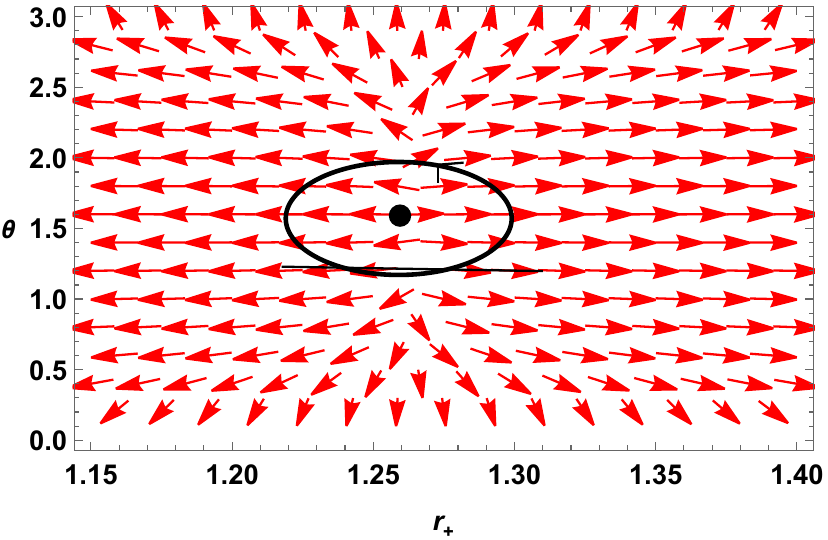}
			\caption{}
			\label{21b}
		\end{subfigure}
		\begin{subfigure}{0.32\textwidth}
			\includegraphics[width=\linewidth]{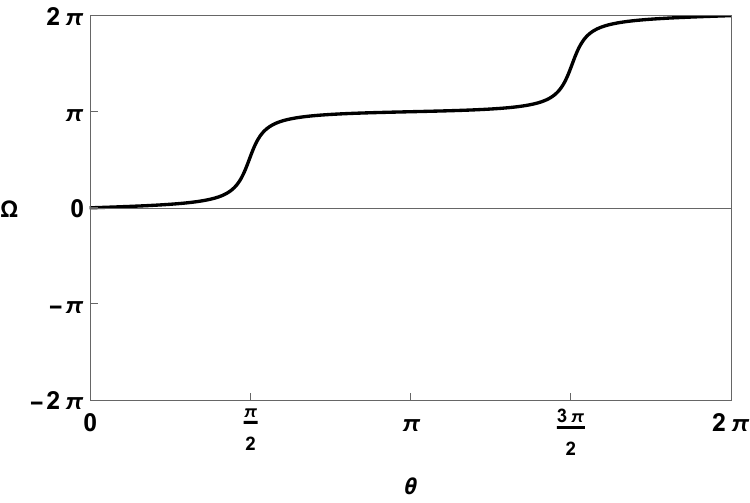}
			\caption{}
			\label{21c}
		\end{subfigure}
		\caption{ Plots for $4D$ HL black hole with spherical horizon in fixed $\zeta$ ensemble at $\zeta=7.2 $ with $ P=0.01$. Figure $\left(a\right)$ shows $\tau$ vs $r_+$ plot,  figure $\left(b\right)$  is the plot of vector field $n$ on a portion of $r_+-\theta$ plane for $\tau=40$ . The red arrows represent the vecor field $n$. The zero point is located at $r_+=1.259. $ In figure $\left(c\right)$, computation of the contour around the zero point $\tau=40$ and $r_+=1.259$ is  shown.}
		\label{21}
	\end{figure}
	
	The horizon radius $r_+$ is plotted  against different values $\tau$ for different set of values of pressure $P$ and $\zeta$ and it is seen that for all values of $P,\zeta$, 4D black hole has only one branch.In Figure.\ref {21a} one such scenario is being demonstrated. Here, we observe a single black hole branch whose points are also the zero points of $\phi$. For $\tau=40$, the zero point is located at $r_+=1.259$. This is also confirmed from the vector plot of $n$ in the $r_+-\theta$ plane as shown in  Figure.\ref {21b}. For finding out the winding number/topological number associated with this zero point, we perform a contour integration around $r_+=1.259$ which is shown in Figure.\ref {21c}. It reveals that the topological charge in this case is equal to $+1$. We have explicitly verified that for any value of pressure and $\zeta$ the topological charge of any zero point on the black hole branch remains constant which is equal to $+1$.The table \ref{t7} shows the same.\\
	
	\begin{table}[ht]
		\caption{Summary of results for $4D$ HL black hole with spherical horizon in fixed $\zeta$ ensemble.} 
		\centering 
		\begin{tabular}{|c| c |c| c |} 
			\hline 
			$\zeta$ & $P$ & No of black hole branches & Topological charge\\ [0.5ex] 
			\hline 
			7.1 & 0.008 & 1 & 1 \\ 
			7.1 & 0.0109 & 1 &1  \\
			7.1 & 0.02 & 1 & 1   \\
			7.5 & 0.008 & 1 & 1 \\
			7.5 & 0.0105 & 1  & 1  \\
			7.5 &  0.0109 & 1  & 1  \\
			7.8 & 0.008 & 1 & 1   \\
			7.8 & 0.0101 & 1& 1   \ \\
			7.8 & 0.0105 & 1 & 1    \   \\ [1ex] 
			\hline 
		\end{tabular}
		\label{t7} 
	\end{table}

	\subsection{For hyperbolic horizon(k=-1)}
	\subsubsection{\textbf{Case I : For D=5}}
For hyperbolic horizon we put $k=-1$ in the expression \ref{phi5} and \ref{eqsz5} which gives
	\begin{equation}
		\phi^r=\frac{\partial\tilde{\mathcal{F}}}{\partial r_+}=\frac{-144 \pi ^2 \zeta ^2 P^2+32 \pi  P^2 r_+^3 \tau -12 P r_+ \tau -24 \pi  P r_+^2+9}{9 P \tau }
	\end{equation}
	and
	\begin{equation}
		\tau=\frac{-3 \left(8 \pi P r_{+}^2+3 -48 \pi ^2 P^2 \zeta ^2\right)}{-4 P r_{+} \left(-3+8 \pi  P r_{+}^2\right)}
	\end{equation} 
	
\begin{figure}[h]
	\centering
	\begin{subfigure}{0.32\textwidth}
		\includegraphics[width=\linewidth]{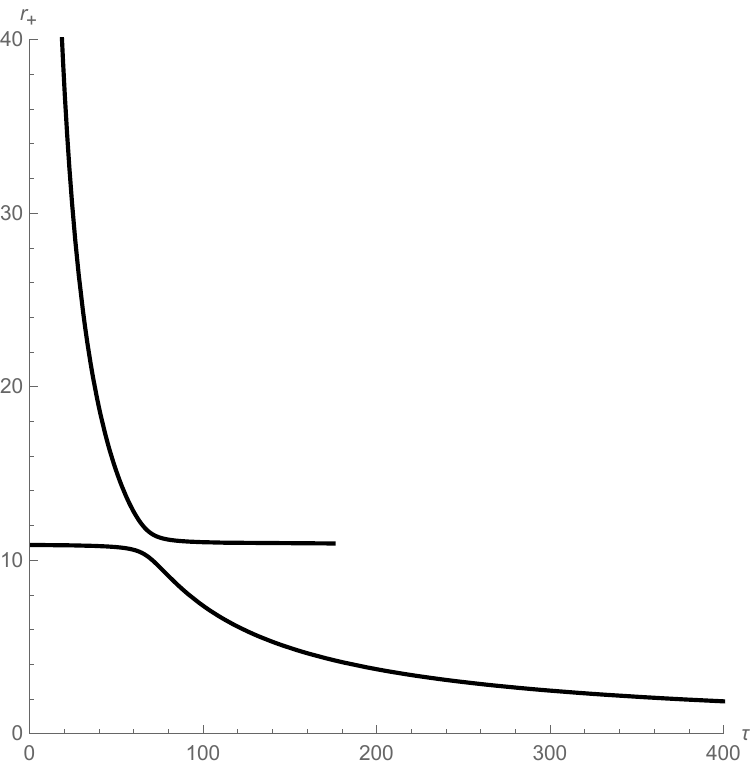}
		\caption{}
		\label{24a}
	\end{subfigure}
	\begin{subfigure}{0.32\textwidth}
		\includegraphics[width=\linewidth]{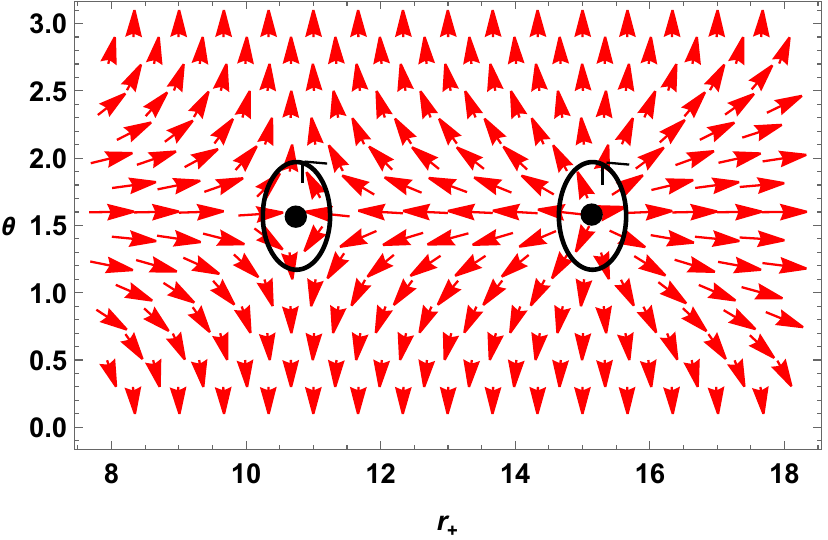}
		\caption{}
		\label{24b}
	\end{subfigure}
	\begin{subfigure}{0.32\textwidth}
		\includegraphics[width=\linewidth]{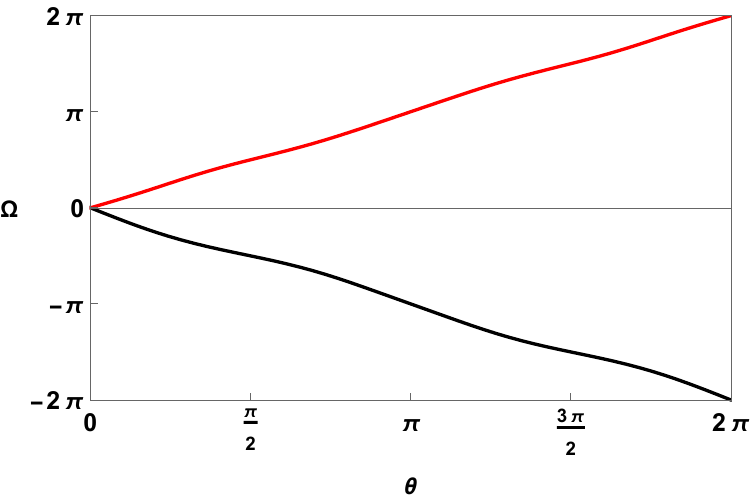}
		\caption{}
		\label{24c}
	\end{subfigure}
	\medskip
	\begin{subfigure}{0.32\textwidth}
		\includegraphics[width=\linewidth]{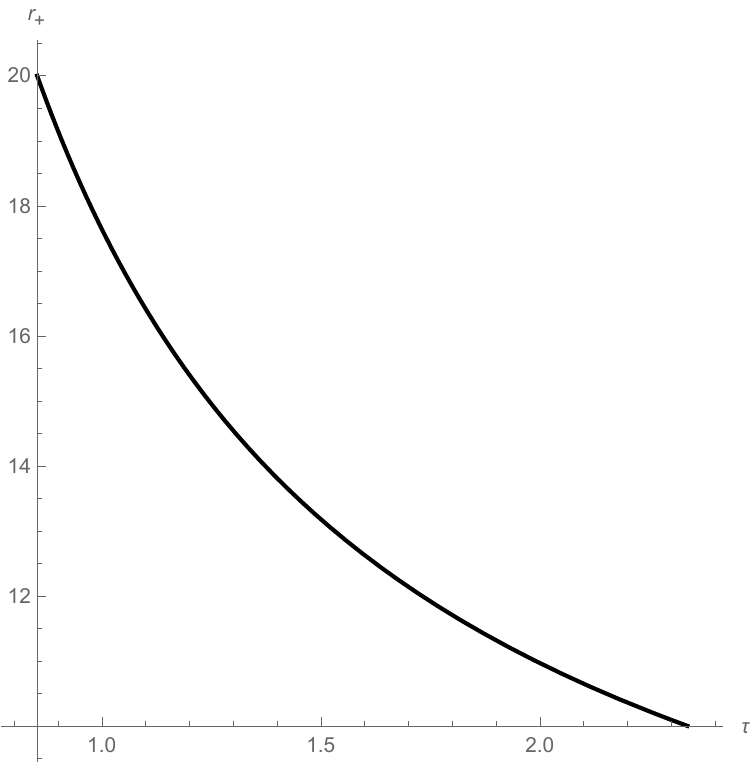}
		\caption{}
		\label{24d}
	\end{subfigure}
	\begin{subfigure}{0.32\textwidth}
		\includegraphics[width=\linewidth]{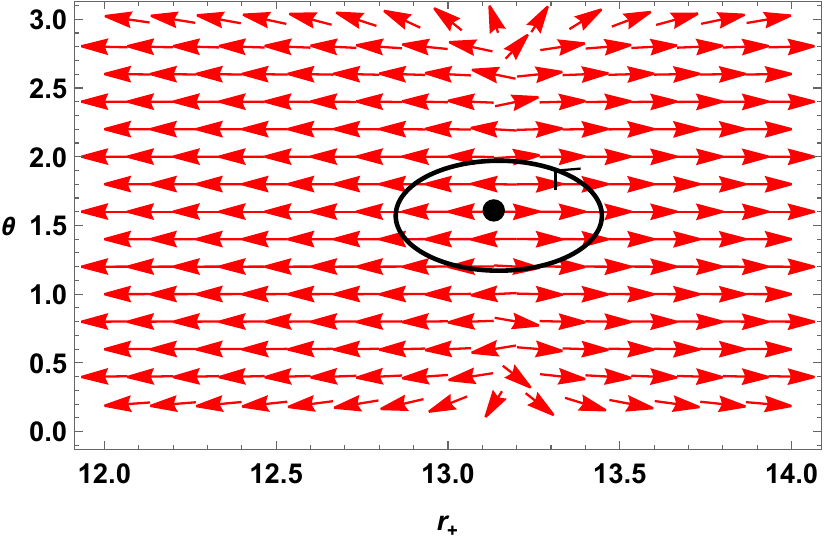}
		\caption{}
		\label{24e}
	\end{subfigure}
	\begin{subfigure}{0.32\textwidth}
		\includegraphics[width=\linewidth]{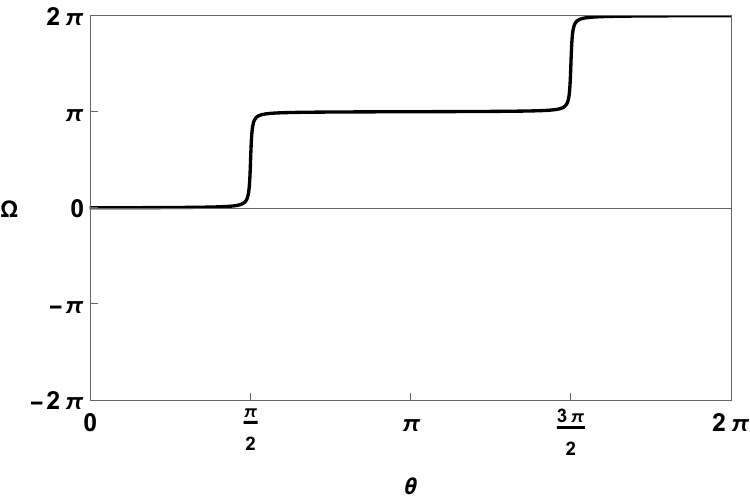}
		\caption{}
		\label{24f}
	\end{subfigure}

	\caption{Plots for $5D$ HL black hole with hyperbolic horizon in fixed $\zeta$ ensemble.Here $\zeta=7.6 $ is kept fixed and pressure is varied.In figure $\left(a\right)$ and $\left(d\right)$ shows $\tau$ vs $r_+$ plot for pressure $P=0.001$ and $P=0.05$ respectively.Figure $\left(b\right)$ and figure $\left(e\right)$ is the plot of vector field $n$ on a portion of $r_+-\theta$ plane .For figure $\left(b\right)$ $\tau$ value is taken to be $\tau=50$ and the zero points are located at $r_+=10.7482$ and   $r_+=15.1483$ for lower and upper allowed region respectively. In figure $\left(c\right)$ computation of the contours around the zero points for $r_+=0.0.7118$ and   $r_+=15.6966$ are shown in black and red colored solid lines respectively.For figure $\left(e\right)$ $\tau$ value is taken to be $\tau=1.5$ and the zero point is located at $r_+=13.1782$ .In figure $\left(f\right)$ computation of the contours around the zero points for $r_+=13.1782$ is shown.}
	\label{24}
\end{figure}

  For 5D HL black hole  with hyperbolic horizon in fixed $\zeta$ ensemble also,we do not get any critical conditions hence there is no phase transition observed.While calculating winding number from the $\tau$ vs $r_+$ curve for a particular value of $\zeta$ and pressure $P$  we will have to take values of $r$ from the allowed range where temperature is positive.\\
\begin{figure}[h]
	\centering
	\begin{subfigure}{0.32\textwidth}
		\includegraphics[width=\linewidth]{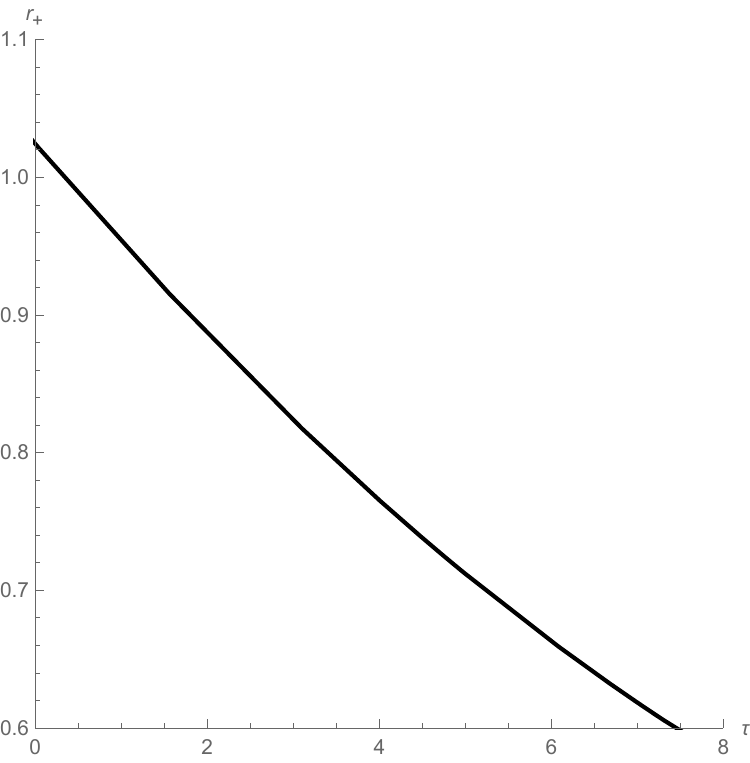}
		\caption{}
		\label{22a}
	\end{subfigure}
	\begin{subfigure}{0.32\textwidth}
		\includegraphics[width=\linewidth]{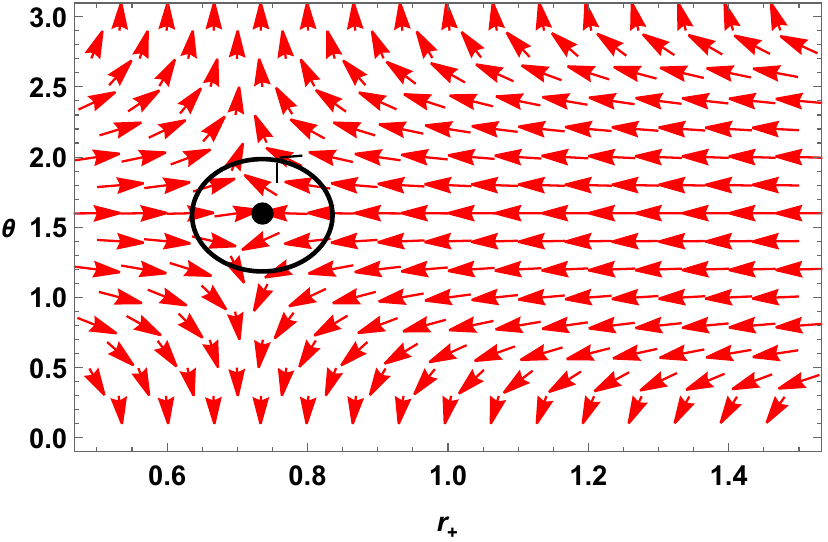}
		\caption{}
		\label{22b}
	\end{subfigure}
	\begin{subfigure}{0.32\textwidth}
		\includegraphics[width=\linewidth]{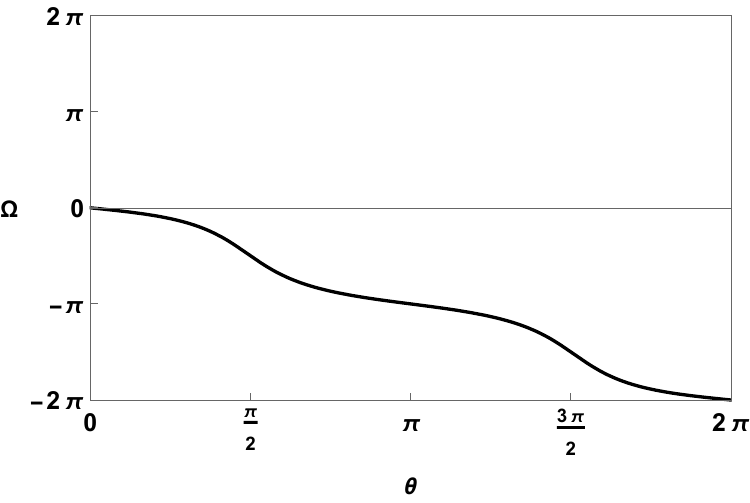}
		\caption{}
		\label{22c}
	\end{subfigure}
	
	\medskip
	
	\begin{subfigure}{0.32\textwidth}
		\includegraphics[width=\linewidth]{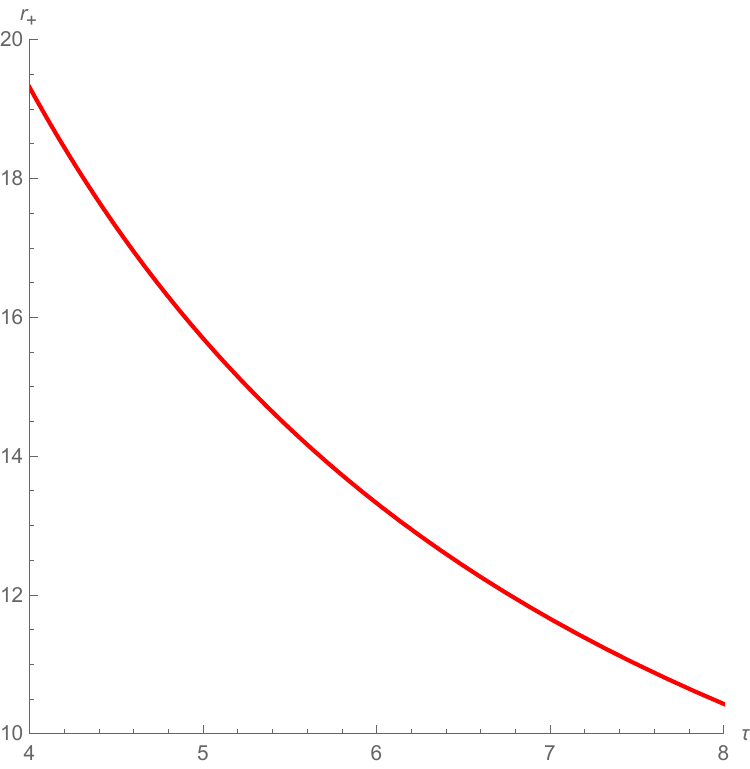}
		\caption{}
		\label{22d}
	\end{subfigure}
	\begin{subfigure}{0.32\textwidth}
		\includegraphics[width=\linewidth]{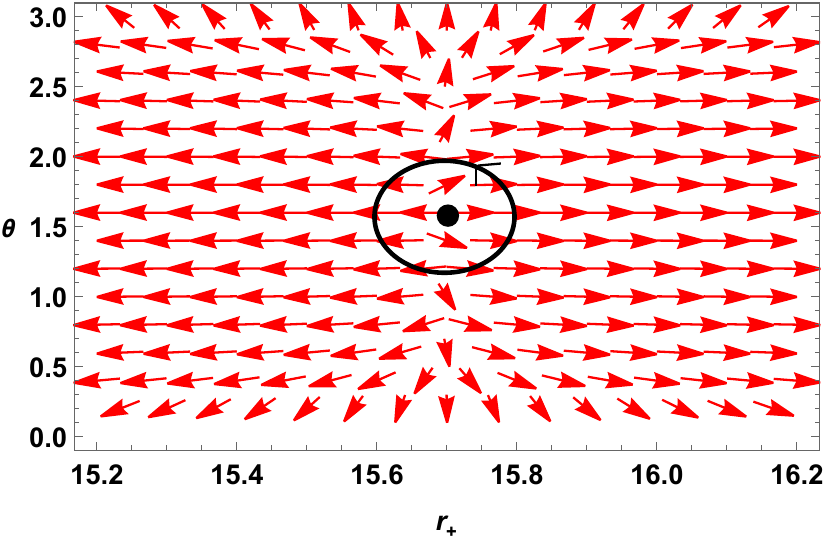}
		\caption{}
		\label{22e}
	\end{subfigure}
	\begin{subfigure}{0.32\textwidth}
		\includegraphics[width=\linewidth]{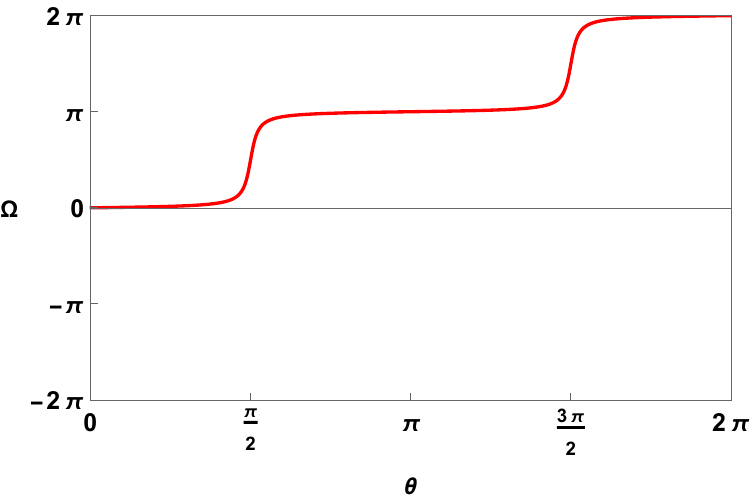}
		\caption{}
		\label{22f}
	\end{subfigure}
	
	\caption{Plots for $5D$ HL black hole with hyperbolic horizon in fixed $\zeta$ ensemble.Here $\zeta=7.6 $ with $ P=0.01$. Figure $\left(a\right)$ and $\left(d\right)$ shows $\tau$ vs $r_+$ plot for two allowed range of values of $r_+$ where temperature is positive.Figure $\left(b\right)$ and figure $\left(b\right)$ is the plot of vector field $n$ on a portion of $r_+-\theta$ plane for $\tau=5$ for both the allowed range.The zero points are located at $r_+=0.0.7118$ and   $r_+=15.6966$ for lower and upper allowed region respectively. In figure $\left(c\right)$ computation of the contours around the zero points for $r_+=0.0.7118$ and   $r_+=15.6966$ are shown in black and red colored solid lines respectively.}
	\label{22}
\end{figure}
It is being observed that at a fixed pressure $P=0.01$ in case of 5D HL black hole with hyperbolic horizon in fixed $\zeta$ ensemble have two types of topological charge $0$ and $1$.At a fixed pressure,for a range of $\zeta$ value we observe two black hole branch in two allowed range of $r_+$ where temperature is positive. For this scenario the total topological charge is equals to $-1+1=0$. One of such scenario is shown in figure \ref{22} where the horizon radius $r_+$ is plotted  against $\tau$  for $\zeta=7.6$ and pressure $P=0.01$.In Figure.\ref{22a} and Figure.\ref{22b} $\tau$ vs $r_+$ plot is represented for two different allowed range of $r_+$ for same set of values of $\zeta$ and $P$.For the case of $\tau=5$, there will two zero points of the vector field $n$ at $r_+=0.2041$, $r_+=2.10093$ as shown in Figure \ref{22b} and figure \ref{22e}. Figure.\ref {22c} and figure \ref{22f} suggests that the winding numbers corresponding to $r_+=0.2041$, $r_+=2.10093$ (represented by the  black and the red colored solid line respectively) are found to be $-1$, and $+1$ respectively. The sum of the winding numbers of the two discontinous branches gives us the total topological charge of the black hole, which in this case equals $1-1=0$.In conclusion of the above study we can say there will be always two allowed range of $r_+$ values for a range of $\zeta$ value at a fixed pressure,for which we will get two branches for some values of $\tau$.For the lower branch we will have winding number $-1$ and for the upper branch the winding number will be $1$.The total topological charge is always $0$.\\
\begin{figure}[h]
	\centering
	\begin{subfigure}{0.32\textwidth}
		\includegraphics[width=\linewidth]{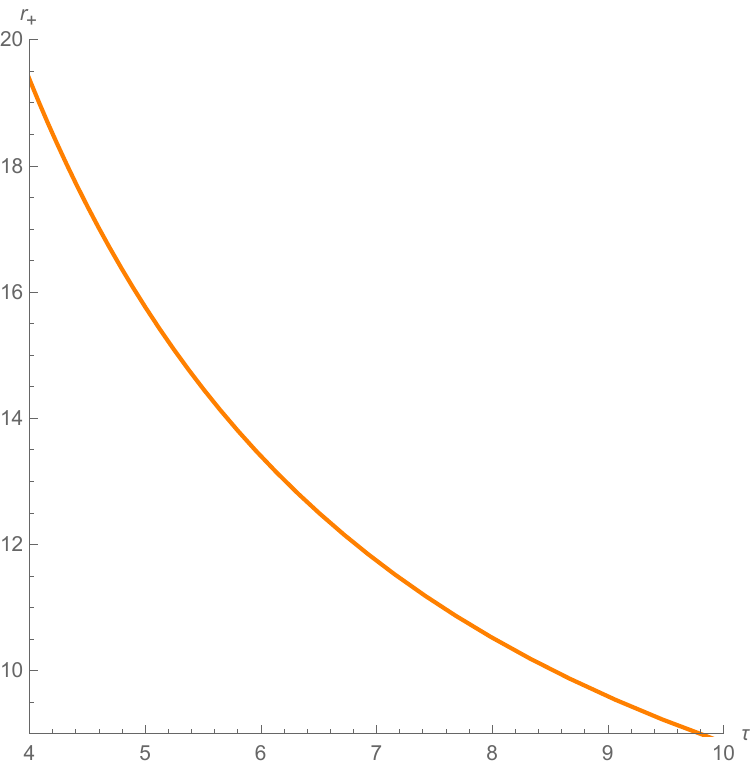}
		\caption{}
		\label{23a}
	\end{subfigure}
	\begin{subfigure}{0.32\textwidth}
		\includegraphics[width=\linewidth]{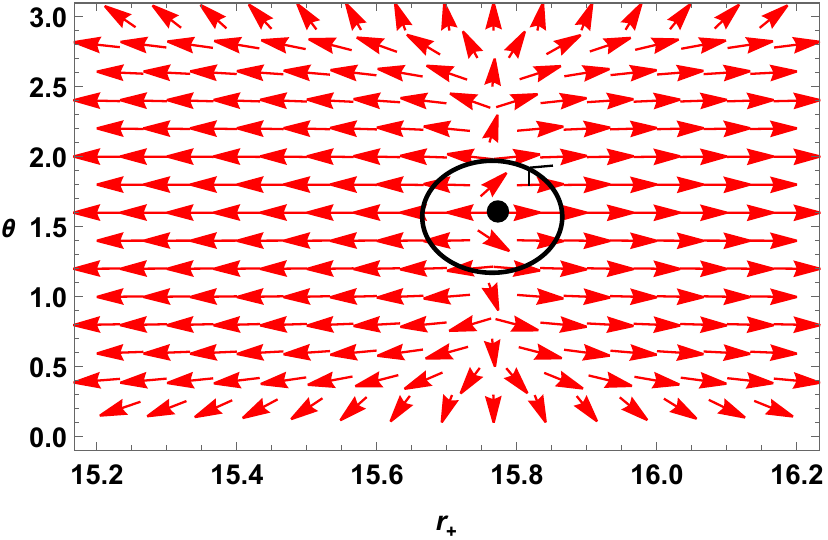}
		\caption{}
		\label{23b}
	\end{subfigure}
	\begin{subfigure}{0.32\textwidth}
		\includegraphics[width=\linewidth]{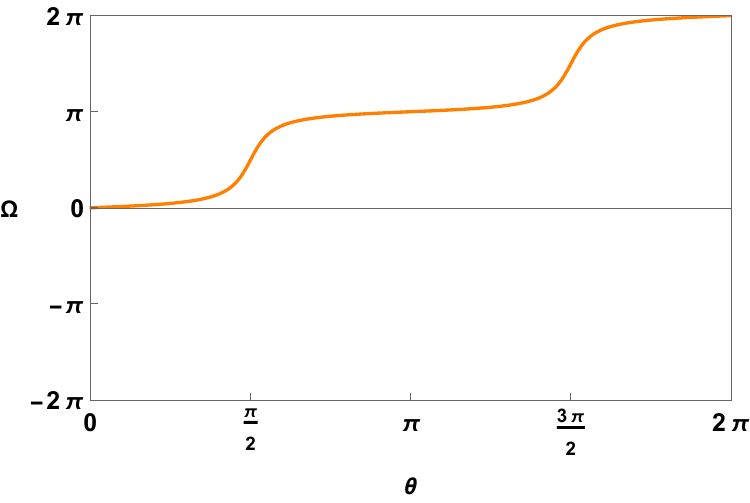}
		\caption{}
		\label{23c}
	\end{subfigure}
	\caption{ Plots for $5D$ HL black hole with hyperbolic horizon in fixed $\zeta$ ensemble,at $\zeta=8 $ with $ P=0.01$. Figure $\left(a\right)$ shows $\tau$ vs $r_+$ plot,  figure $\left(b\right)$  is the plot of vector field $n$ on a portion of $r_+-\theta$ plane for $\tau=5$ . The zero point is located at $r_+=15.7648 $. In figure $\left(c\right)$, computation of the contours around the zero point $r_+= 15.7648$ is shown in orange colored solid lines.}
	\label{23}
\end{figure}
We now consider a higher range of values of $zeta$ for the same value of pressure. Taking $\zeta=8$,horizon radius $r_+$ is plotted  against $\tau$ considering the allowed range for $r_+$ in Figure.\ref {23a}. Here, we observe only one black hole branches.For $\tau=5$, the zero point is located at $r_+= 15.7648$.  The vector plot of $n$ in the $r_+-\theta$ plane  in  Figure.\ref {23b} demonstrates the same. As shown in Figure.\ref {15c}, the winding number corresponding to $r_+= 15.7648$ (orange colored line) are found to be $+1$. Hence, the total topological charge  equals $1$.\\
Above analysis is followed by plotting the horizon radius $r_+$  against $\tau$ for different values pressure $P$ keeping $\zeta$ value fixed at $\zeta=7.6$.The number of black hole branch along with the total topological charge is changed when pressure is slightly varied and $\zeta$ is kept fixed.Figure \ref{24} confirms the same where variation of pressure is shown and $\zeta$ is kept fixed at $\zeta=7.6$.It is seen that when the pressure is lowered from $P=0.01$ to $P=0.001$ then the number of branch(Figure \ref{24a}) and total topological remains the same(Figure \ref{24c}) .When pressure is slightly increased to $P=0.05$ then total topological charge becomes zero. place(Figure \ref{24d}).However the total topological charge remains the same i.e $1$(Figure \ref{40f}).Variations of the topological charge and number of black hole branches with $\zeta$ and $P$ are shown in Table \ref{t8}


\begin{table}[ht]
	\caption{Summary of results for $5D$ HL black hole with hyperbolic horizon in fixed $\zeta$ ensemble.} 
	\centering 
	\begin{tabular}{|c| c |c| c|} 
		\hline 
		$\zeta$ & $P$ & No of black hole branches & Topological charge  \\ [0.5ex] 
		\hline 
		7.2 & 0.035 & 1 & 1 \\ 
		7.2 & 0.096 & 2 & -1+1=0 \\
		7.8 & 0.008 & 2 & -1+1=0 \\
		7.8 &  0.07 & 1  & 1   \\
		8 & 0.0058 & 2 & -1+1=0   \\
		8 & 0.05 & 1& 1    \\
		      [1ex] 
		\hline 
	\end{tabular}
	\label{t8} 
\end{table}
\newpage

\subsubsection{\textbf{Case II : For D=4}}

For 4D HL black hole, we put $k=-1$ in the expression \ref{phi4} and \ref{eqsz4} which gives
\begin{equation}
	\phi^r=\frac{4 \pi  P \left(12 r_+^2 \tau -8 \pi  r_+ \phi ^2+\tau  \zeta ^2\right)}{3 \tau }+\frac{3 (8 \pi  r-\tau )}{16 \pi  P r_+^2 \tau }-\frac{64 \pi ^2 P r_+ \zeta ^2 \log (r_+)}{3 \tau }-\frac{8 \pi  r_+}{\tau }-2
\end{equation}
and
\begin{equation}
	\tau=\frac{8 \pi  r \left(64 \pi ^2 P^2 r_{+}^2 \zeta ^2+128 \pi ^2 P^2 r_{+}^2 \zeta ^2 \log (r_{+})+48 \pi  P r_{+}^2-9\right)}{64 \pi ^2 P^2 r_{+}^2 \left(12 r_{+}^2+\zeta ^2\right)-96 \pi  P r_{+}^2-9}
\end{equation} 

\begin{figure}[h]
	\centering
	\begin{subfigure}{0.32\textwidth}
		\includegraphics[width=\linewidth]{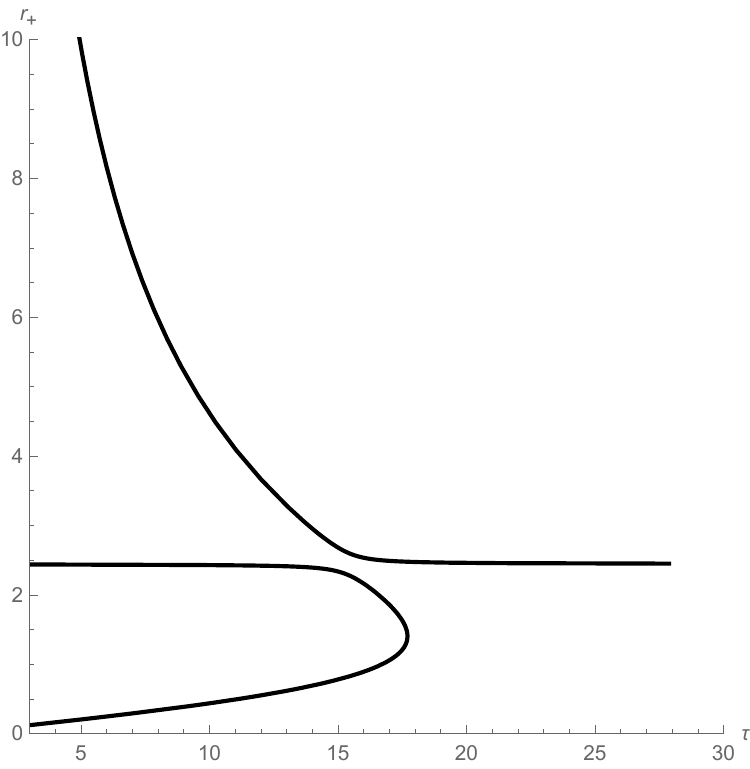}
		\caption{}
		\label{25a}
	\end{subfigure}
	\begin{subfigure}{0.32\textwidth}
		\includegraphics[width=\linewidth]{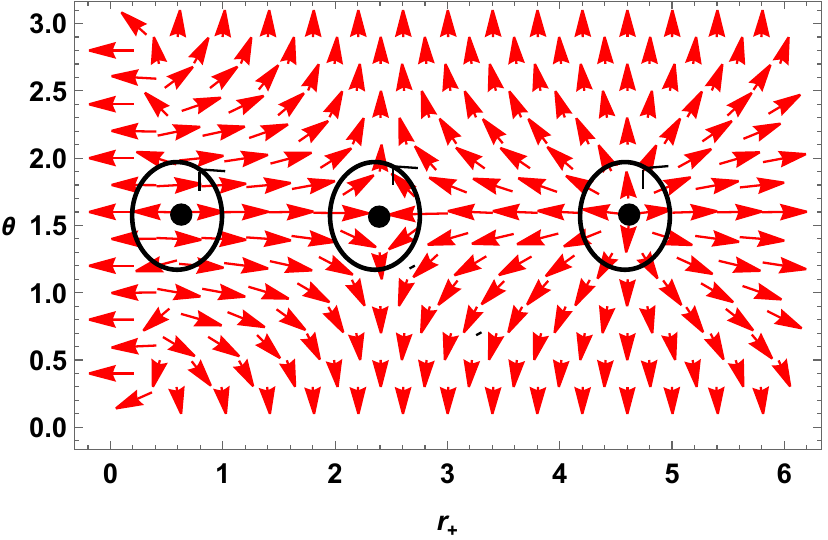}
		\caption{}
		\label{25b}
	\end{subfigure}
	\begin{subfigure}{0.32\textwidth}
		\includegraphics[width=\linewidth]{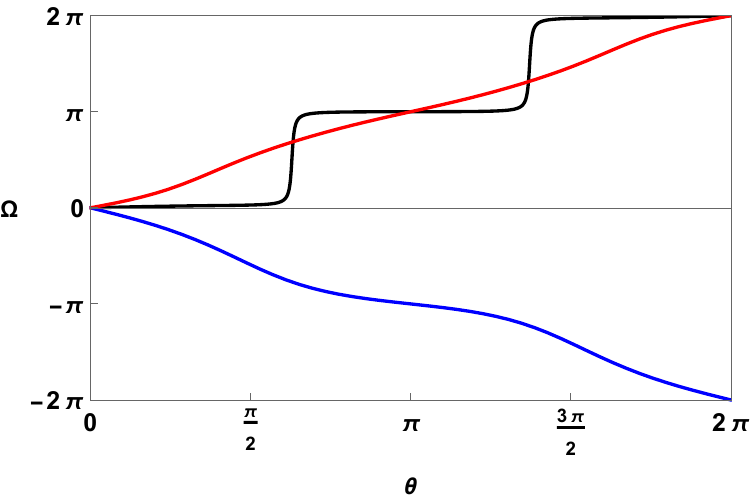}
		\caption{}
		\label{25c}
	\end{subfigure}
	
	\caption{Plots for $4D$ HL black hole with hyperbolic horizon in fixed $\zeta$ ensemble.Here $\zeta=0.2 $ with $ P=0.01$. Figure $\left(a\right)$ and $\left(d\right)$ shows $\tau$ vs $r_+$ plot for two allowed range of values of $r_+$ where temperature is positive.Figure $\left(b\right)$ is the plot of vector field $n$ on a portion of $r_+-\theta$ plane for $\tau=10$ for both the allowed range.The zero points are located at $r_+=0.5934,2.3551$ and   $r_+=4.581$ for lower and upper allowed region respectively. In figure $\left(c\right)$ computation of the contours around the zero points for $r_+=0.5934,2.3551$ and   $r_+=4.581$ are shown in black,blue and red colored solid lines respectively.}
	\label{25}
\end{figure}

\begin{figure}[h]
	\centering
	\begin{subfigure}{0.32\textwidth}
		\includegraphics[width=\linewidth]{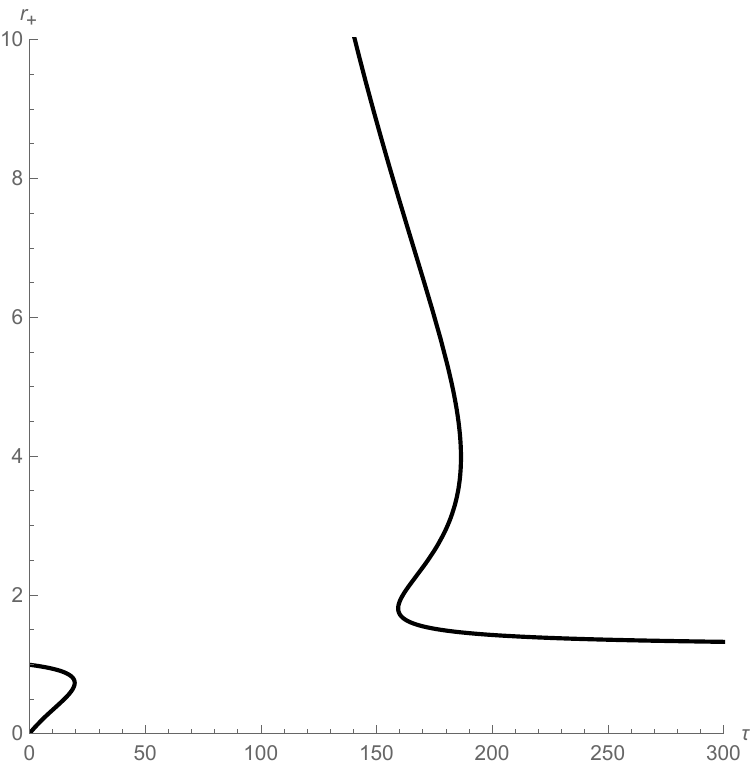}
		\caption{}
		\label{26a}
	\end{subfigure}
\begin{subfigure}{0.32\textwidth}
	\includegraphics[width=\linewidth]{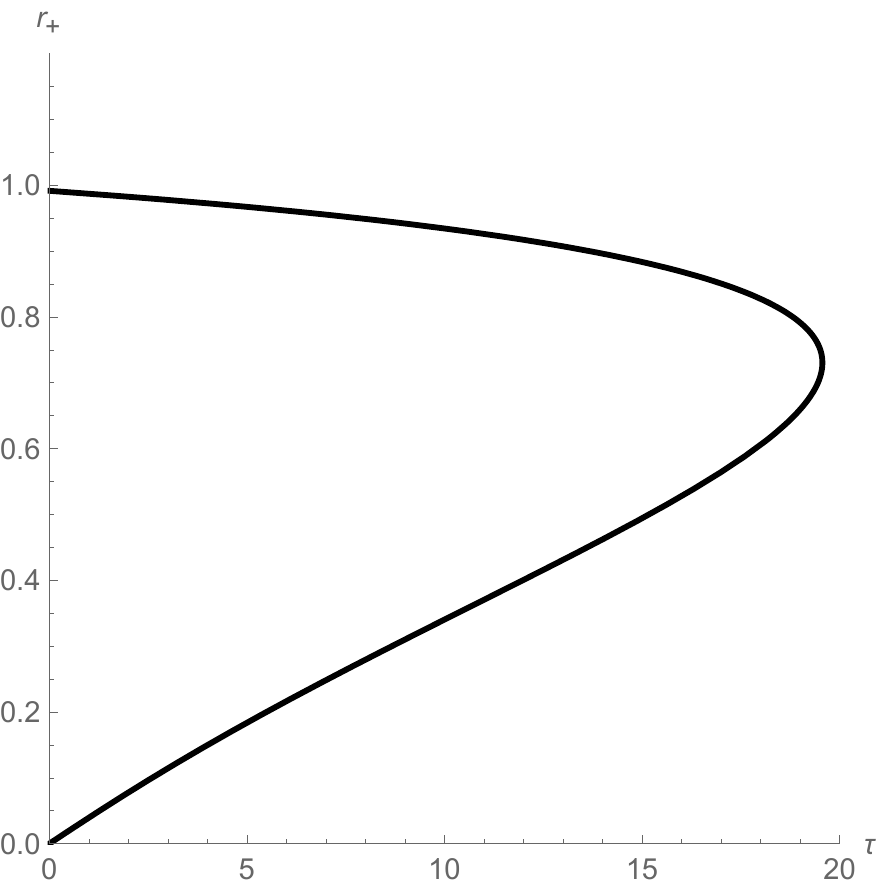}
	\caption{}
	\label{26b}
\end{subfigure}
	\begin{subfigure}{0.32\textwidth}
		\includegraphics[width=\linewidth]{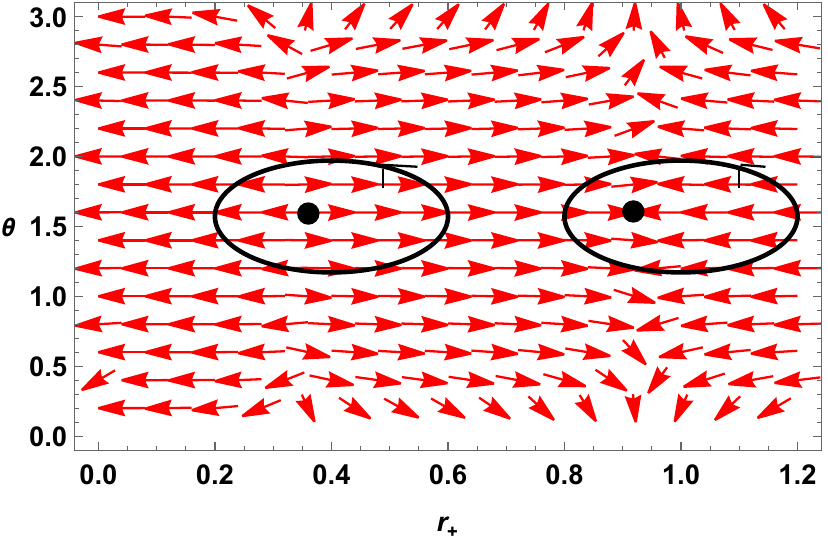}
		\caption{}
		\label{26c}
	\end{subfigure}
\medskip
\begin{subfigure}{0.32\textwidth}
	\includegraphics[width=\linewidth]{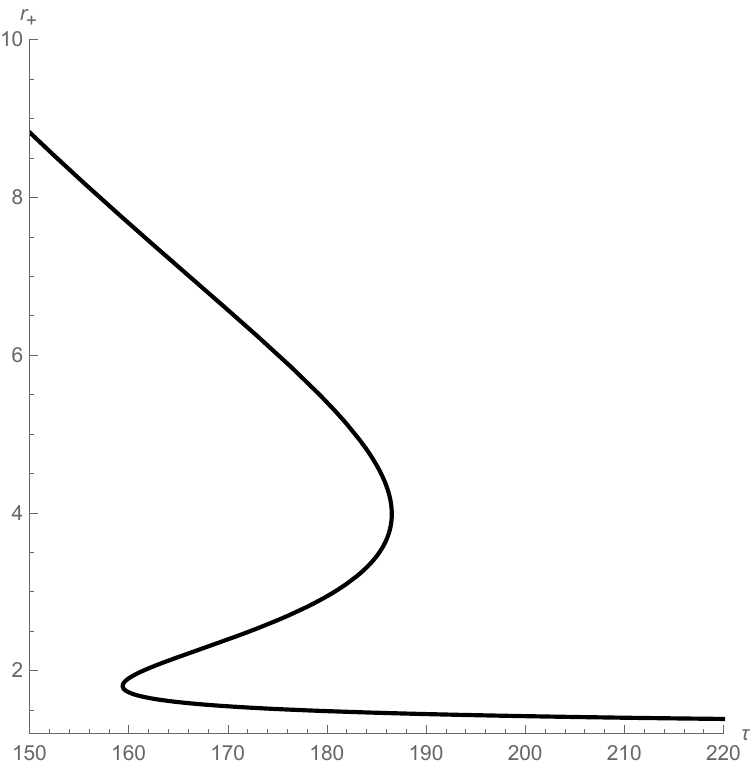}
	\caption{}
	\label{26d}
\end{subfigure}
\begin{subfigure}{0.32\textwidth}
	\includegraphics[width=\linewidth]{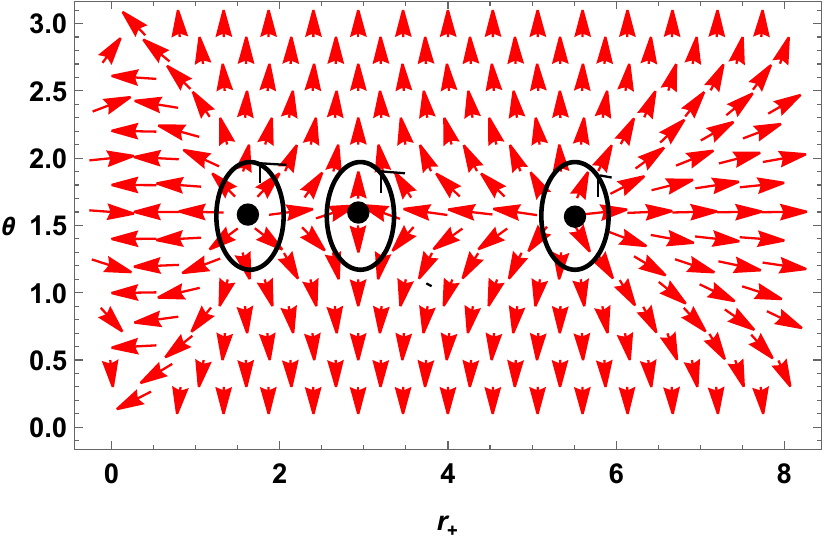}
	\caption{}
	\label{26e}
\end{subfigure}
	\begin{subfigure}{0.32\textwidth}
		\includegraphics[width=\linewidth]{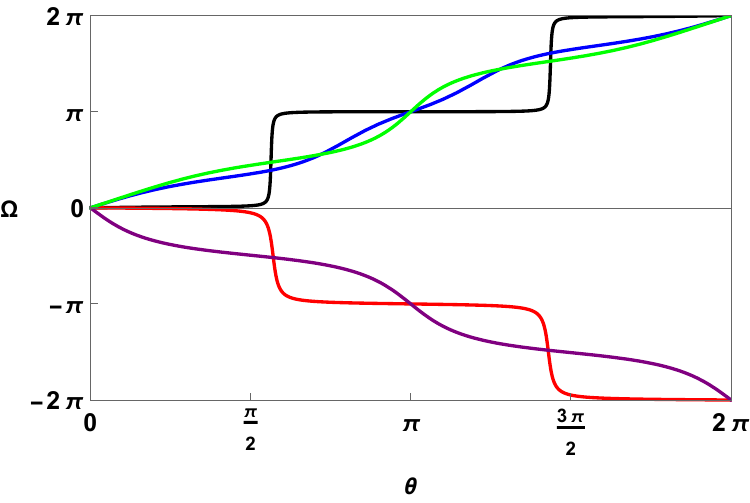}
		\caption{}
		\label{26f}
	\end{subfigure}
	\caption{ Plots for $4D$ HL black hole with hyperbolic horizon in fixed $\zeta$ ensemble,at $\zeta=11.1 $ with $ P=0.01$.Figure $\left(a\right)$  is the $\tau$ vs $r_+$ plot for the two allowed region where temperature is positive. Figure $\left(b\right)$ shows $\tau$ vs $r_+$ plot for the lower allowed branch($0\leq r_+\leq1.2$) and figure (c) is the corresponding vector field $n$ on a portion of $r_+-\theta$ plane for $\tau=10$ . Figure $\left(d\right)$ shows $\tau$ vs $r_+$ plot for the upper allowed branch($1.2\leq r_+\leq 300$) and figure (d) is the corresponding vector field $n$ on a portion of $r_+-\theta$ plane for $\tau=180$. In figure $\left(e\right)$, computation of the contours around the zero points are shown in different colored solid lines.Red and black colored solid line corresponds to figure (c) and rest are related to figure (e)}
	\label{26}
\end{figure}

\begin{figure}[h]
	\centering
	\begin{subfigure}{0.32\textwidth}
		\includegraphics[width=\linewidth]{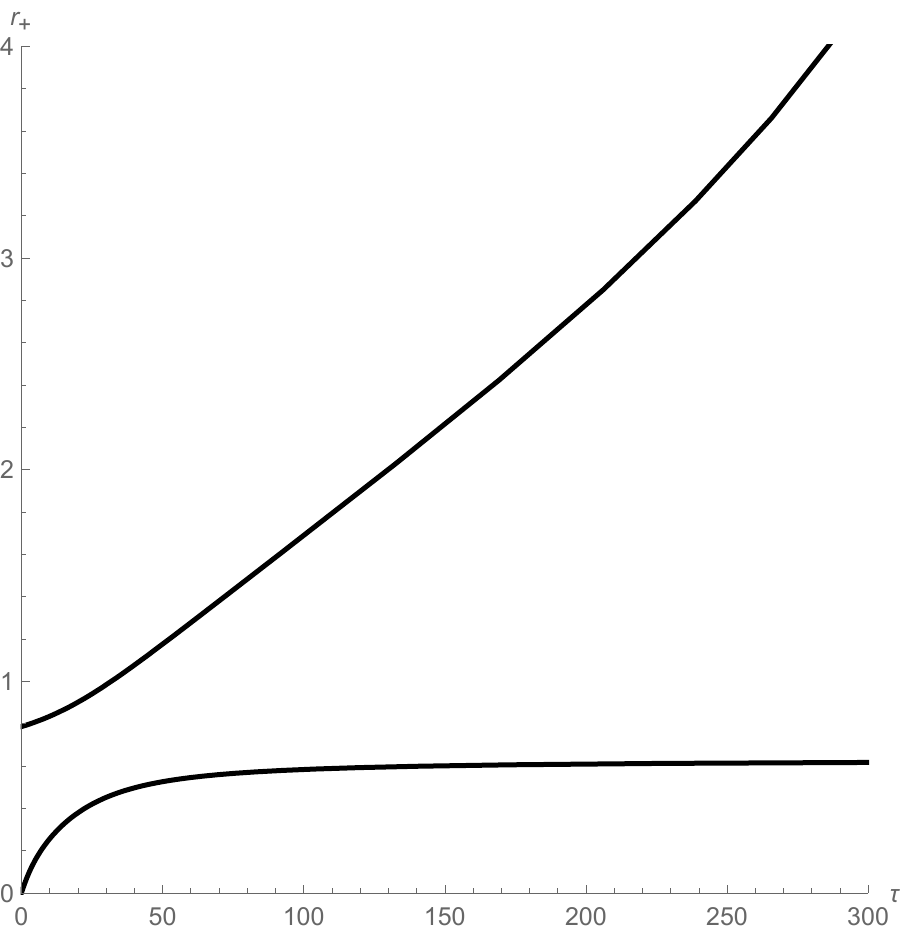}
		\caption{}
		\label{27a}
	\end{subfigure}
	\begin{subfigure}{0.32\textwidth}
		\includegraphics[width=\linewidth]{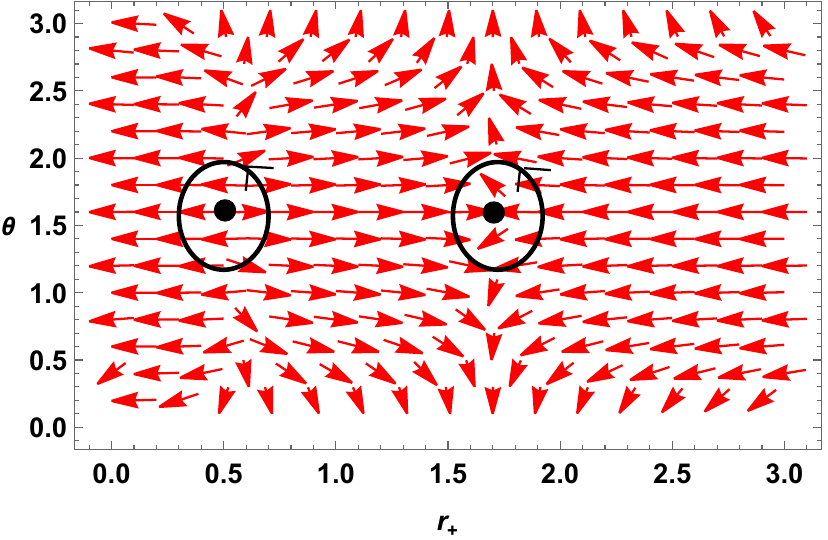}
		\caption{}
		\label{27b}
	\end{subfigure}
	\begin{subfigure}{0.32\textwidth}
		\includegraphics[width=\linewidth]{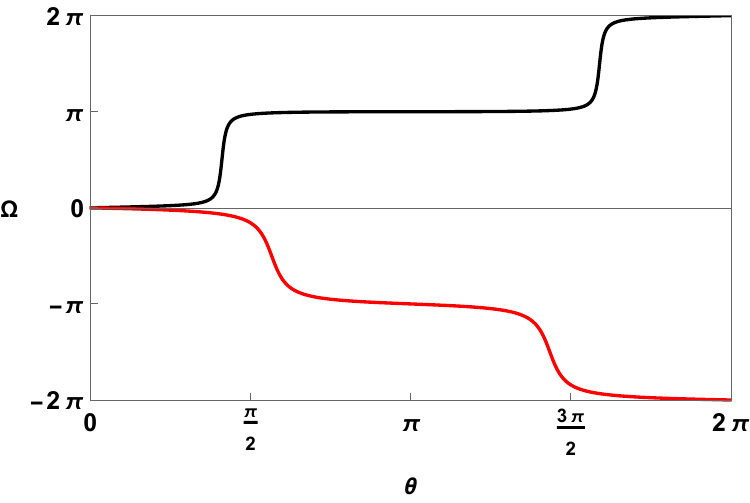}
		\caption{}
		\label{27c}
	\end{subfigure}
	
	\caption{Plots for $5D$ HL black hole with hyperbolic horizon in fixed $\zeta$ ensemble.Here $\zeta=7.6 $ is kept fixed and pressure is varied.In figure $\left(a\right)$ and $\left(d\right)$ shows $\tau$ vs $r_+$ plot for pressure $P=0.001$ and $P=0.05$ respectively.Figure $\left(b\right)$ and figure $\left(e\right)$ is the plot of vector field $n$ on a portion of $r_+-\theta$ plane .For figure $\left(b\right)$ $\tau$ value is taken to be $\tau=50$ and the zero points are located at $r_+=10.7482$ and   $r_+=15.1483$ for lower and upper allowed region respectively. In figure $\left(c\right)$ computation of the contours around the zero points for $r_+=0.0.7118$ and   $r_+=15.6966$ are shown in black and red colored solid lines respectively.For figure $\left(e\right)$ $\tau$ value is taken to be $\tau=1.5$ and the zero point is located at $r_+=13.1782$ .In figure $\left(f\right)$ computation of the contours around the zero points for $r_+=13.1782$ is shown.}
	\label{27}
\end{figure}

For 4D HL black hole  with hyperbolic horizon in fixed $\zeta$ ensemble,we do not get any critical conditions hence there is no superfluid $\lambda$ phase transition observed.While calculating winding number from the $\tau$ vs $r_+$ curve for a particular value of $\zeta$ the positive temperature condition needs to be keep in mind in this case also.It is being observed that at a fixed pressure $P=0.01$ in case of 4D HL black hole with hyperbolic horizon in fixed $\zeta$ ensemble have two types of topological charge $0$ and $1$.For very high value of $\zeta$ we observe topological charge is equal to zero.Two of such scenario where topological charge is one,is shown in figure \ref{25} and figure \ref{26} where the horizon radius $r_+$ is plotted  against $\tau$  for $\zeta=0.2$ and $\zeta=11.1$ respectively keeping  pressure constant at $P=0.01$\\
In Figure.\ref{25a} $\tau$ vs $r_+$ plot is represented for two different allowed range of $r_+$ for same set of values of $\zeta$ and $P$.For the case of $\tau=10$, there will three zero points of the vector field $n$.Figure.\ref {25c}  suggests that the  sum of the winding numbers of the two discontinous branches gives us the total topological charge of the black hole, which in this case equals $1-1+1=1$.
An interesting case is represented in figure.\ref{26a} where $\tau$ vs $r_+$ plot is represented for two different allowed range of $r_+$ for $\zeta=11.1$(which is beyond our considered range) and $P=0.01$.As figure \ref{26a} suggests in one of the two region where temperature is positive$(1.2 \leq r_+ \leq 300)$ we observe two black hole branches.For $\tau=10$, there will be two zero points on the vector field $n$ for this case(figure \ref{26c}).Figure.\ref {26f} suggests that the  sum of the winding numbers for this region gives us the total topological charge equals to $1-1=0$.On the other allowed region we observe three black hole branches i.e small,intermediate and large black hole branches. For $\tau=180$, there will three zero points on the vector field $n$ for this case(figure \ref{26e}).Figure.\ref {26f} suggests that the  sum of the winding numbers for this region gives us the total topological charge equals to $1-1+1=1$.
We now consider a higher range of values of $zeta$ for the same value of pressure. Taking $\zeta=20$,horizon radius $r_+$ is plotted  against $\tau$ considering the allowed range for $r_+$ in Figure.\ref {27a}. Here, we observe two black hole branches.For $\tau=100$, the contour integration is operated around the points $r_+= 0.5$ and $r_+=1.75$. As shown in Figure.\ref {27c}, the winding number corresponding to $r_+= 0.5$ (black colored line) are found to be $+1$ and the winding number corresponding to $r_+= 1.75$ (red colored line) are found to be $-1$  . Hence, the total topological charge  equals $1-1=0$.\\

\newpage
Above analysis is followed by investigating the effect of variation of pressure pressure on the number of black hole branch along with the total topological charge.In this case we see similar pattern as that of 5D HL black hole with hyperbolic horizon which is : when the pressure is lowered ,the number of branch and total topological remains the same and when pressure is increased sufficiently then total topological charge becomes zero.Variations of the topological charge and number of black hole branches with $\zeta$ and $P$ are shown in Table \ref{t9}.
\newpage
\begin{table}[ht]
	\caption{Summary of results for $4D$ HL black hole with hyperbolic horizon in fixed $\zeta$ ensemble.} 
	\centering 
	\begin{tabular}{|c| c |c| c|} 
		\hline 
		$\zeta$ & $P$ & No of black hole branches & Topological charge  \\ [0.5ex] 
		\hline 
		0.2 & 0.01 & 3 & 1-1+1=1 \\ 
		0.2 & 0.05 & 2 & 1-1+1=1 \\
		0.2 & 1 & 2 & -1+1=0 \\
		11.1 &  0.01 & 5 & 1-1+1-1+1=1   \\
		11.1 & 0.005 & 2 & -1+1=0   \\
		14    & 0.01&4&-1+1-1+1=0\\
		[1ex] 
		\hline 
	\end{tabular}
	\label{t9} 
\end{table}

	\section{Conclusion}        
	In this paper, we studied the thermodynamic topology of four and five dimensional Horava Lifshitz (HL) black holes, also referred to as superfluid black holes. In our analysis, the Horava Lifshitz (HL) black holes are treated as topological defects in their thermodynamic spaces. The winding numbers at those defects were calculated in order to understand the topology of these spaces. We primarily focused our study in two different ensembles: fixed $\epsilon$ ensemble and fixed $\zeta$ ensemble, where $\epsilon$ is a parameter of the HL black holes and $\zeta$ is its conjugate parameter. In the fixed $\epsilon$ ensemble, three different horizon types were worked out: the spherical horizon for $k=+1$, the flat horizon for $k=0$, and  the hyperbolic horizon for $k=-1$. In the fixed $\zeta$ ensemble, two different horizon types were worked out : the spherical horizon for $k=+1$, and  the hyperbolic horizon for $k=-1$. Fixed $\zeta$ ensemble could not be defined in case of flat horizon with $k=0$. \\

We find that in the fixed $\epsilon$ ensemble for 5D HL black holes with spherical horizon, the total topological charge is 1 for $\epsilon<1$, and it becomes 0 when $\epsilon \geq 1$. Notably, a superfluid $\lambda$ transition is observed at $\epsilon=0.9428$. In the $0\leq\epsilon<1$ range, variations in the number of branches in the $\tau-r$ curve are found. For $\epsilon<0.9428$, only one branch is visible, while for $\epsilon > 0.9428$, three branches are observed. However, the total topological charge remains unchanged and remains +1 for both $\epsilon<0.9428$ and $\epsilon>0.9428$ cases. In the General Relativity (GR) limit ($\epsilon=1$) and for values $\epsilon \geq 1$, the total topological charge is 0.\\

Similar observations are made for 4D HL black holes with a spherical horizon in the fixed $\epsilon$ ensemble. For $\epsilon<0.9785$, one branch is observed, whereas for $\epsilon$ greater than $0.9785$ but less than $1$, three branches appear. The total topological charge is +1 for both $\epsilon<0.9785$ and $\epsilon>0.9785$ cases. In the GR limit ($\epsilon=1$) and for values $\epsilon \geq 1$, the total topological charge is 0.\\

For both 5D and 4D HL black holes with a flat horizon ($k=0$), the total topological charge is always 1 for all values of $\epsilon$ and $P$.\\

In the case of a 5D HL black hole with a hyperbolic horizon, the total topological charge is zero for $\epsilon<1$ at all values of $P$. In the GR limit, the total topological charge changes to 1.\\

However, for a 4D HL black hole with a hyperbolic horizon, the total topological charge is always 1 (even in the GR limit) at all values of $P$.\\

One notable finding in the fixed $\epsilon$ ensemble is that the change in pressure has no significant impact on determining the overall thermodynamic topology. It primarily depends on the parameter $\epsilon$.\\

Moving to the fixed $\zeta$ ensemble, for a 5D black hole with a spherical horizon, the total topological charge is either +1 or 0. Interestingly, pressure $P$ becomes a crucial parameter in determining the overall thermodynamic topology of HL black holes in the fixed $\zeta$ ensemble. For a fixed pressure, we observed a range of $\zeta$ values resulting in a topological charge of $1$ and another range resulting in a charge of 0. Exploring further, we observed that, for the same $\zeta$ value, increasing pressure led to changes in the total topological charge. The number of branches in the $\tau-r_{+}$ curve also changes based on the values of $P$ and $\zeta$, but the total topological number remains either 1 or 0.\\

In contrast to its $5D$ counterpart, $4D$ HL black hole with a spherical horizon consistently maintains a total topological charge of $+1$ at all values of $\zeta$ and pressure. We did not observe the superfluid $\lambda$ phase transition property in this ensemble.\\

In both 5D and 4D HL black holes with a hyperbolic horizon in the fixed $\zeta$ ensemble, the total topological charge takes values $1$ or $0$, depending on the combination of $\zeta$ and $P$ values.\\

Hence, from our findings, we conclude that, the thermodynamic topology of $D=4,5$ HL black holes is highly influenced by the type of horizon, choice of ensemble, pressure, and the parameters, $\epsilon$ or $\zeta$. The roles of thermodynamic parameters and pressure in determining the thermodynamic topology of HL black holes change based on the ensemble we work in and the space time dimension of the black hole.\\

Extending this work to other black hole systems and modified gravity theories could yield more generalized conclusions regarding the relationship between topological number and horizon topology in black holes. We plan to address these issues in our future work.

\end{document}